\pdfoutput=1

\documentclass[11pt,twoside,a4paper,cmspaper,final,collab]{cms-tdr}
\def\svnVersion{471839P}\def\cmsCernNoTag{CERN-EP-2017-286}\def\cmsCernDate{\today}\def\cmsMessage{Published in the Journal of High Energy Physics as \href{http://dx.doi.org/10.1007/JHEP08(2018)011}{\doi{10.1007/JHEP08(2018)011.}}}

\begin{document}\cmsNoteHeader{TOP-17-005}

\hyphenation{had-ron-i-za-tion}
\hyphenation{cal-or-i-me-ter}
\hyphenation{de-vices}
\RCS$Revision: 463933 $
\RCS$HeadURL: svn+ssh://svn.cern.ch/reps/tdr2/papers/TOP-17-005/trunk/TOP-17-005.tex $
\RCS$Id: TOP-17-005.tex 463933 2018-06-09 14:58:46Z alverson $
\newlength\cmsFigWidth
\ifthenelse{\boolean{cms@external}}{\setlength\cmsFigWidth{0.85\columnwidth}}{\setlength\cmsFigWidth{0.4\textwidth}}
\ifthenelse{\boolean{cms@external}}{\providecommand{\cmsLeft}{top\xspace}}{\providecommand{\cmsLeft}{left\xspace}}
\ifthenelse{\boolean{cms@external}}{\providecommand{\cmsRight}{bottom\xspace}}{\providecommand{\cmsRight}{right\xspace}}
\newlength\cmsTabSkip\setlength{\cmsTabSkip}{2ex}

\newcommand{\MT}{\ensuremath{M_\text{T}}\xspace}
\newcommand{\Njets}{\ensuremath{N_\text{j}}\xspace}
\newcommand{\Nbjets}{\ensuremath{N_\text{b}}\xspace}
\newcommand{\mll}{\ensuremath{M(\ell\ell)}\xspace}
\newcommand{\mZ}{\ensuremath{M(\cPZ)}\xspace}
\newcommand{\WZ}{\ensuremath{\PW\cPZ}\xspace}
\newcommand{\ttZ}{\ensuremath{\ttbar\cPZ}\xspace}
\newcommand{\ttVV}{\ensuremath{\ttbar\mathrm{VV}}\xspace}
\newcommand{\ttX}{\ensuremath{\cPqt(\cPaqt)\mathrm{X}}\xspace}
\newcommand{\ttW}{\ensuremath{\ttbar\PW}\xspace}
\newcommand{\ttH}{\ensuremath{\ttbar\PH}\xspace}
\newcommand{\ptCor}{\ensuremath{\pt^\text{cor}}\xspace}
\newcommand{\BDT}{\ensuremath{\textit{D}}\xspace}
\newcommand{\cuW}{\ensuremath{\bar{c}_{\text{uW}}}\xspace}
\newcommand{\cuG}{\ensuremath{\bar{c}_{\text{uG}}}\xspace}
\newcommand{\cuB}{\ensuremath{\bar{c}_{\text{uB}}}\xspace}
\newcommand{\cHu}{\ensuremath{\bar{c}_{\text{Hu}}}\xspace}
\newcommand{\cH}{\ensuremath{\bar{c}_{\PH}}\xspace}
\newcommand{\tcthreeG}{\ensuremath{\widetilde{c}_{\text{3G}}}\xspace}
\newcommand{\cthreeG}{\ensuremath{\bar{c}_{\text{3G}}}\xspace}
\newcommand{\ctwoG}{\ensuremath{\bar{c}_{\text{2G}}}\xspace}
\newcommand{\OuW}{\ensuremath{\mathcal{O}_{\text{uW}}}\xspace}
\newcommand{\OuG}{\ensuremath{\mathcal{O}_{\text{uG}}}\xspace}
\newcommand{\OuB}{\ensuremath{\mathcal{O}_{\text{uB}}}\xspace}
\newcommand{\OHu}{\ensuremath{\mathcal{O}_{\text{Hu}}}\xspace}
\newcommand{\OH}{\ensuremath{\mathcal{O}_{\PH}}\xspace}
\newcommand{\OtthreeG}{\ensuremath{\mathcal{O}_{\widetilde{\text{3G}}}}\xspace}
\newcommand{\OthreeG}{\ensuremath{\mathcal{O}_{\text{3G}}}\xspace}
\newcommand{\OtwoG}{\ensuremath{\mathcal{O}_{\text{2G}}}\xspace}

\cmsNoteHeader{TOP-17-005}
\title{Measurement of the cross section for top quark pair production in association with a W or Z boson in proton-proton collisions at \texorpdfstring{$\sqrt {s}=13\TeV$}{sqrt(s) = 13 TeV}}

\date{\today}

\abstract{
A measurement is performed of the cross section of top quark pair production in
association with a $\PW$ or $\PZ$ boson using
proton-proton collisions at a center-of-mass energy of 13\TeV
at the LHC.  The data sample corresponds to an integrated
luminosity of 35.9\fbinv, collected by the CMS
experiment in 2016. The measurement is performed in the same-sign dilepton,
three- and four-lepton final states. The production cross sections are
measured to be $\sigma(\ttW)=
0.77^{+0.12}_{-0.11}\stat^{+0.13}_{-0.12}\syst\unit{pb}$
and $\sigma(\ttZ)=0.99^{+0.09}_{-0.08}\stat^{+0.12}_{-0.10}\syst\unit{pb}$. The expected (observed) signal
significance for the \ttW production in same-sign dilepton channel is found to be 4.5\,(5.3) standard deviations, while for the \ttZ production in three- and four-lepton channels both
the expected and the observed significances are found to be in excess of 5 standard deviations.
The results are in agreement with the standard model
predictions and are used to constrain the Wilson
coefficients for eight dimension-six operators describing new interactions that would modify
\ttW and \ttZ production.}

\hypersetup{
pdfauthor={CMS Collaboration},
pdftitle={Measurement of the cross section for top quark pair production in association with a W or Z boson in proton-proton collisions at sqrt(s) = 13 TeV},
pdfsubject={CMS},
pdfkeywords={CMS, physics, top}}

\maketitle

\section{Introduction}
\label{sec:Introduction}
\begin{figure}[!hbtp]
\centering
\includegraphics[width=0.4\textwidth]{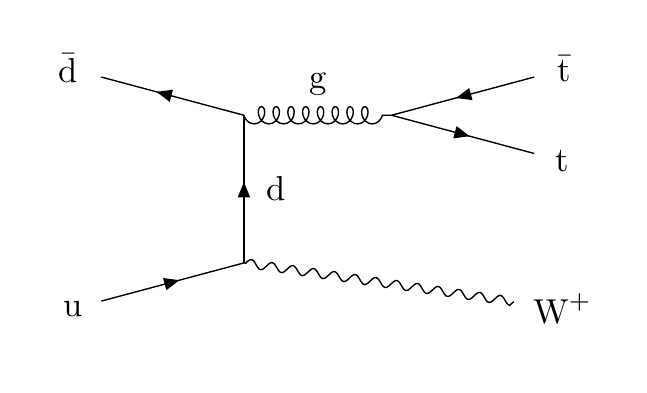}
\includegraphics[width=0.4\textwidth]{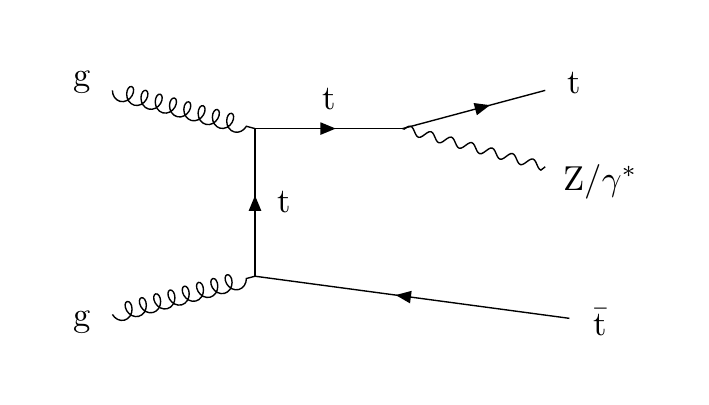}
\caption{Representative leading-order Feynman diagrams for $\ttW$ and $\ttZ$ production at the LHC.}
\label{fig:ttV_Fey}
\end{figure}

 The 13\TeV center-of-mass energy of proton-proton (pp) collisions at the LHC opens the possibility for studying the processes at larger mass scales than previously explored in the laboratory. The top quark-antiquark pair (\ttbar) produced in association with a $\PW$ (\ttW) or $\PZ$ (\ttZ) boson is among the most massive signatures that can be studied with high precision.
The theoretical cross sections at next-to-leading order (NLO) in quantum chromodynamics (QCD) for \ttW and \ttZ production at $\sqrt{s} = 13\TeV$ are about 3--4 times higher than those at 8 TeV~\cite{deFlorian:2016spz}. This, coupled with the higher integrated luminosity collected at  13\TeV collisions, allows for a much more accurate study of these processes. Precise measurements of the production cross section for \ttW and \ttZ  are of particular interest
because these topologies can receive sizeable contributions from new physics
(NP) beyond the standard model (SM)~\cite{Bylund2016,Englert2016}. Furthermore, these processes form dominant backgrounds to several searches for NP, as well as to the measurements of SM processes, such as \ttbar production in association with the Higgs boson (\ttH). In addition, \ttZ production is the most sensitive process for directly measuring the coupling of the top quark to the $\PZ$ boson. The Feynman diagrams for the dominant production mechanisms of these processes are shown in Fig.~\ref{fig:ttV_Fey}, to which the charge-conjugate states should be added.

The \ttZ cross section was measured by the CMS collaboration at $\sqrt{s} = 7\TeV$ with
a precision of $\approx$ 50\%~\cite{PRL-110-172002}.
At $\sqrt{s} = 8\TeV$ CMS used multivariate techniques in events containing two, three, or four charged leptons to measure the \ttW and \ttZ cross sections with a precision of 30 and 25\%, respectively~\cite{EPJC-C74-2014-9, JHEP-1601-2016-096}. The \ttZ process was observed with a significance of 6.4 standard deviations, and evidence for \ttW production was found with a significance of 4.8 standard deviations.
The ATLAS Collaboration analyzed events containing two and three charged leptons for its \ttW measurement, and using two, three, and four charged leptons for the \ttZ channel, achieving a similar precision \cite{Aad:2015eua}.
In a more recent publication, the ATLAS Collaboration reported the first measurement of the \ttW and \ttZ production cross sections at $\sqrt{s} = 13\TeV$~\cite{ATLAS:2016ttVarticles} with a significantly smaller data set than the one considered here.

In this paper we present measurements of the $\ttZ$ and $\ttW$ production cross sections at $\sqrt{s} = 13\TeV$ with a data set corresponding to an integrated luminosity of 35.9\fbinv.
The measurements are performed using events in which at least one of the $\PW$ bosons, originating from a top quark decay, further decays to a charged lepton and a neutrino, and the associated $\PW$ or $\PZ$ boson decays to a charged lepton and a neutrino or a charged lepton pair, where the charged lepton ($\ell$) refers to an electron or a muon. The contribution from $\tau$ leptons are included through their decays to electrons and muons.
The analysis is performed in three exclusive final states, in which events with two leptons of same charge, denoted as same-sign (SS) dileptons, are used to extract the \ttW signal, while events with  three or four charged leptons that include a lepton pair of opposite charge and same flavor (OSSF) are used to measure the \ttZ signal yield. In addition to the individual \ttW and \ttZ cross section measurements, a fit is performed in all three final states to simultaneously extract these cross sections. Furthermore, the results are interpreted in the context of an effective field theory to constrain the Wilson coefficients~\cite{PhysRev.179.1499}, which parameterize the strength of new physics interactions, for a set of selected dimension-six operators that might signal the presence of NP contributions in \ttW and \ttZ production.

\label{sec:cms}
\section{The CMS detector}

The central feature of the CMS apparatus is a superconducting solenoid of 6\unit{m} internal diameter, providing a magnetic field of 3.8\unit{T}. Within the solenoid volume are a silicon pixel and strip tracker, a lead tungstate crystal electromagnetic calorimeter (ECAL), and a brass and scintillator hadron calorimeter (HCAL), each composed of a barrel and two endcap sections. Forward calorimeters extend the pseudorapidity ($\eta$) coverage provided by the barrel and endcap detectors. Muons are detected in gas-ionization chambers embedded in the steel magnetic flux-return yoke outside the solenoid.  A more detailed description of the CMS detector, together with a definition of the coordinate system used and the relevant kinematic variables, can be found in Ref.~\cite{Chatrchyan:2008zzk}. Events of interest are selected using a two-tiered trigger system~\cite{Khachatryan:2016bia}. The first level, composed of custom hardware processors, uses information from the calorimeters and muon detectors to select events, while the second level
selects events by running a version of the full event reconstruction software optimized for fast processing on a farm of computer processors.

\section {Event and object selection}
\label{sec:objects}

\label{sec:Samples}
Events are selected by online triggers that require the presence of at least one electron or muon, with transverse momentum, \pt, greater than 27 or 24\GeV, respectively. The selection efficiencies for the signal and background events that pass all requirements are found to be greater than 95 and 98\% for the dilepton analysis and for the three- and four-lepton analyses, respectively.

The Monte Carlo (MC) simulations are used to estimate some of the backgrounds, as well as to calculate the selection efficiencies for the \ttZ and \ttW signal events.
The simulated events for the $\PW \Pgg^{*}$, $\PW \PW$, $\cPqt \PW \PZ$, and for pairs of top quarks associated with a pair of bosons (\ttVV, where $\text{V} = \PW$, $\PZ$, or $\PH$) processes, are performed at leading order (LO) in QCD, and
for \ttZ, \ttW, $\cPqt\PZ \Pq$, $\cPqt\PH \Pq$, $\cPqt\PH \PW$, $\PW \PZ$, $\PW \PW \PZ$, $\PW \PZ \PZ$, $\PZ\PZ \PZ$, $\ttbar\Pgg^{*}$, and $\PZ \Pgg^*$ final states at NLO in QCD using the \MGvATNLO v2.2.2 or v2.3.3~\cite{Alwall:2014hca}.
The NLO \POWHEG~v2~\cite{powheg2} generator is used for the production of the \ttH~\cite{Hartanto:2015uka} and $\qqbar$$\to \PZ\PZ$~\cite{Melia:2011tj,Nason:2013ydw} processes, while the $\Pg \Pg \to \PZ \PZ$ process is generated at LO in QCD with \MCFM v7.0~\cite{Campbell:2010ff}.
The simulated samples of $\PZ \PZ$ events are scaled to the cross sections calculated at next-to-next-to-leading order (NNLO) in QCD for $\qqbar \to \PZ \PZ$~\cite{Cascioli:2014yka} (using a scaling $K$ factor of 1.1) and for $\Pg \Pg \to \PZ \PZ$ at NLO ~\cite{Caola:2015psa} (using $K=1.7$).
The NNPDF3.0LO~\cite{Ball:2014uwa} parton distribution functions (PDFs) are used for the simulation generated at LO and the NNPDF3.0NLO~\cite{Ball:2014uwa} PDF for those generated at NLO.
Parton showering, hadronization, and the underlying event are simulated using \PYTHIA~v8.212~\cite{Sjostrand:2007gs,Sjostrand:2014zea} with the CUETP8M1 tune~\cite{Skands:2014pea,CMS-PAS-GEN-14-001}.
The double counting of the partons generated with \MGvATNLO and those with \PYTHIA is removed using
the MLM~\cite{Alwall:2007fs} and the {\sc FxFx}~\cite{Frederix:2012ps} matching schemes, in the LO and NLO generated events, respectively.
All events are processed through a simulation of the CMS detector based on \GEANTfour~\cite{Geant} and are reconstructed with the same algorithms as used for data. Simultaneous $\Pp \Pp$ collisions in the same or nearby bunch crossings, referred to as pileup (PU), are also simulated. The PU distribution used in simulation is chosen to match the one observed in the data.

The theoretical cross sections for the  \ttW and \ttZ signal processes are computed at NLO in QCD using \MGvATNLO and found to be $0.628\pm 0.082$ and $0.839\pm 0.101$\unit{pb}~\cite{deFlorian:2016spz}, respectively. These values are used to normalize the expected signal yields, as well as to rescale the measured signal strengths to obtain the final cross sections. In the calculation for \ttZ, the cross section corresponds to a phase space where the invariant mass of all pairs of leptons is required to be greater than 10\GeV.

Event reconstruction uses the CMS particle-flow (PF) algorithm~\cite{Sirunyan:2017ulk} for particle reconstruction and identification.  Because of PU, there can be far more than one collision vertex reconstructed per event. The reconstructed vertex for which the sum of the \pt of the physics objects is largest is chosen to be the primary pp interaction vertex. The physics objects here are the objects obtained by a jet finding algorithm~\cite{Cacciari:2008gp,Cacciari:fastjet1} applied to all charged tracks associated with this vertex, plus the missing transverse momentum (\ptmiss), which is computed as the magnitude of the vector sum of the \pt of all PF candidates.

Leptons are required to have $\pt>10\GeV$ and $\abs{\eta} < 2.5\,(2.4)$ for electrons (muons) and must be isolated from the other particles produced in the collision. A relative isolation parameter, $I_\text{rel}$, is determined by a cone-based algorithm.
For each electron (muon) candidate, a cone of $\Delta R = \sqrt{\smash[b]{(\Delta \eta)^2 + (\Delta \phi)^2}} = 0.3\,(0.4)$ is constructed around the track direction at the event primary vertex, where $\Delta \eta$ and $\Delta \phi$ are the respective differences in pseudorapidity and azimuthal angle (in radians) relative to the lepton track.
The scalar sum of the \pt of all PF particles within this cone is calculated, excluding the lepton candidate and any charged particle not originating from the selected primary vertex. Exclusion of such particles removes the PU contribution from the charged particles, and a correction is therefore still required for the neutral component. The average energy density deposited by neutral particles in the event, computed with the \FASTJET~\cite{Cacciari:fastjet1,Cacciari:fastjet2} program, is therefore subtracted from the neutral component to the sum of the \pt of particles in the cone. The quantity $I_\text{rel}$ is then defined as the ratio of this corrected sum to the $\pt$ of the lepton candidate. An electron candidate is selected if $I_\text{rel}<0.1$ for all three analyses, while a muon candidate is selected if $I_\text{rel}<0.25$ for the three- and four-lepton analyses, and  if $I_\text{rel}<0.15$ for the SS dilepton analysis.
Consistency of the origination of the lepton from the primary vertex is enforced by requiring lepton transverse and longitudinal displacements from
the primary vertex to be less than 0.05 and 0.1\unit{cm}, respectively. Additionally, the transverse impact parameter is required to be smaller than 4 standard deviations, where the impact parameter is the minimum spatial distance between the lepton trajectory and the primary vertex.

Jets are reconstructed by clustering PF candidates using the anti-\kt algorithm~\cite{Cacciari:2008gp} with a distance parameter $R=0.4$. The influence of PU is mitigated through a charged-hadron subtraction technique,
which removes the energy of charged hadrons not originating from the primary vertex~\cite{CMS-PAS-JME-14-001}. Jets are calibrated in simulation, and separately in data, accounting for energy deposits of neutral particles from PU and any nonlinear detector response. Calibrated jets with $\pt> 30\GeV$ and $\abs{\eta}<2.4$ are selected for the analysis. Furthermore, jets formed with fewer than three PF candidates or with electromagnetic or hadronic energy fractions greater than 99\% are vetoed. A selected jet can also overlap with selected leptons and lead thereby to some double counting. To prevent such cases, jets that are found within a cone of $\Delta R=0.4$ around any of the signal leptons are  removed from consideration.

A multivariate b tagging discriminator~\cite{Chatrchyan:2012jua,CMS-PAS-BTV-15-001} is used to identify jets that originate from the hadronization of b quarks (b jets).
The selection criteria used in this analysis gives about 1\% rate for tagging light-quark or gluon jets as b jets and a corresponding b tagging efficiency of around 70\%, depending on the jet \pt and $\eta$.

\section{Event selection}
\label{sec:eventselection}

\label{sec:TwoLepton}
\subsection{SS dilepton analysis}

We measure the production rate of \ttW events in the decay channel that yields exactly two leptons with the same charge.
Requiring the same electric charge for the two leptons retains only one third of the signal in the dilepton final state. However, this selection significantly  improves the signal-to-background ratio, as SS lepton pairs are produced in SM processes with relatively small cross sections. The main backgrounds to this analysis originate from misreconstruction effects:
misidentification of leptons from heavy-quark decays, hereafter called nonprompt leptons to distinguish them from prompt leptons originating from $\PW$ and $\PZ$ boson decays, and mismeasurement of the charge of one of the leptons in events with an oppositely charged lepton pair.

We select events with two SS leptons ($\Pgm\Pgm$, $\Pgm\Pe$, $\Pe\Pe$), requiring the $\pt$ of both leptons to be above 25\GeV. To avoid inefficiencies due to the trigger selection in the $\Pe \Pe$ channel, the electron with higher $\pt$ is  required to have $\pt > 40$\GeV.  Events containing additional leptons passing looser identification and isolation requirements are vetoed. These loose identification and isolation criteria are the same as used to estimate the nonprompt background in data (see Section~\ref{sec:backgrounds}). The invariant mass of the two leptons must be greater than 12\GeV to suppress Drell--Yan (DY) and quarkonium processes. To suppress $\PZ\to\EE$ events, the invariant mass of the two electrons is required to lie outside the 15\GeV window around the $\PZ$ boson mass $\mZ$~\cite{PDG2016}, followed by the requirement that $\ptmiss > 30\GeV$.

In order to distinguish these backgrounds from the signal, a multivariate analysis (MVA) has been developed. The MVA has been trained using the \ttW signal and the main background process, using events with at least two jets, one or more of which are identified as b jets.  Among the observables examined as inputs to the MVA training, the following are found to provide the best discrimination between the signal and background: the number of jets, \Njets, the number of b jets, \Nbjets, the scalar sum of \pt of the jets, \HT, \ptmiss, the highest-\pt (leading) and the lowest-\pt (trailing) lepton \pt, the invariant mass calculated using \ptmiss and \pt of each lepton, \MT, the leading and next-to-highest-\pt (subleading) jet \pt, and the separation $\Delta R$ between the trailing lepton and the nearest selected jet.

\begin{figure}
        \centering
                \includegraphics[width=0.45\textwidth]{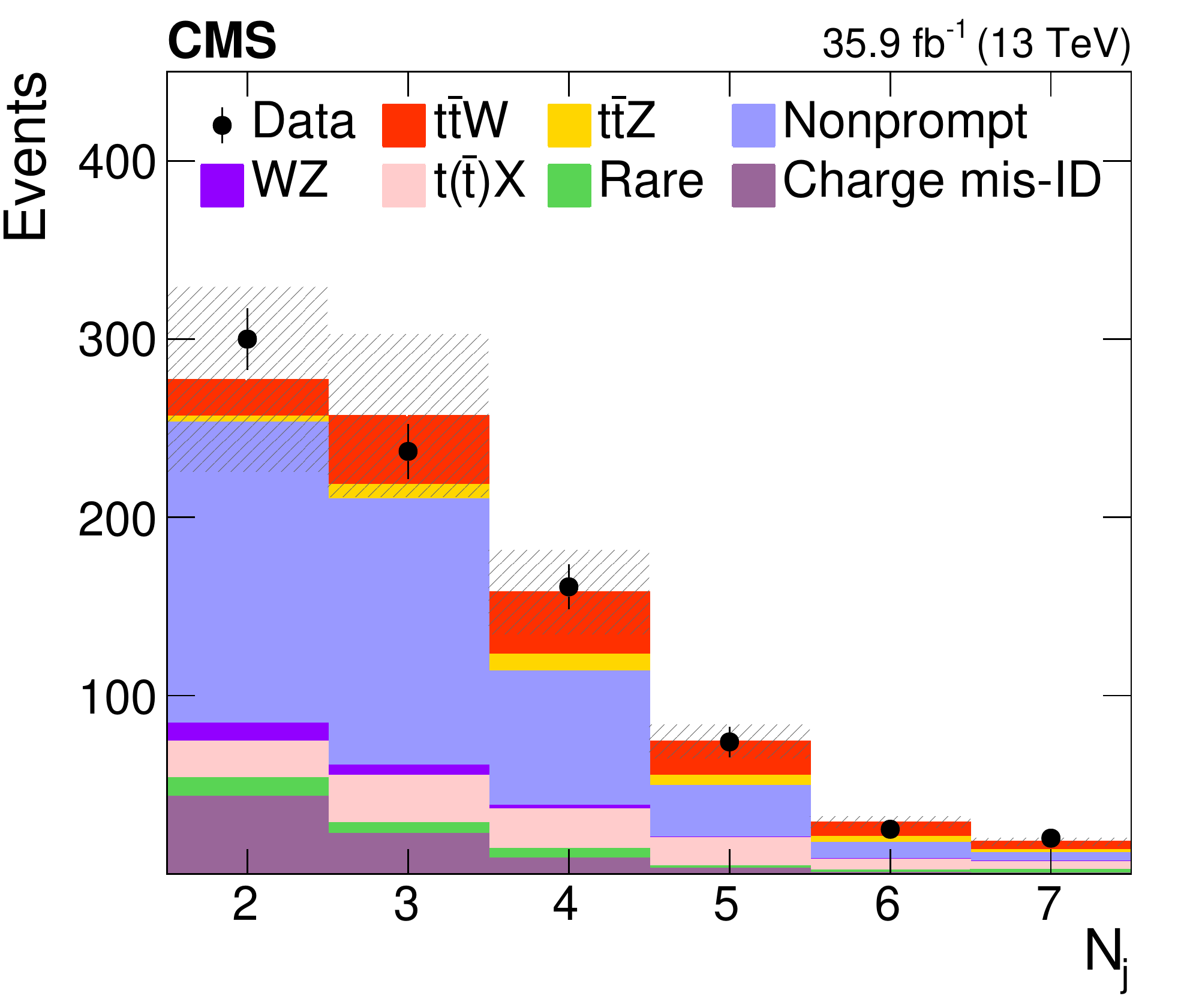}
                 \includegraphics[width=0.45\textwidth]{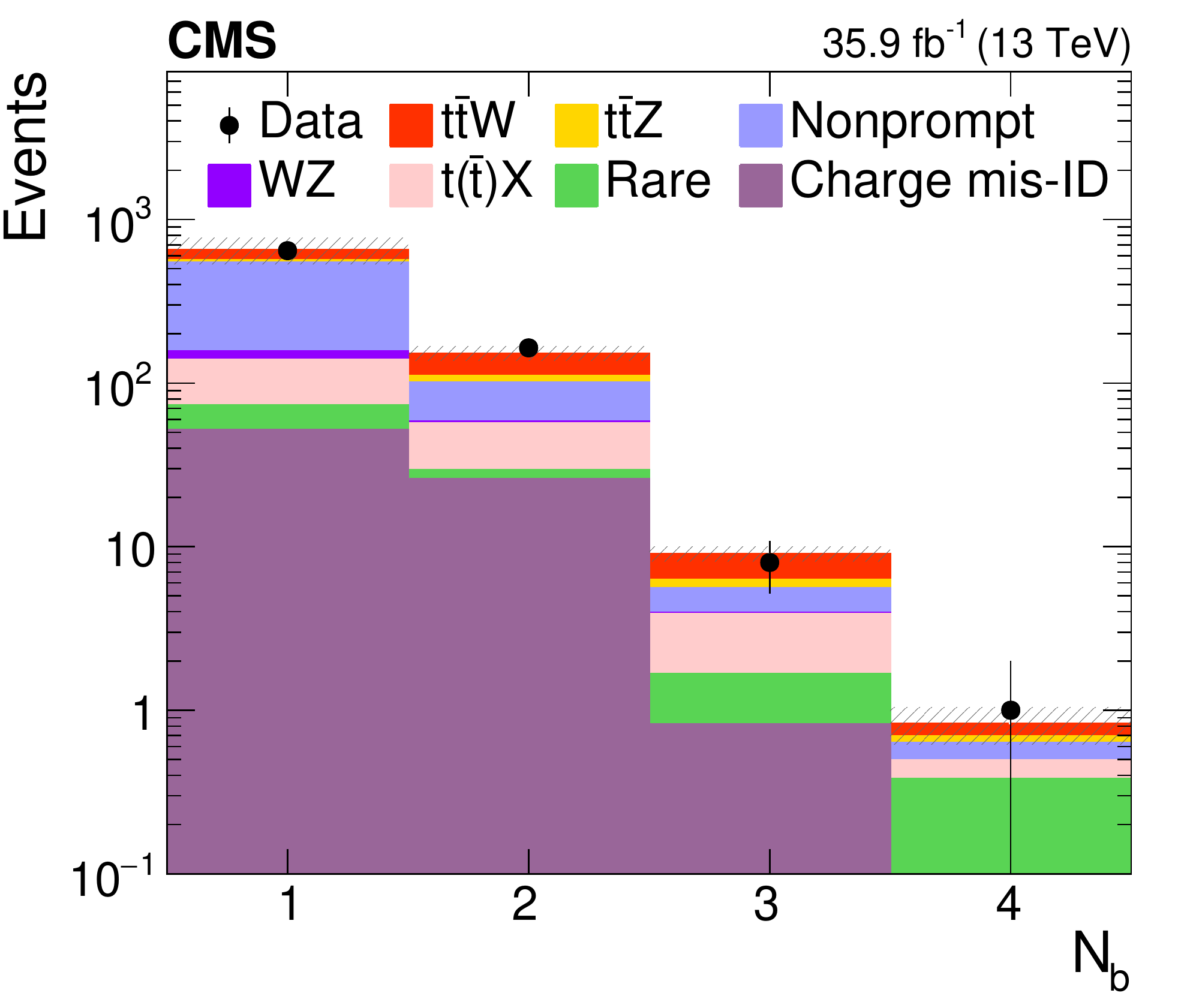}\\
                 \includegraphics[width=0.45\textwidth]{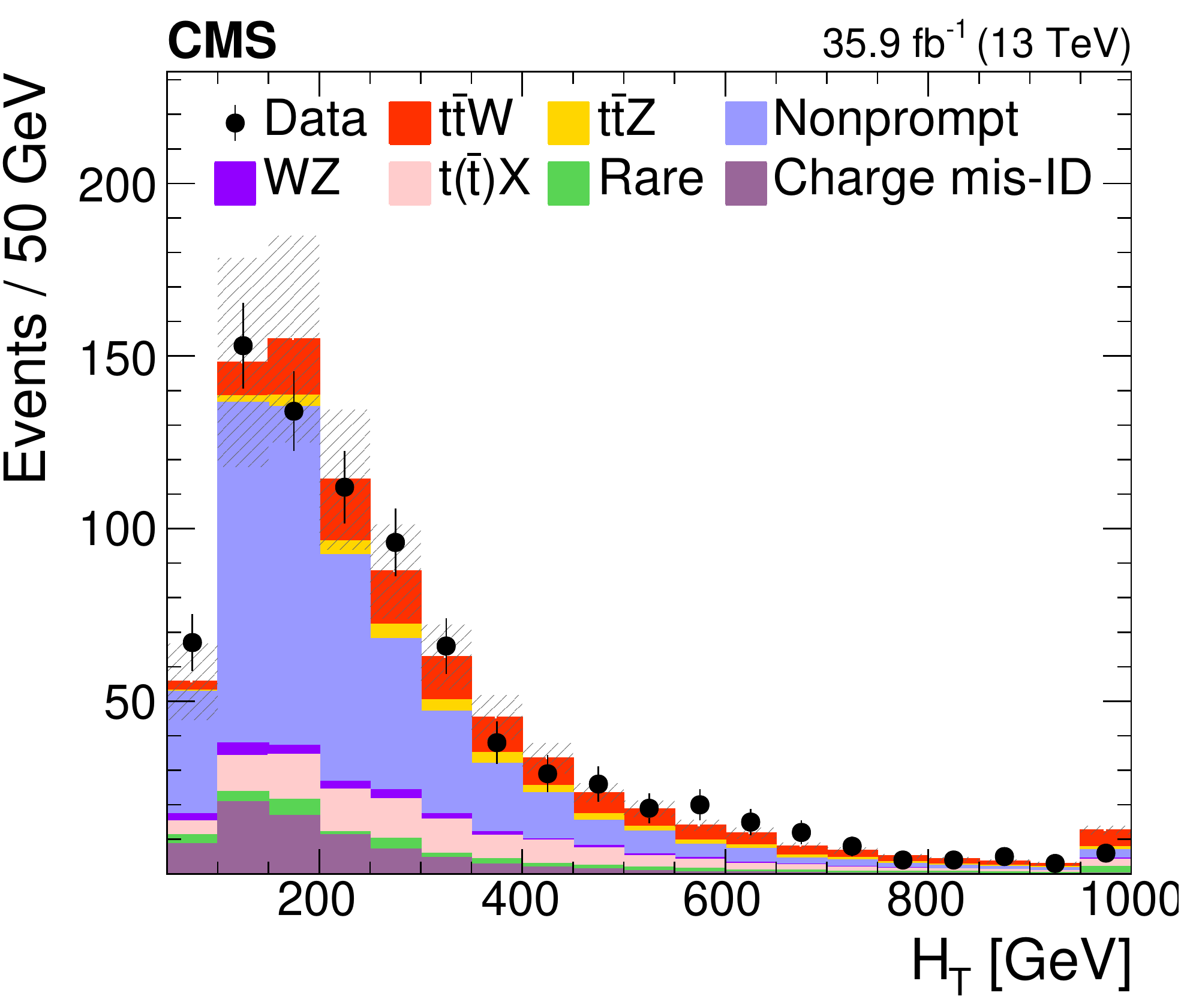}
                 \includegraphics[width=0.45\textwidth]{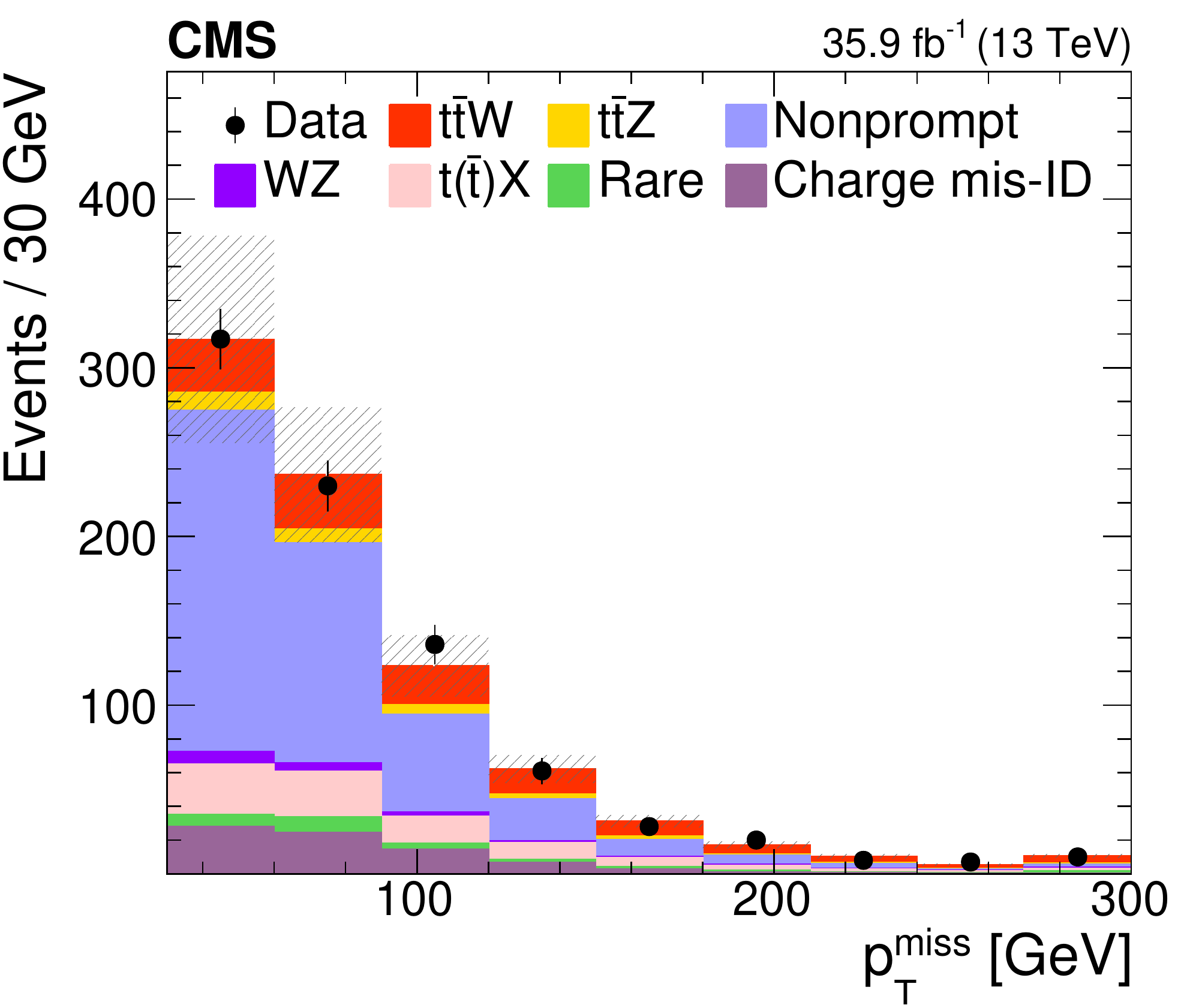}\\
                 \includegraphics[width=0.45\textwidth]{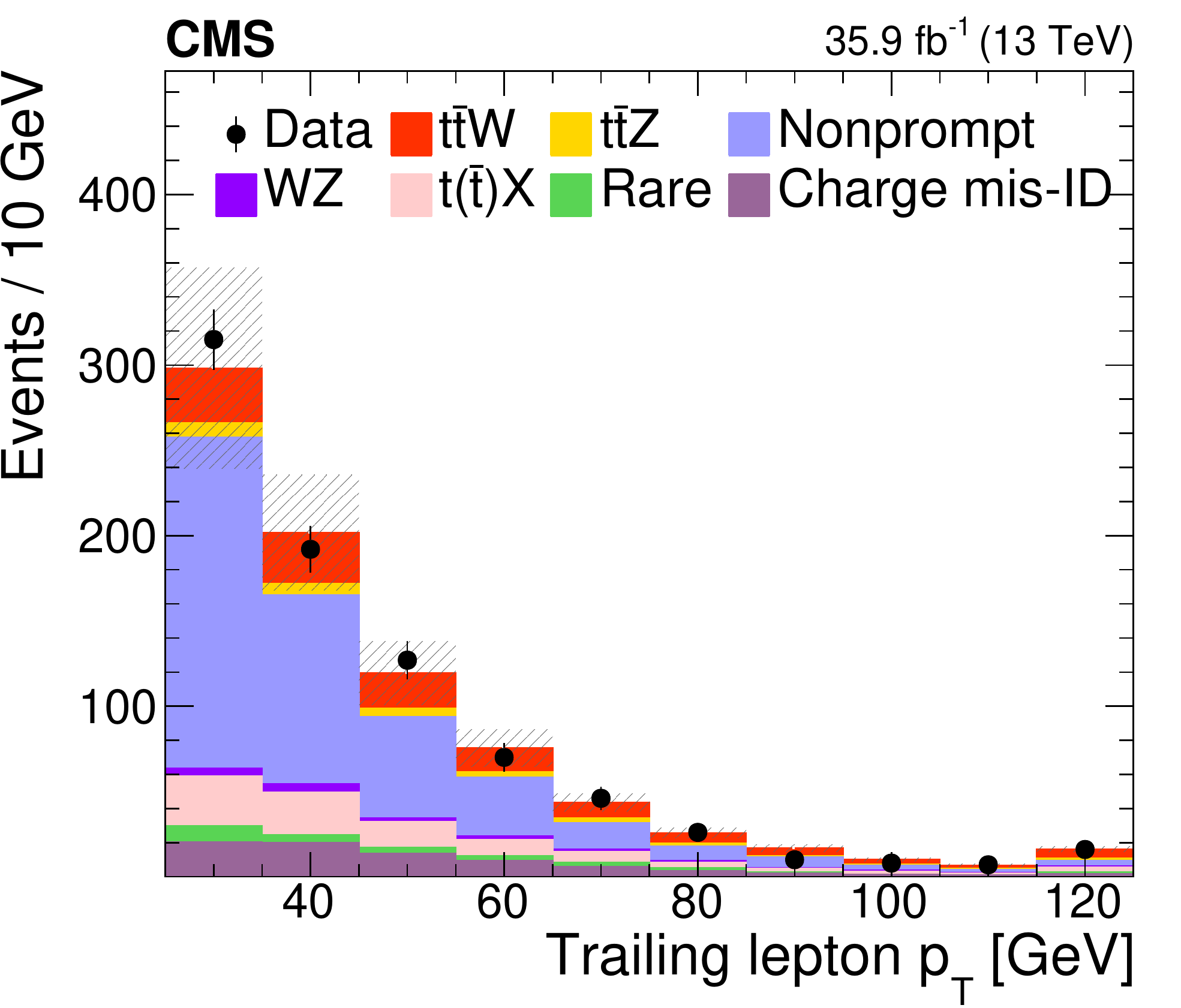}
                 \includegraphics[width=0.45\textwidth]{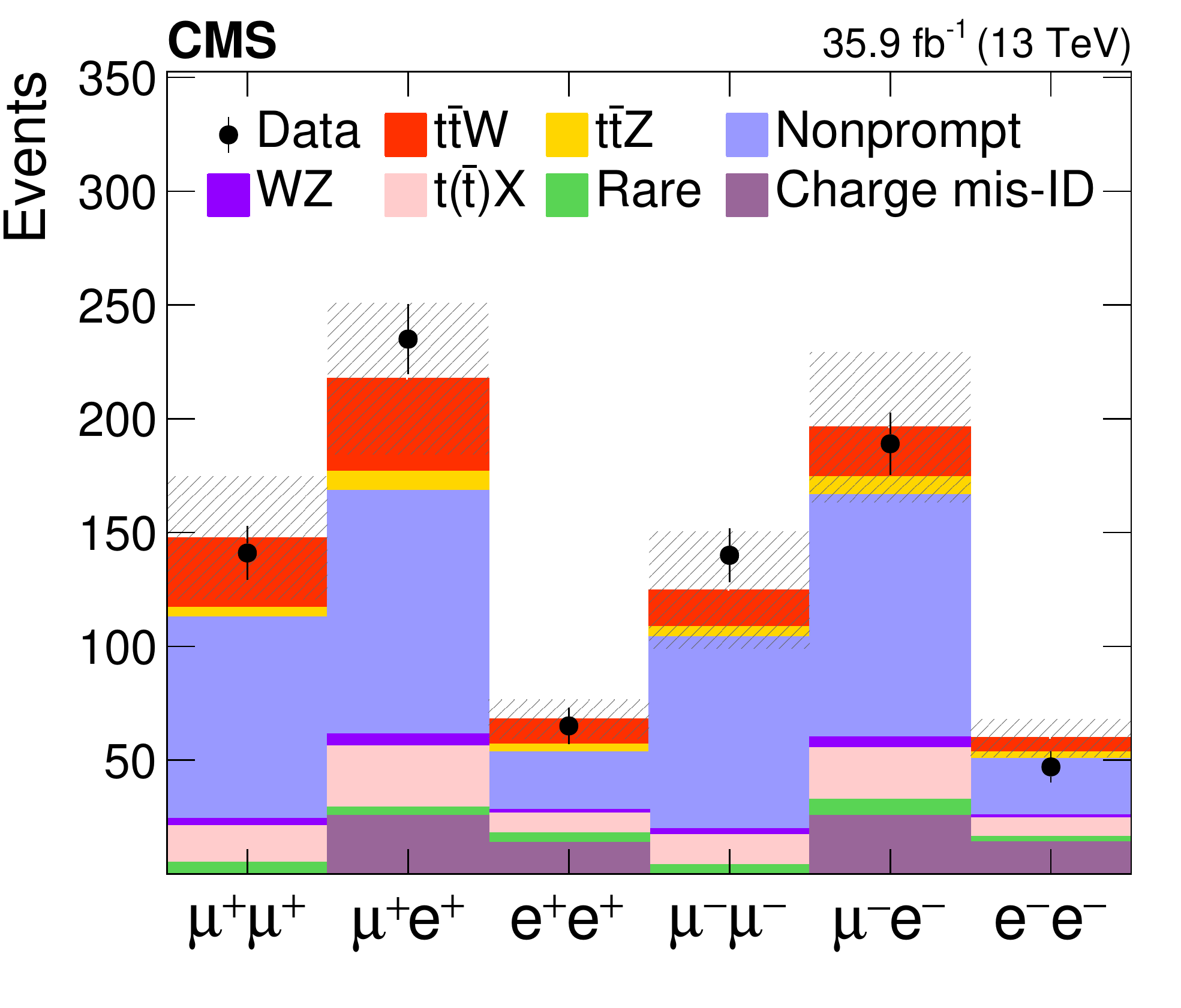}
                 \caption{Distributions of different variables in data from the SS dilepton analysis, compared to the MC generated expectations. From left to right: jet and b jet multiplicity (upper), \HT and \ptmiss(center), trailing lepton \pt and event yields in each lepton-flavor combination (lower). The expected contributions from the different background processes are stacked, as well as the expected contribution from the signal. The shaded band represents the total uncertainty in the prediction of the background and the signal processes. See Section~\ref{sec:backgrounds} for the definition of each background category. }
                \label{fig:MVAvariables}
\end{figure}

\begin{figure}[h]
\centering
\centering
\includegraphics[width=0.49\textwidth]{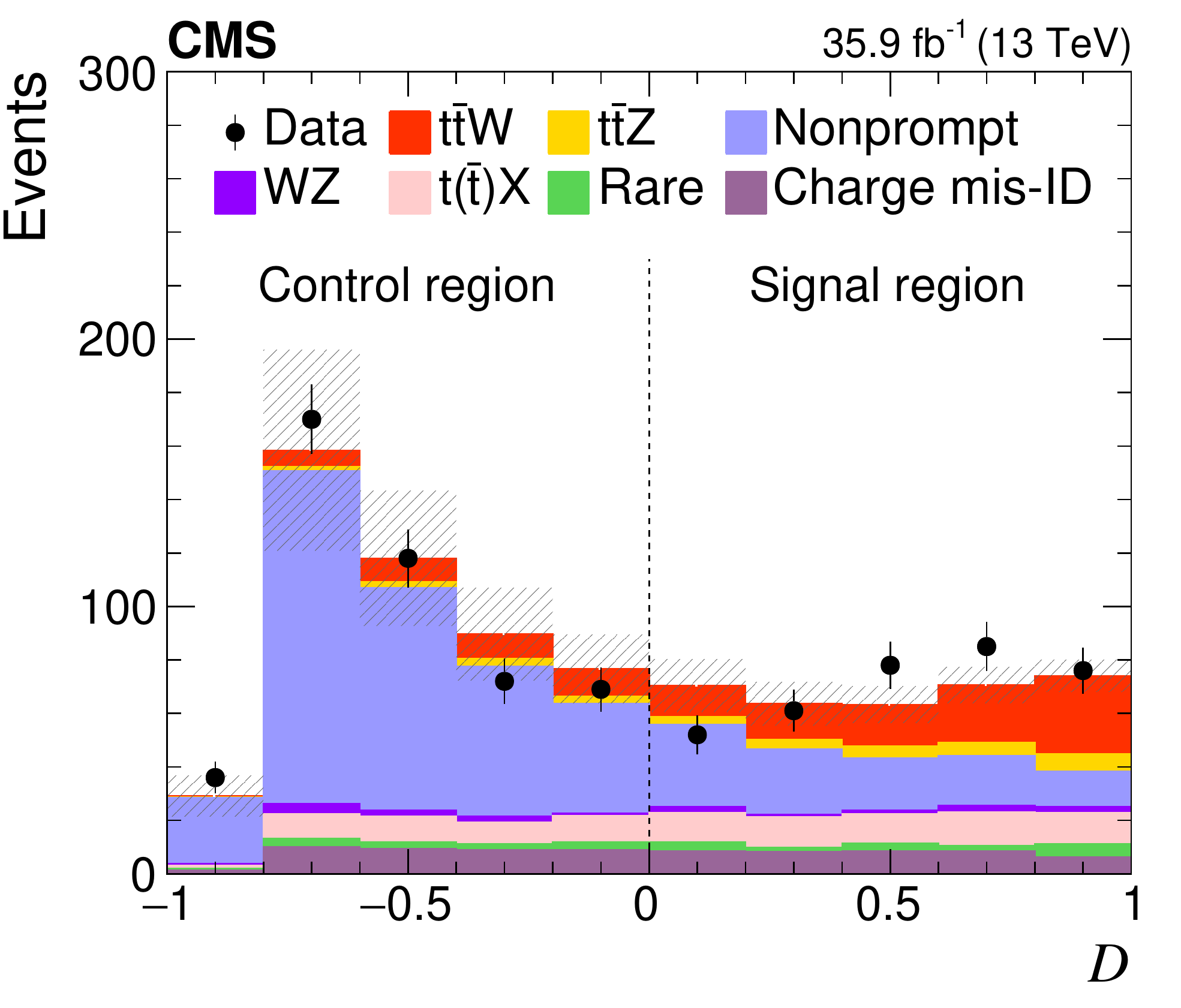}
\caption{Distribution of the boosted decision tree classifier \BDT for background and signal processes in the SS dilepton analysis. The expected contribution from the different background processes, and the signal as well as the observed data are shown. The shaded band represents the total uncertainty in the prediction of the background and the signal processes. See Section~\ref{sec:backgrounds} for the definition of each background category.}
\label{fig:BDTGoutput}
\end{figure}

A boosted decision tree classifier with gradient boosting~\cite{Hocker:2007ht} is used as the MVA discriminant, and simulated events are split into equal training and testing samples. Figure~\ref{fig:MVAvariables} shows the kinematic distributions of variables used in the MVA, and Fig.~\ref{fig:BDTGoutput} displays the output of the boosted decision tree classifier (\BDT) for all background sources and the signal, scaled to the integrated luminosity of the analyzed data samples.

Events with $\BDT > 0$ are selected to suppress the background from nonprompt leptons, and, for final signal extraction, they are split into two categories: $0< \BDT <0.6$ and $\BDT > 0.6$. These values are optimized to achieve the best expected sensitivity for \ttW. Furthermore, the number of jets and b jets are also used to form five exclusive event categories that maximize signal significance. The categories are formed using events with $\Njets = 2$, 3, and $>$3. The latter two categories are further split according to the number of b jets, $\Nbjets = 1$ and $\Nbjets > 1$. Events with $\BDT < 0$ are also used in the signal extraction procedure to constrain the uncertainties in the nonprompt lepton background.

Each of these categories is further split into two sets according to the total charge of the leptons: $\ell^+\ell^+$ or $\ell^-\ell^-$. This increases the sensitivity to the charge-asymmetric production of the signal ($\ttW^+$ vs. $\ttW^-$) resulting from the pp nature of the collision at the LHC, while the main backgrounds yield charge-symmetric dileptons. In total, we form 20 exclusive signal regions.

\label{sec:ThreeLepton}
\subsection{Three-lepton analysis}

The production rate of \ttZ events is measured in the final state with three leptons.

We select events that contain exactly three leptons ($\Pgm\Pgm\Pgm$, $\Pgm\Pgm\Pe$, $\Pgm\Pe\Pe$, or $\Pe\Pe\Pe$), requiring the leading, subleading, and trailing lepton \pt to be above 40, 20, and $10 $\GeV, respectively.  To reduce backgrounds from multilepton processes that do not contain a $\PZ$ boson, we require at least one OSSF lepton pair with invariant mass, \mll, consistent with the $\PZ$ boson hypothesis, namely $|\mll - \mZ| <10\GeV$.

Signal events are expected to have at least four jets, two of which originate from b quarks. When the events pass the jet and b jet requirements defined in the previous section, one obtains a sample of events enriched in signal, with minimal background contribution.  However, nearly 70\% of the signal events fail the requirement of having four jets with two of them identified as b jets. We therefore make use of lower jet and b jet multiplicities to form nine exclusive event categories to include a larger fraction of the signal events. These nine categories are formed using events with $\Njets = 2$, 3, and $> 3$, where each jet multiplicity gets further split according to the b jet multiplicity, $\Nbjets = 0$, 1, and $> 1$.

Despite the larger background contamination, the $\Njets = 3$ categories, especially in bins with larger \Nbjets,  improve the signal sensitivity, as this category  recovers signal efficiency for the jets that fall outside the acceptance.
The  $\Njets = 2$ category provides a background-dominated region that helps to constrain the background uncertainties.  We use all nine signal regions to extract the signal significance and the cross section.

\label{sec:FourLepton}
\subsection{Four-lepton analysis}

In addition to the three-lepton final state, events with four leptons are exclusively analyzed for the measurement of the $\ttbar$Z production rate.

The \ttZ events in this channel are characterized by the presence of two b jets, \ptmiss, and four leptons, two of which form an OSSF pair consistent with the $\PZ$ boson mass.  The event selection is optimized to obtain high signal efficiency in simulation in order to profit from low expected background yields. Events with exactly four leptons that pass the lepton identification and isolation requirements described in Section~\ref{sec:Samples} are selected. The leading lepton must have $\pt > 40\GeV$ and the $\pt$ of the remaining three leptons must exceed 10\GeV.  The sum of the lepton charges must be zero, and the invariant mass of any lepton pair is required to be greater than 12\GeV.  At least one OSSF lepton pair with an invariant mass $|\mll - M({\PZ})| <20\GeV$ must be present in the event. Events with $\Pgm\Pgm\Pgm\Pgm$, $\Pe\Pe\Pe\Pe$, and $\Pgm\Pgm\Pe\Pe$ final states, in which a second OSSF lepton pair consistent with the $\PZ$ boson mass is found, are rejected. Events containing two jets are selected and split into two categories for signal extraction: one with zero b jets and the other with at least one b jet.

\section{Background predictions}
\label{sec:backgrounds}

\subsection{Background due to nonprompt leptons}

Nonprompt leptons, i.e.\xspace leptons from heavy-flavor hadron decay, misidentified hadrons, \allowbreak muons from light meson decays, or electrons from unidentified photon conversions, are strongly rejected by the identification and the isolation criteria applied on electrons and muons.  Nonetheless, a residual background from such leptons leaks into the analysis selection. Such backgrounds are mainly expected from \ttbar production, in which one or two of the leptons originate from the leptonic $\PW$ boson decays and an additional nonprompt lepton comes from the semileptonic decays of a b hadron, as well as from $\PZ \to \ell \ell$ events containing an additional misidentified lepton. These backgrounds are estimated using a data-based technique.
From a control sample in data, we calculate the probability for
a loosely identified nonprompt lepton to pass the full set of tight requirements, designated as the tight-to-loose ratio. For loose leptons we choose a relaxed isolation requirement, $I_\text{rel} < 1$, and additional electron identification requirements on the variables that distinguish prompt  electrons from hadrons and photons which are misidentified as electrons. The tight-to-loose ratios are measured in a data control sample of QCD multijet events that are enriched in nonprompt leptons. This control sample consists of events with a single lepton and at least one jet, where the lepton and jets are separated by $\Delta R > 1$. We suppress the prompt lepton contamination, mostly from W+jets, by requiring $\ptmiss < 20\GeV$ and $\MT < 20\GeV$, where \MT is the transverse mass constructed using $\ptmiss$ and the selected lepton. The residual prompt lepton contamination is subtracted using estimates from MC simulation. This subtraction is relevant only for the high-\pt leptons, and its effect on the total estimated background does not exceed a few percent. These tight-to-loose ratios are parametrized as functions of the $\eta$ of the leptons and $\ptCor$, with the latter calculated through corrections to lepton \pt as a function of the energy in the isolation cone. This definition has no impact on the \pt of the leptons that pass the isolation requirement, but modifies the \pt of those that fail, and extract thereby a more accurate value of true \pt~\cite{SUS_15_008}.
The tight-to-loose ratios are then used together with the observed number of events in sideband regions. These sideband regions contain events that pass full event criteria in each analysis region, except that at least one of the leptons
passes the loose selection but does not pass the tight selection. Each event in this region is assigned a weight as a function of the \pt and $\eta$ of the loose lepton to account for the probability of the lepton to pass the tight selection.

\begin{figure}[h]
\centering
\includegraphics[width=.45\textwidth]{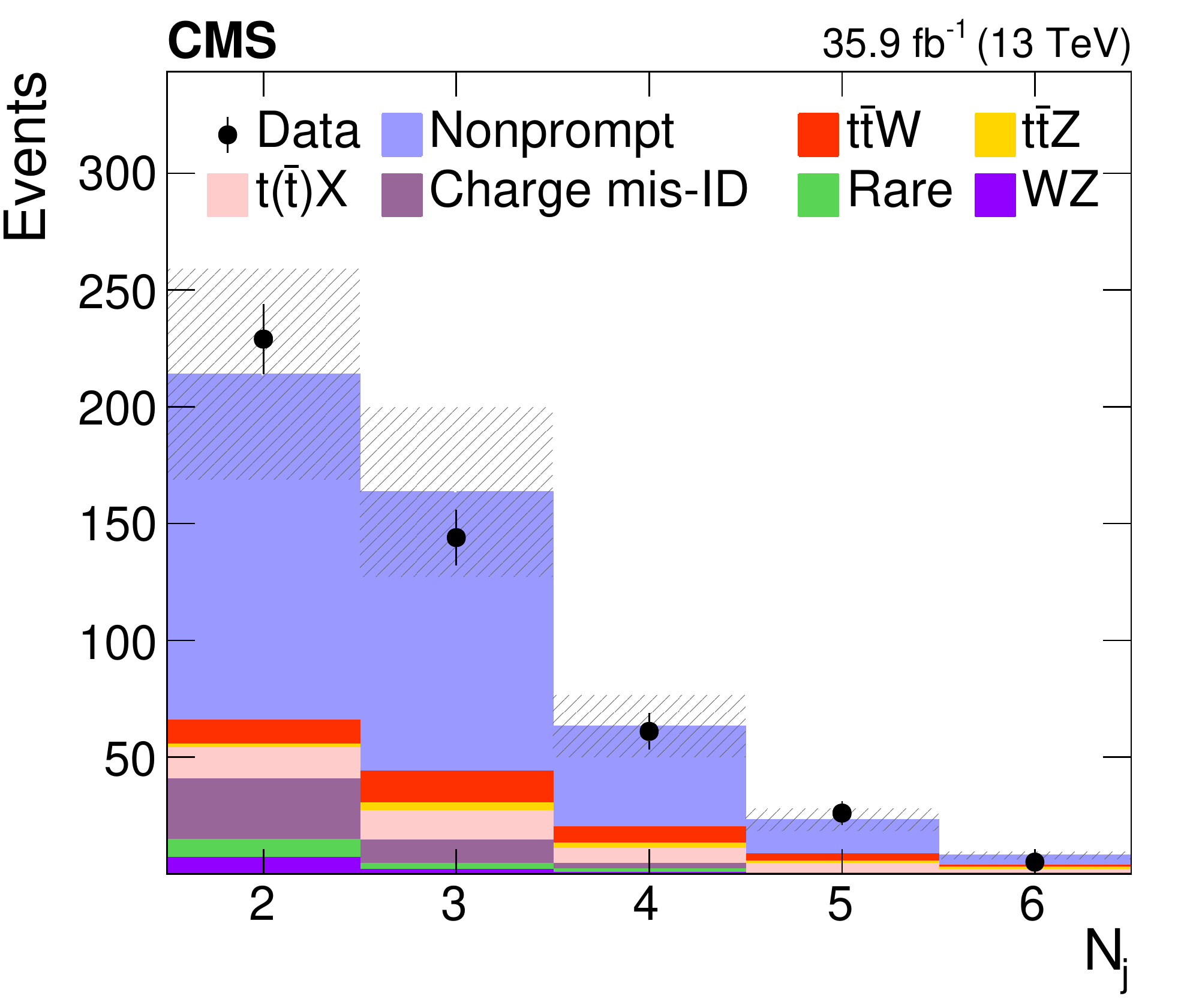}
\includegraphics[width=.45\textwidth]{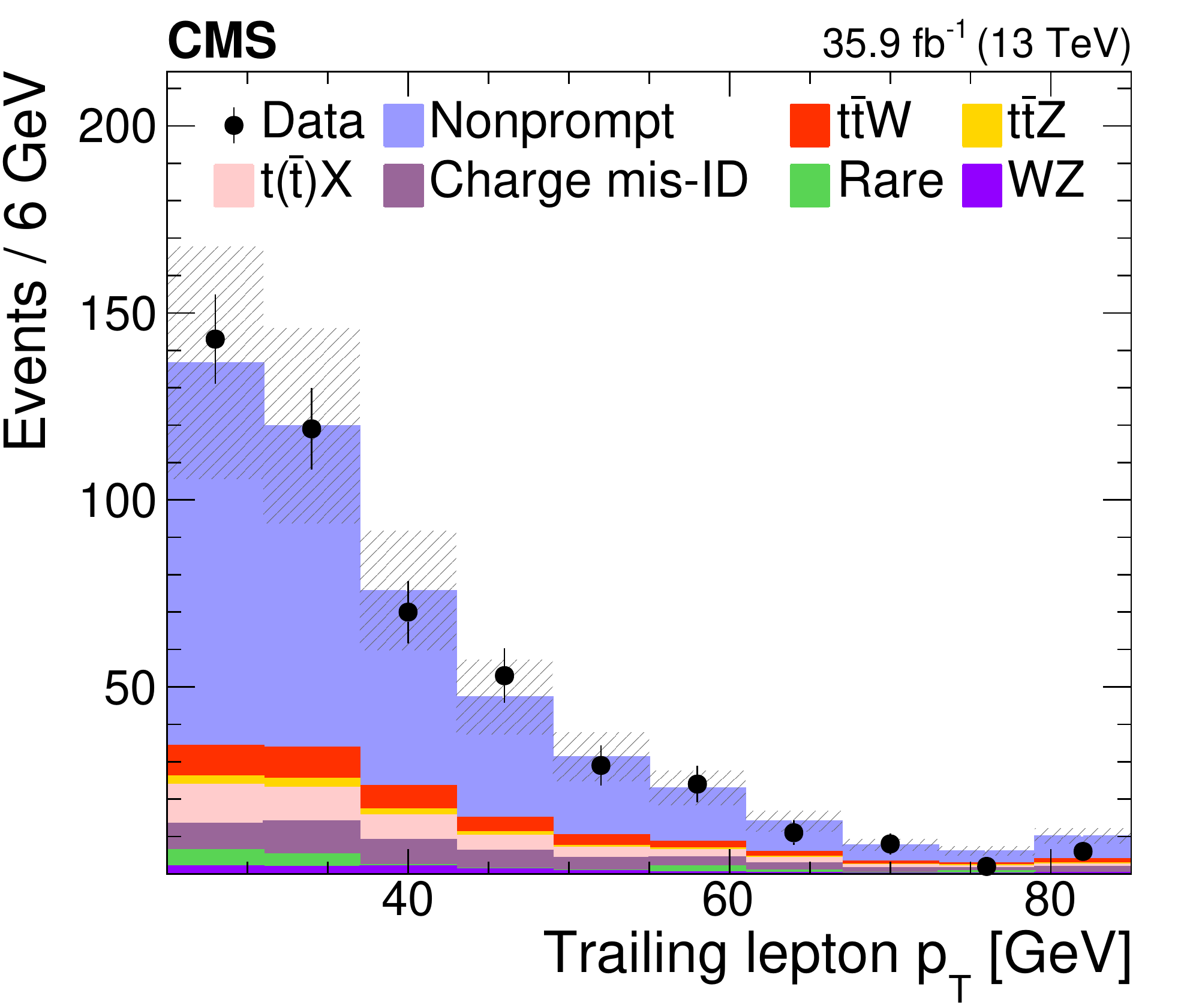}
\caption{Distributions of the predicted and observed yields versus \Njets (left) and $\pt$ of the trailing lepton (right) in control regions enriched with nonprompt lepton background in the SS dilepton channel. The shaded band represents the total uncertainty in the prediction of the background and the signal processes. See Section~\ref{sec:backgrounds} for the definition of each background category.}
\label{figures:ttbar2L_background}
\end{figure}

\begin{figure}[h]
\centering
\includegraphics[width=.45\textwidth]{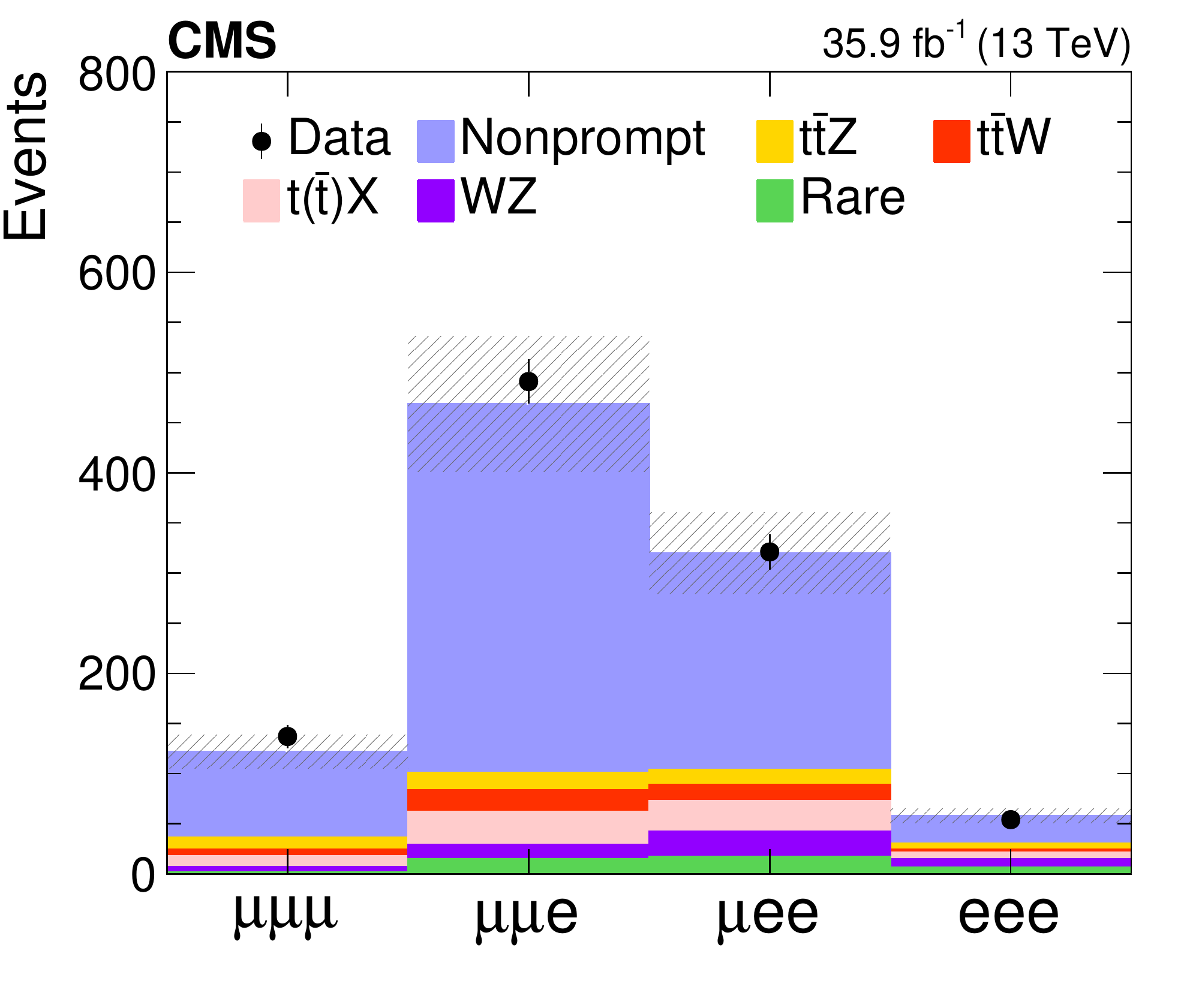}
\includegraphics[width=.45\textwidth]{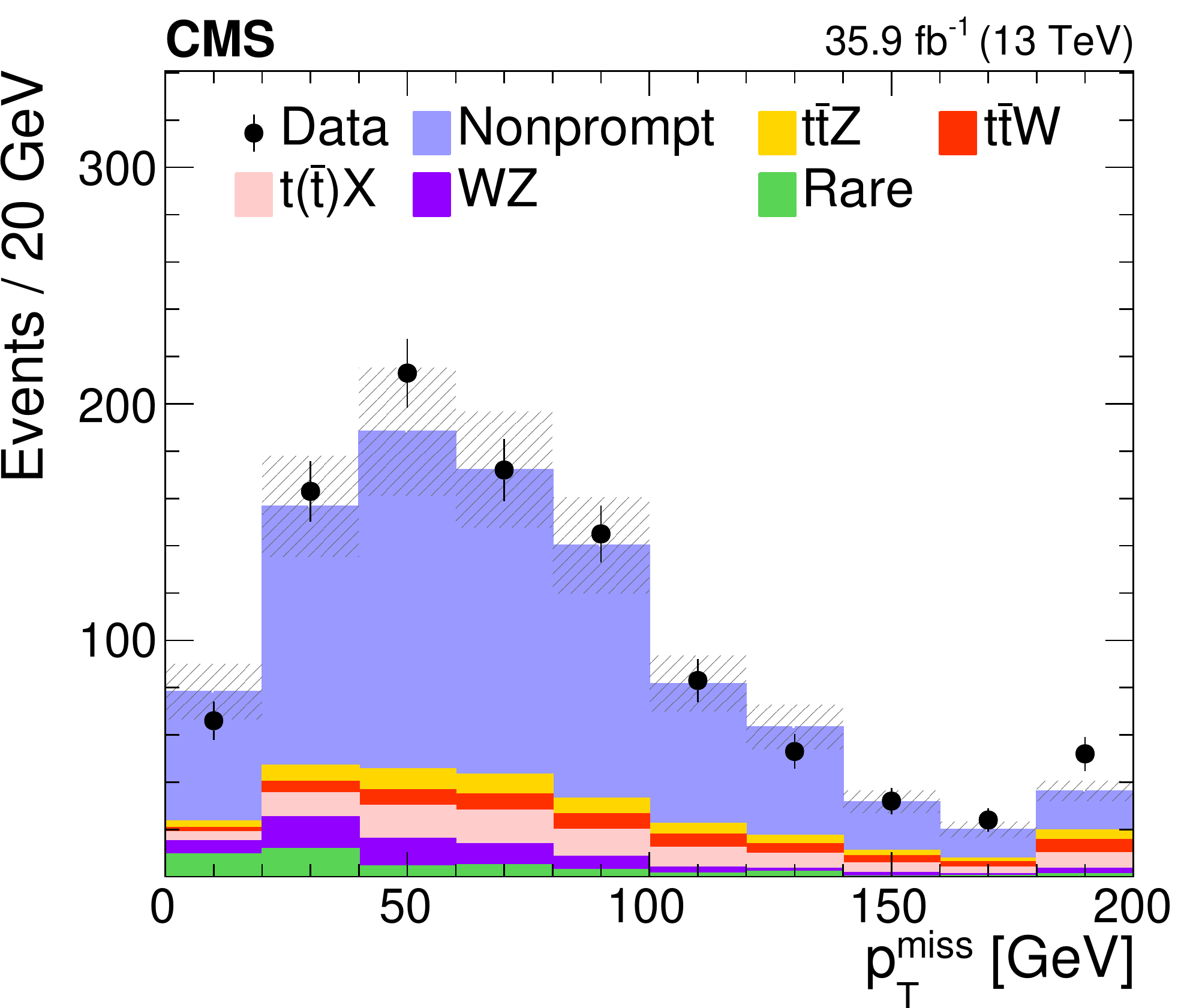}
\includegraphics[width=.45\textwidth]{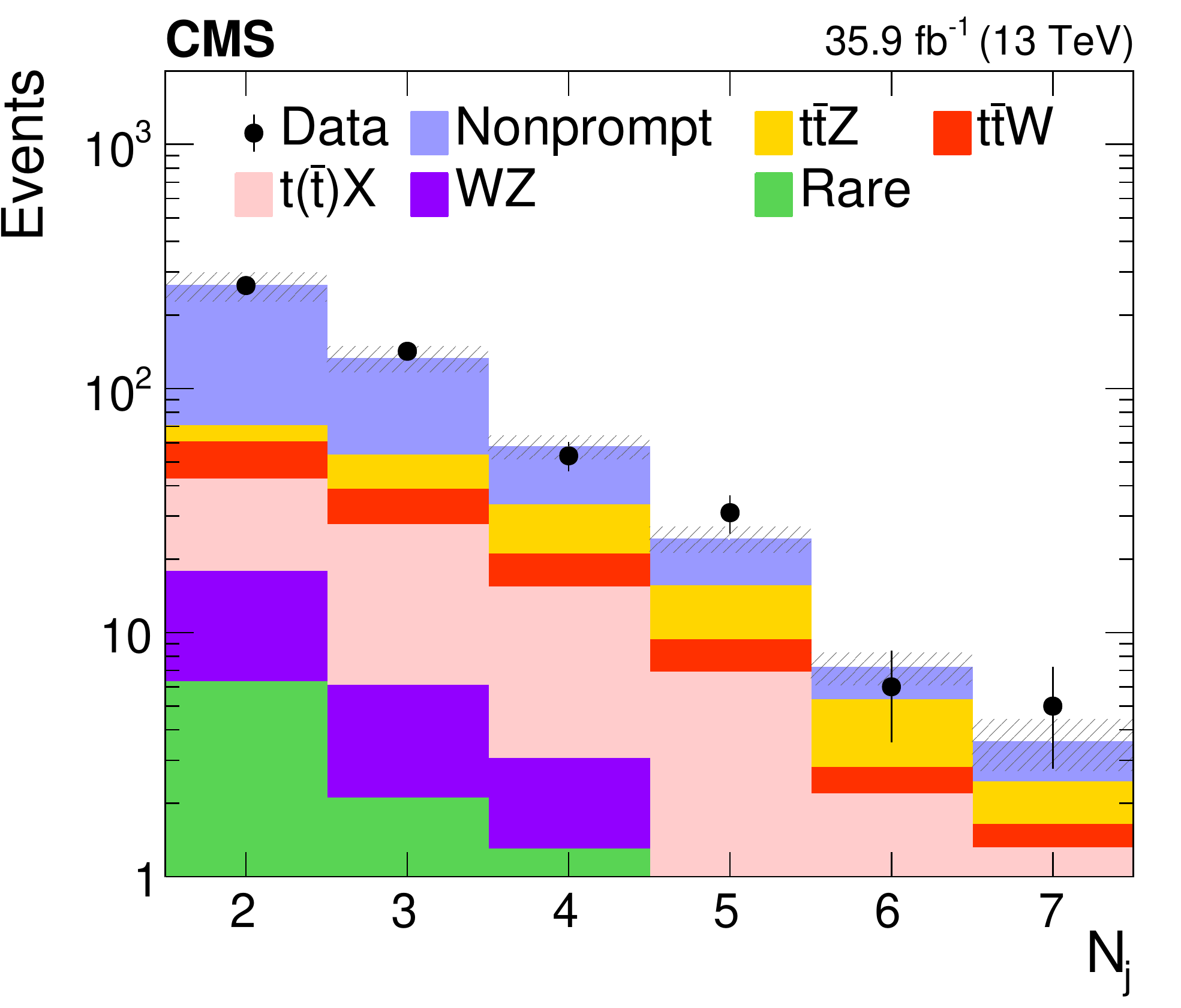}
\includegraphics[width=.45\textwidth]{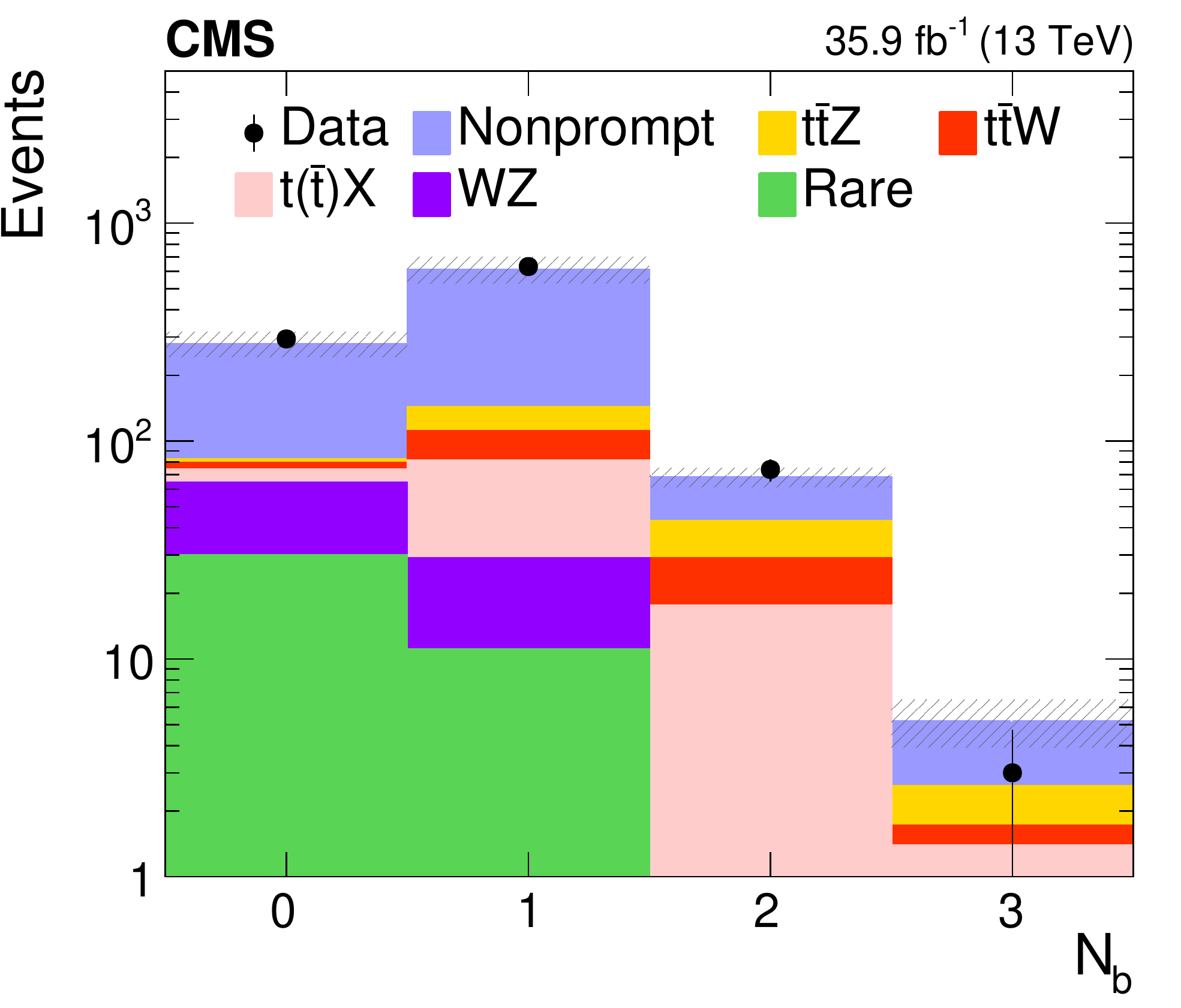}
\caption{Distributions of the predicted and observed yields versus different three-lepton channels, \ptmiss (upper panels), and jet and b jet multiplicity (lower panels) in control regions enriched with nonprompt lepton background. The shaded band represents the total uncertainty in the prediction of the background and the signal processes. See Section~\ref{sec:backgrounds} for the definition of each background category.}
\label{figures:ttbar3L_background}
\end{figure}

We validate this technique using simulated events. The tight-to-loose ratios are first measured for electrons and muons in simulated multijet events, and  applied in simulated $\ttbar$ and Z+jets events in the same way as in data, to extract predictions for the nonprompt background contribution.  These predictions agree very well with the observed yields in simulation, not only for the integral yields, but also for distributions in all kinematic variables used to form the analysis regions, including the boosted decision tree output \BDT.  Additionally, data control regions used in the signal-extraction regions and enriched in processes with nonprompt leptons, are formed to check any other potential sources of mismodeling.
For the SS dilepton channel, we use the region with $\BDT< 0$.  Figure~\ref{figures:ttbar2L_background} shows the predicted background and observed data yields versus \Njets and the \pt of the trailing lepton. Events in this region are also used in the signal extraction procedure for \ttW.  The potential systematic effects for the extrapolation from $\BDT<0$ to $\BDT>0$ are studied in simulation and found to be negligible compared to other sources of uncertainty. For the three-lepton channel this control region is defined by either the absence of an OSSF lepton pair, or by the presence of an OSSF, with its invariant mass being at least 10\GeV away from $M(\PZ)$, and with at least one b jet present. This region is dominated by \ttbar  events in which both $\PW$ bosons decay leptonically  and an additional nonprompt lepton is present. Figure~\ref{figures:ttbar3L_background} shows the predicted and observed yields versus the flavor of the leptons, \ptmiss, \Njets, and \Nbjets.  Both of these control regions show very good agreement between predicted and observed yields and for kinematic distributions that are relevant for the signal extraction.

Based on the extensive aforementioned validation in both data and simulated control samples, we conclude that a systematic uncertainty of 30\% is appropriate for the prediction of the background from nonprompt leptons. The statistical uncertainties due to the limited number of observed events in the sideband regions of data are taken into account, and are often found to be larger or comparable to the systematic uncertainty.

\subsection{Background induced by the mismeasurement of the lepton charge}

The charge mismeasurement rate for muons is negligible and background is significant only for the channels with at least one electron.  This background is estimated with a partially data-based approach. The opposite-charge $\Pe \Pe$ or $\Pe \Pgm$ data events passing the full kinematic selection are weighted by the \pt- and $\eta$-dependent electron-charge misidentification probabilities. These probabilities are obtained from MC simulation. The charge mismeasurement rate in simulation is validated through a comparison with data.  It is measured in DY events in MC and in data, where events are selected when the two SS electrons have an invariant mass that falls within a $\PZ$ boson mass window, $76 < \mll < 106\GeV$. The measured electron charge misidentification rates in data and in DY simulation are in good agreement and vary from $4\times10^{-5}$ for low-\pt electrons in the barrel region
to $4\times10^{-3}$ for high-\pt electrons in the endcap.

The process contributing to this category of background in signal regions is primarily \ttbar production. Based on the agreement in the charge mismeasurement rate between data and MC events, and the simulation studies of charge misidentification rate comparison between \ttbar and DY MC events, we assign a 20\% systematic uncertainty in the estimation of this background~\cite{SUS-16-035}.

\subsection{Background due to $\WZ$ production}

Kinematic distributions for the background from \WZ events are taken from simulation. This background has the highest expected yields in the analysis region with no b-tagged jets. The data used for this analysis contain a substantial number of \WZ events that can be isolated and compared with the MC predictions. We define a control region in a subset of the data with the following requirements: we select events with three leptons, with the same \pt thresholds as the ones used in the \ttZ selection, that have two leptons forming an OSSF pair with $|\mll- \mZ | <  10 \GeV$, less than two jets, and no b-tagged jets. Additionally \ptmiss is required to be greater than 30\GeV, and $\MT$, constructed using this \ptmiss and the lepton not used in the \mll calculation, is required to be greater than 50\GeV.

\begin{figure}[h]
\centering
\includegraphics[width=.40\textwidth]{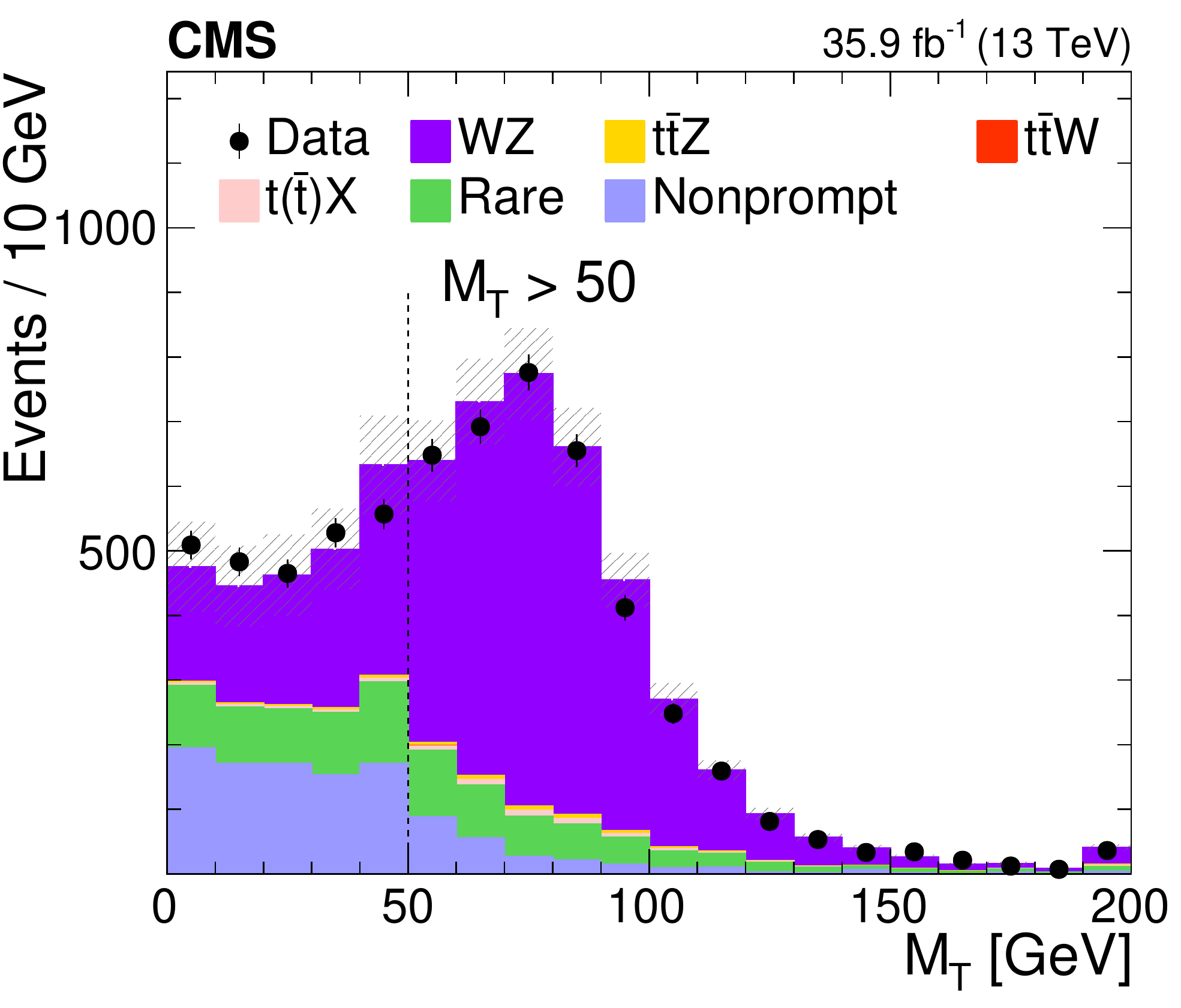}
\includegraphics[width=.40\textwidth]{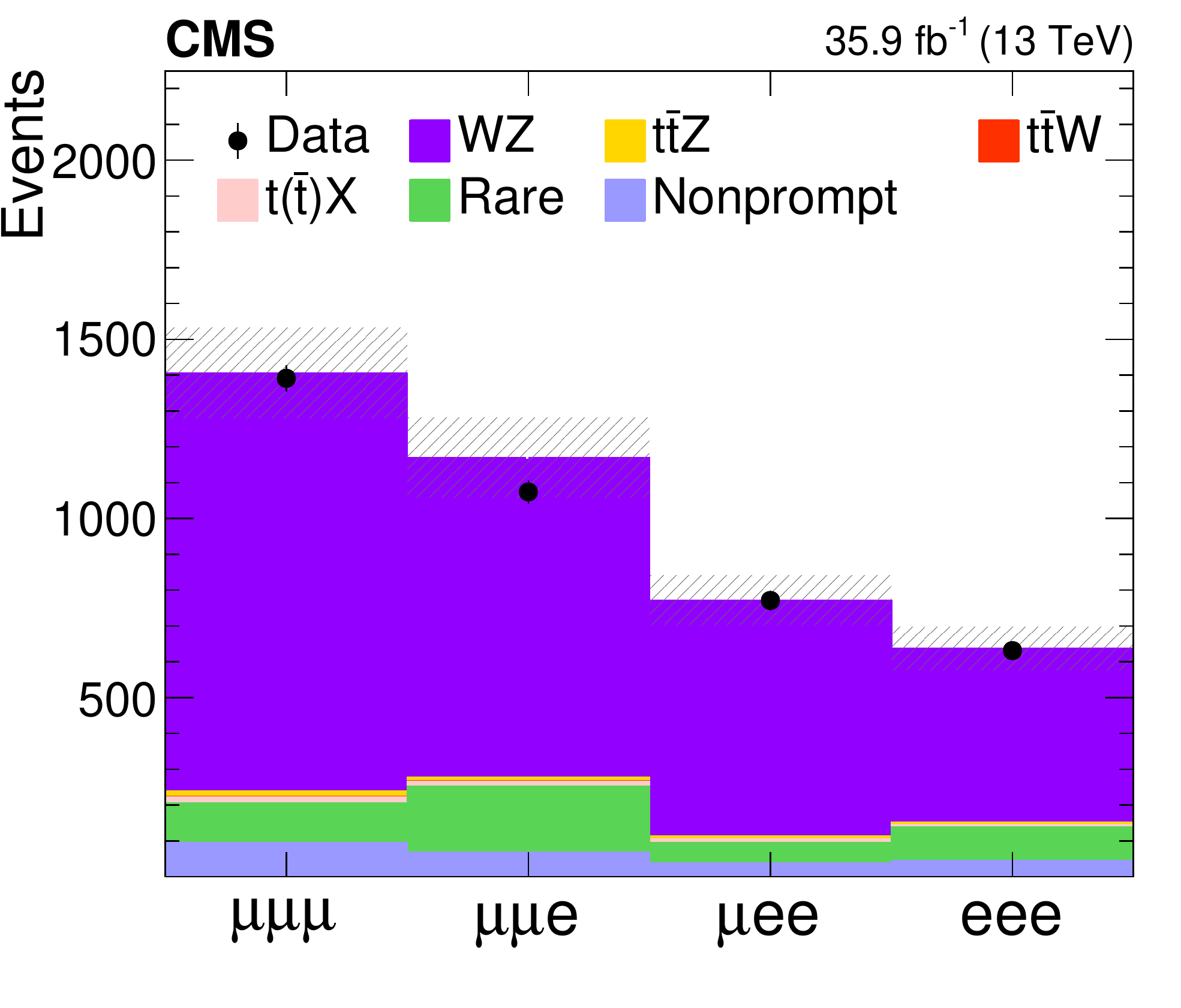}
\includegraphics[width=.40\textwidth]{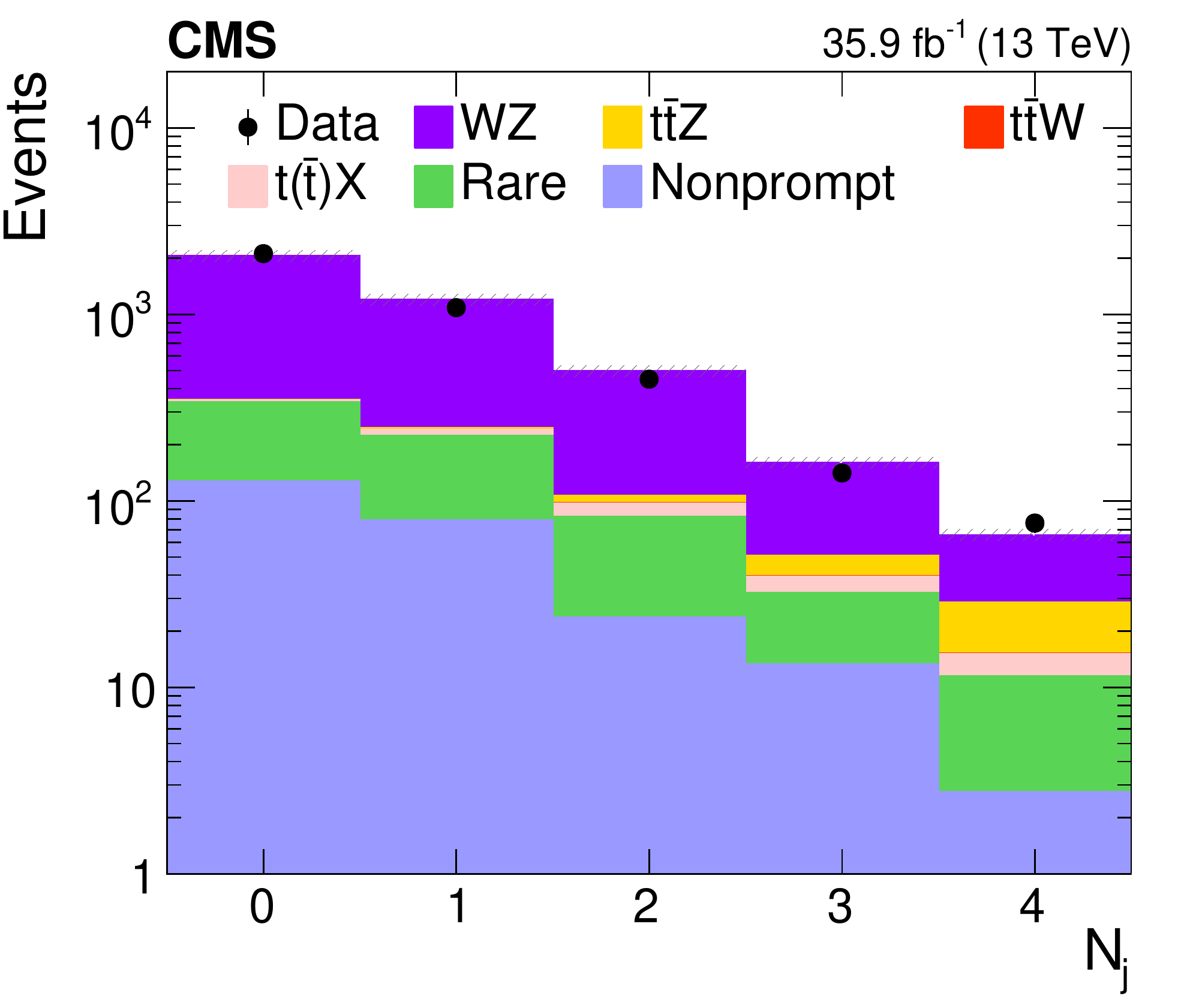}
\includegraphics[width=.40\textwidth]{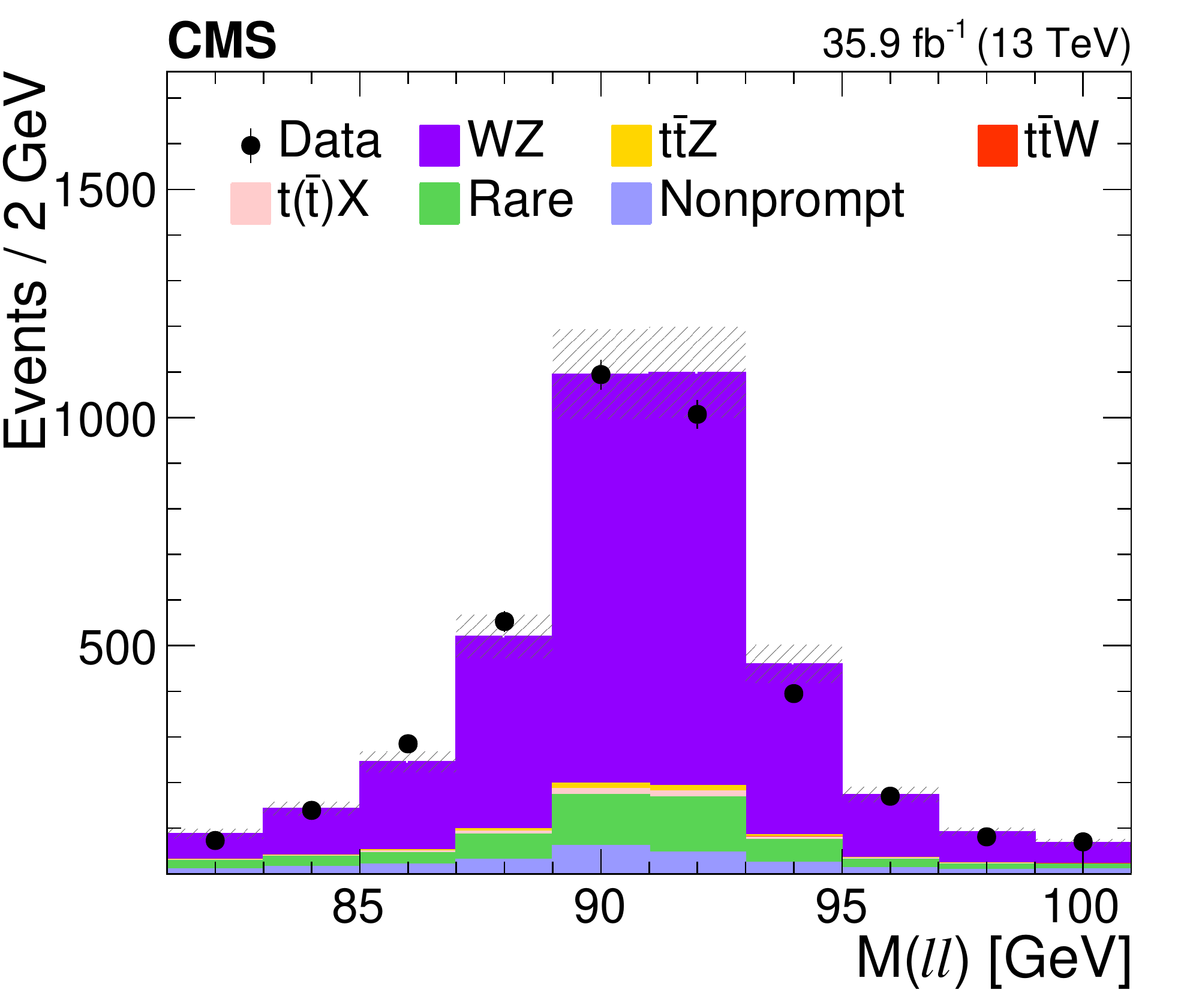}
\caption{Distributions of the predicted and observed yields versus \MT (upper left),  lepton flavor (upper right), jet multiplicity (lower left), and the reconstructed invariant mass of the $\PZ$ boson candidates (lower right) in the \WZ-enriched control region. The requirements on \MT and \Njets are removed for the distributions of these variables. The shaded band represents the total uncertainty in the prediction of the background and the signal processes.}
\label{figures:WZ_background}
\end{figure}

This selection provides a data sample that is expected to be 85\% pure in \WZ events. Figure~\ref{figures:WZ_background} shows the number of events as a function of \MT, lepton flavor,  \Njets, and \mll. The expected background from nonprompt leptons is measured from data using the method described above. The other background contributions are obtained from simulation. We observe overall agreement between data and the total expectation in
all four-lepton channels and also in the kinematic distributions. The ratio of the total observed yield to the predicted one is found to be $0.94 \pm 0.07$, where the uncertainty reflects only statistical sources. With this level of agreement between the data and MC prediction, we proceed without applying any corrections to the \WZ prediction obtained from the simulation. The statistical uncertainty in the ratio is propagated to the final prediction. We also study possible mismodeling of the \WZ + heavy-flavor background at large b jet multiplicities. We find that the \WZ contribution at high b jet multiplicities is mainly caused by the misidentification of light-flavor jets as b jets.
The fraction of WZ events containing at least one b quark is predicted by the simulation to vary between 5 and 15\% across all of the analysis categories. We apply scale factors to take into account the differences in b tagging efficiencies and misidentification probabilities between data and simulation~\cite{Chatrchyan:2012jua,CMS-PAS-BTV-15-001}.  Once all the corrections are applied, we check the agreement between data and $\PZ$+jets simulated events as a function of \Nbjets in OSSF dilepton events consistent with the $\mZ$. Based on this study, we assign a 10\% systematic uncertainty to the \WZ background estimate, which covers the differences between data and expectations found in the control region. For the three-lepton analysis, an additional 20\% uncertainty is introduced for regions with $\Njets > 3$.  Other systematic uncertainties associated with the extrapolation from this control region to high \Njets or \Nbjets regions, such as jet energy scale and b tagging uncertainties, are considered separately.

\subsection{ Background due to \ttX and other rare SM processes}
The background events containing either multiple bosons or top quark(s) in association with a W, Z, or a Higgs boson are estimated from simulation scaled by their NLO cross section and normalized to the integrated luminosity.
The backgrounds that have at least one top quark in final state, i.e. \ttH, $\cPqt \PW \PZ$, $\cPqt \Pq \PZ$, $\cPqt \PH \Pq$, $\cPqt \PH \PW$, \ttVV, and $\ttbar\ttbar$, are denoted as \ttX, while all others, i.e. $\PW \PW$, $\PZ \PZ$, $\PW\Pgg^{*}$, $\PZ\Pgg^{*}$, $\PW \PW \PW$, $\PW \PW \PZ$, $\PW \PZ \PZ$, $\PZ \PZ \PZ$, are grouped into the rare SM processes category.

For background yields in the $\ttX$ category, we studied the theoretical and systematic uncertainties separately. The theoretical uncertainties for the inclusive cross section are around 10\%~\cite{Campbell:2013yla, Frixione:2015zaa, Alwall:2014hca}. Using the simulations, we study the effect of the changes made in renormalization and factorization scales ($\mu_\mathrm{R}$ and $\mu_\mathrm{F}$), as well as the uncertainties from choice in PDF in the phase-space region relative to this analysis. From these studies we deduce an additional theoretical uncertainty of 2\%.
On the experimental side, to account for the differences in the lepton-selection efficiencies,
b jet identification efficiencies, mistagging rate between the simulation and the data, we apply scale factors to the  predictions obtained from simulations, and assign systematic uncertainties associated with these scale factors. These experimental uncertainties are estimated in each analysis category (see Section~\ref{sec:Systematic}) and are applied in addition to the above-mentioned 10\% uncertainty in the yield.

\begin{figure}[h]
    \centering
    \includegraphics[width=0.38\textwidth]{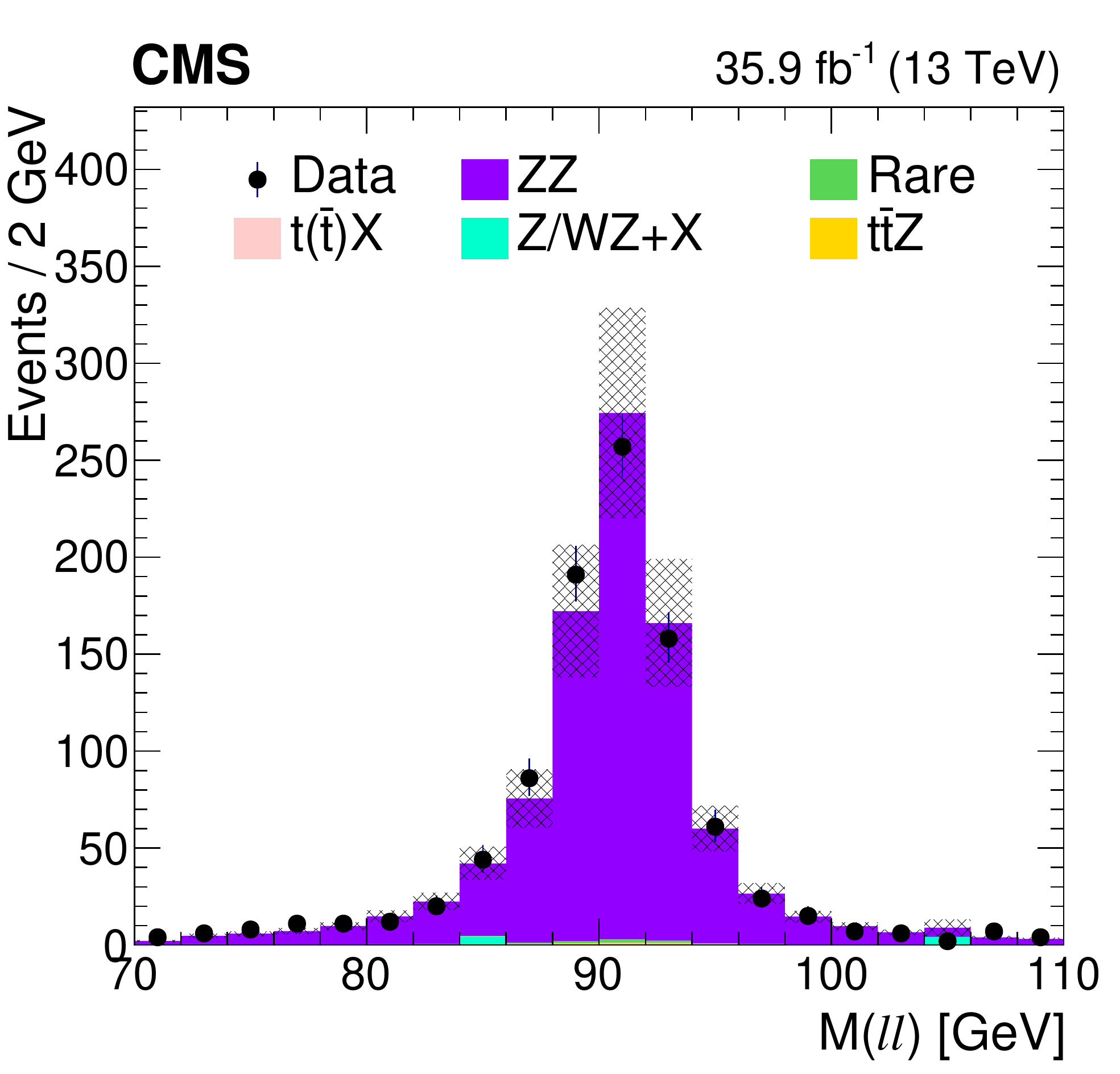}
    \includegraphics[width=0.38\textwidth]{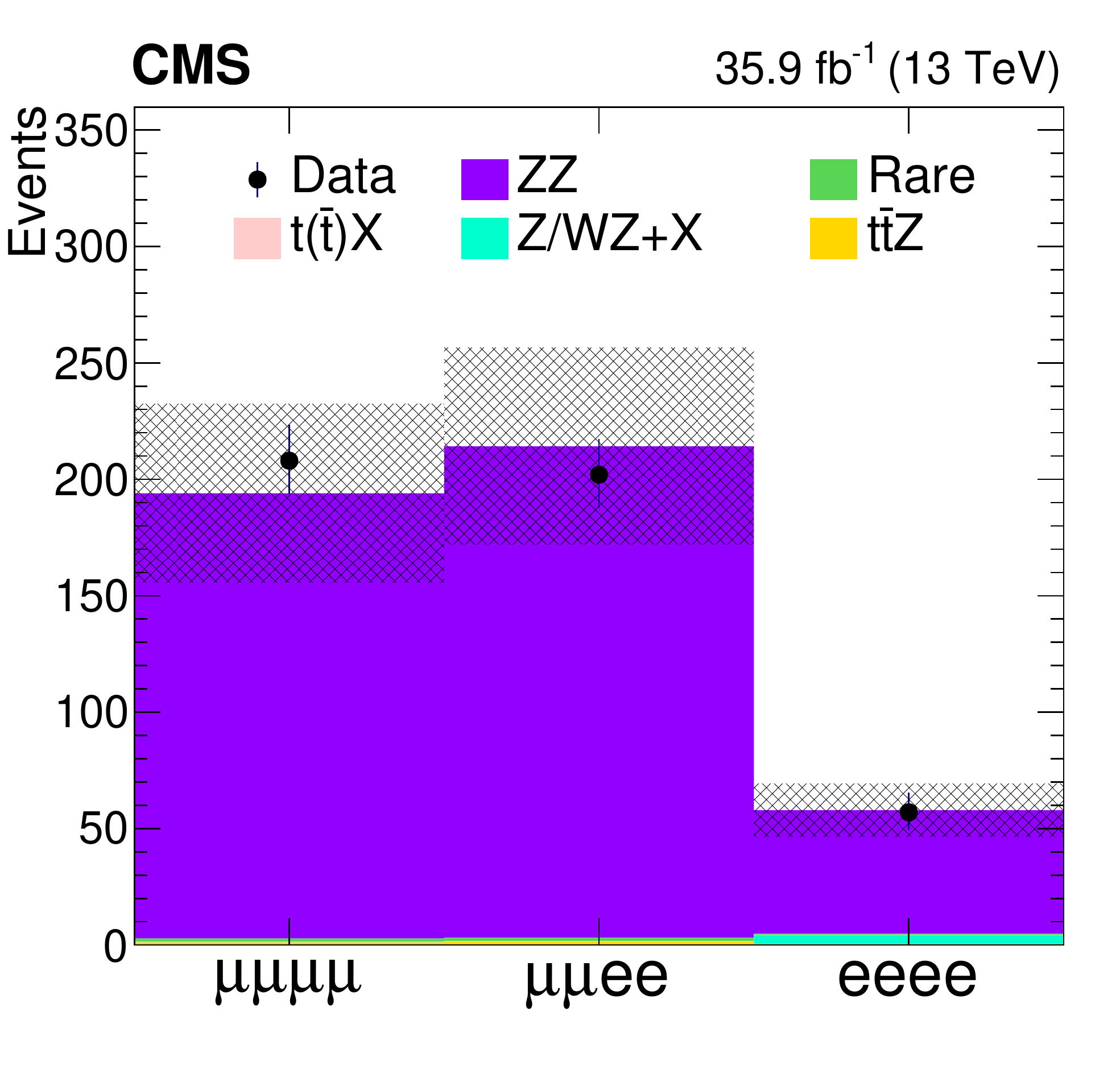}\\
    \includegraphics[width=0.38\textwidth]{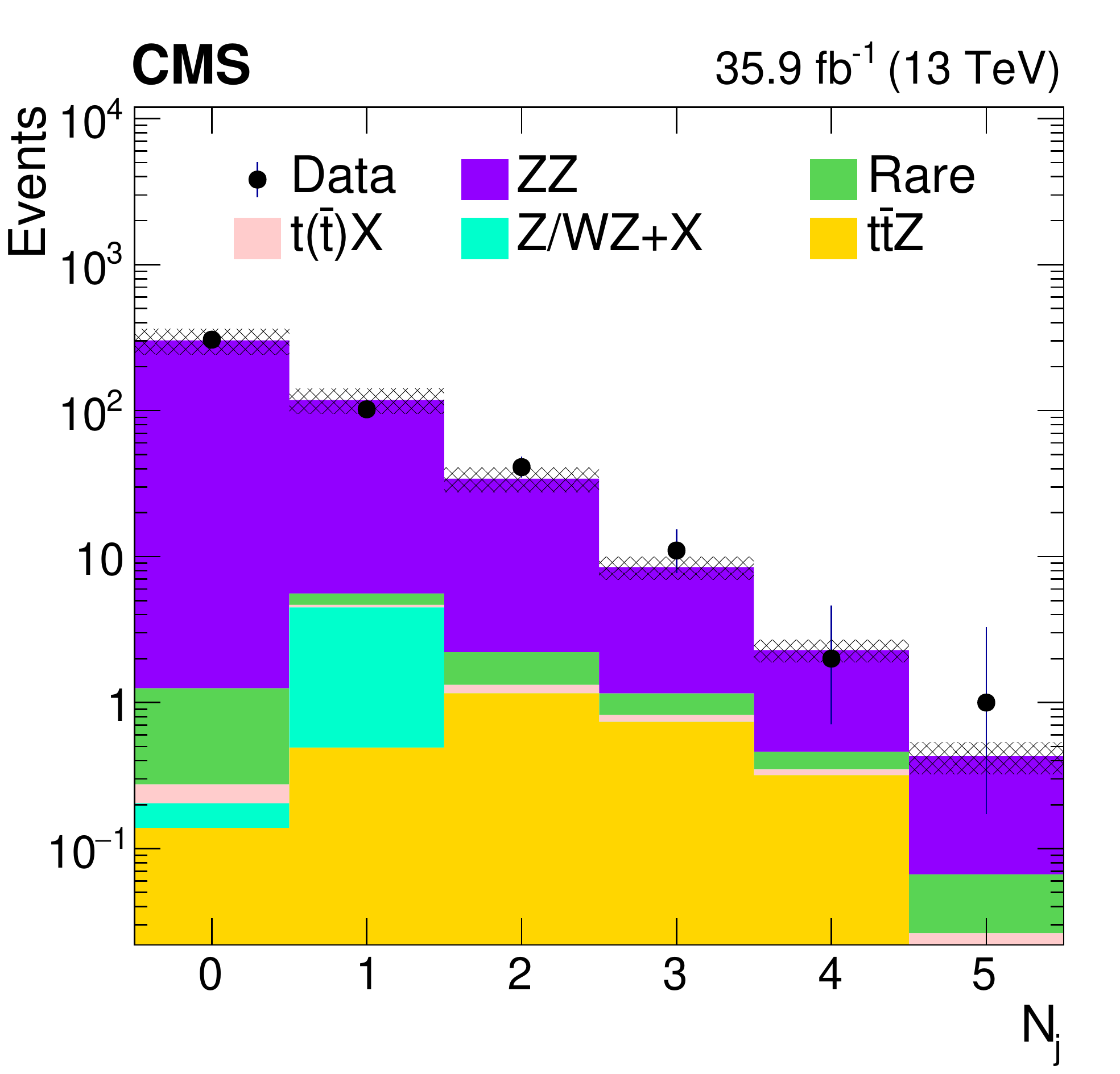}
    \includegraphics[width=0.38\textwidth]{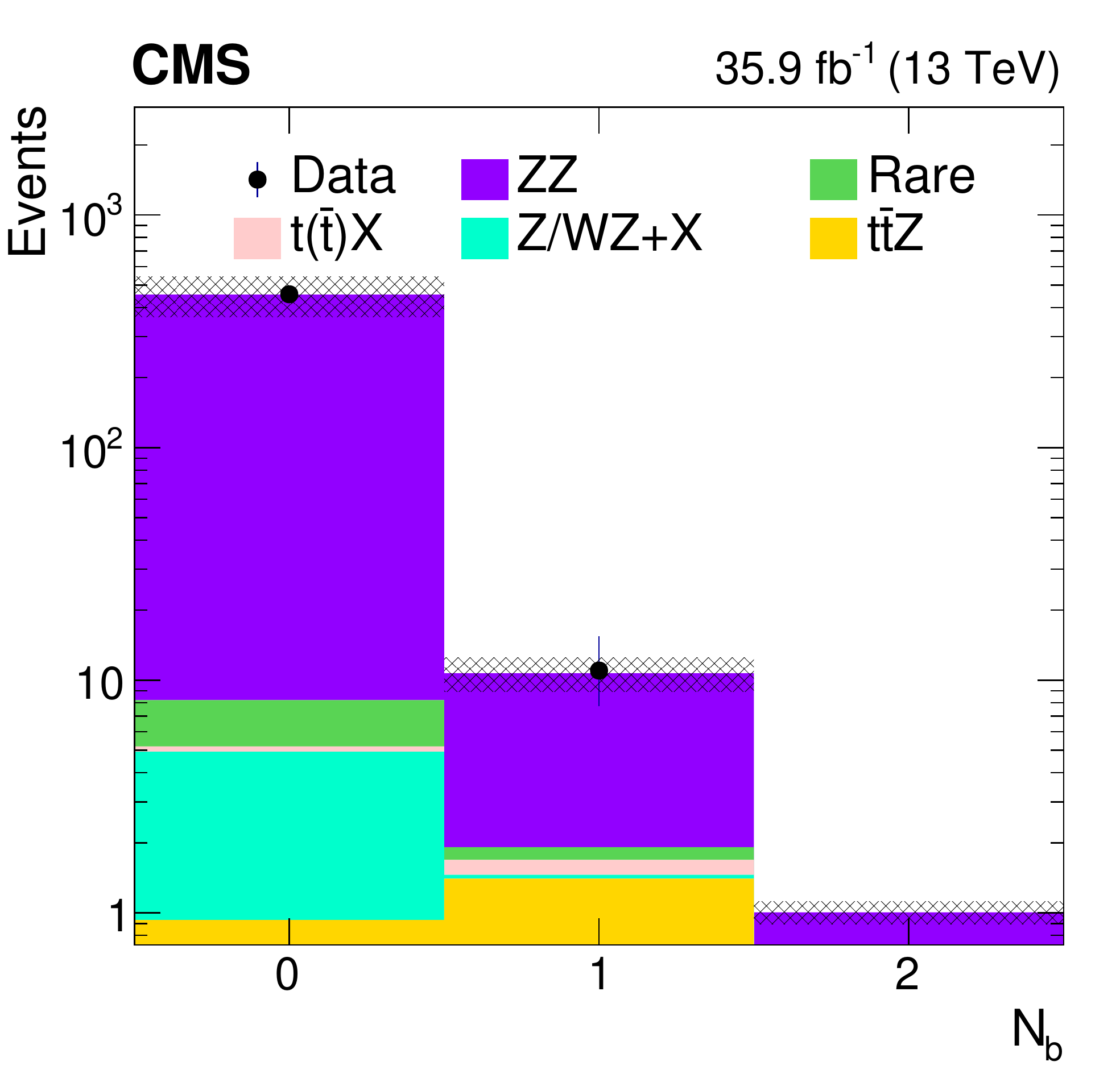}

    \caption{Comparison of data with MC predictions for the mass of the Z boson candidate (upper left), event yields (upper right), jet multiplicity (lower left) and b jet multiplicity (lower right) in a $\PZ\PZ$-dominated background control region. The shaded band represents the total uncertainty in the prediction of the background and the signal processes.}
    \label{fig:controlPlots4L}
\end{figure}

The rate for the backgrounds from rare SM processes, except $\PZ \PZ$, are assigned an overall 50\% systematic uncertainty. This is motivated by the fact that these processes are not yet measured at the LHC and the uncertainties associated with the absence of higher-order effects might be large in the phase-space region relevant to this analysis. For the $\PZ \PZ$ background, the consistency between data and simulation is validated in a $\PZ \PZ$-dominated background region. The events are selected following the first four steps mentioned in Section~\ref{sec:FourLepton}, in the given selection sequence, requiring two OSSF lepton pairs with an invariant mass within a 20\GeV window of $\mZ$. The distributions of the expected and observed data yields in this $\PZ \PZ$ enriched control region are shown in Fig. \ref{fig:controlPlots4L}. The $\PZ \PZ$ control region, which is better than 95\% pure in $\PZ \PZ$ events, shows good agreement between data and simulation in events with extra jets.  Based on this study in the four-lepton control region, as well as considering the studies done for the $\PW \PZ$ background at high jet multiplicities, we assign a 20\%
systematic uncertainty. Additional experimental uncertainties, as previously described for the \ttX and $\PW \PZ$ backgrounds, are also applied to the $\PZ \PZ$ background.

\section{Signal acceptance and systematic uncertainties}
\label{sec:Systematic}
The uncertainty in the integrated luminosity is 2.5\%~\cite{CMS-PAS-LUM-17-001}.
Simulated events are reweighted according to the distribution of the true number of interactions at each bunch crossing.
The uncertainty in the total inelastic $\Pp \Pp$ cross section, which affects the PU estimate, is 5\%~\cite{ATLAS:2016pu} and it leads to a 1--2\% uncertainty in the expected yields.

 We measure the trigger efficiencies in a data sample independent from the one used for the signal selection, as well as in simulation. These efficiencies are measured for each channel separately and parametrized as a function of lepton \pt and $\eta$.  The overall efficiency for the SS dilepton channel is higher than 95\% and that for the three- and four-lepton analyses is greater than 98\%. The trigger efficiencies measured in simulation agree within 1\% with the measurements in data, with an exception of the SS dimuon channel, in which the difference reaches 3\%. The event yields in simulation are therefore scaled to match the trigger efficiencies in data. The systematic uncertainty due to this scaling is 2--4\% depending on the channel.

Reconstructed lepton selection efficiencies are measured using a ``tag-and-probe" method~\cite{Chatrchyan:2012xi,Khachatryan:2015hwa} in bins of lepton \pt and $\eta$, and are higher than 65\,(96)\% for electrons (muons). These measurements are performed separately in data and in simulation. The differences between these two measurements are typically around 1--4\% per
lepton, which corresponds to 3--7\% for all leptons in the event. The systematic uncertainties related to this source vary between 2 and 7\%.

Uncertainties in the jet energy calibrations are estimated by shifting the energy of jets in the simulation up and down by one standard deviation. Depending on $\pt$ and $\eta$,
the uncertainty in jet energy scale changes by 2--5\%~\cite{JME-13-003,CMS-PAS-JME-16-004}. For the signal and backgrounds modelled through simulation, the resulting uncertainty is determined by the observed differences in yields with and without the shift in jet energies.
The same technique is used to calculate the uncertainties caused by the jet energy resolution, for which the uncertainty is found to be 1--6\%.
These uncertainties are also propagated in the \ptmiss variable, and the resulting uncertainty in signal selection is found to be around 1\%.
The b tagging efficiency in the simulation is corrected using scale factors determined from data~\cite{Chatrchyan:2012jua,CMS-PAS-BTV-15-001}.
These contribute with an uncertainty of about 2--5\% on the predicted yields, which depend on \pt, $\eta$ and jet and b-tag multiplicity.

To estimate the theoretical uncertainties due to $\mu_\mathrm{R}$ and $\mu_\mathrm{F}$ choices, each of these parameters is varied independently up and down by a factor of 2, ignoring the anti-correlated variations. For the acceptance uncertainties, the envelope of the results is used as an uncertainty in each search bin, and found not to exceed 2\%. The different replicas in the NNPDF30 PDF set~\cite{Ball:2014uwa} are used to estimate the corresponding uncertainty in acceptance, which is typically less than 1\%.

The theoretical uncertainty in the cross sections for top quark (pair) production in association with a Higgs boson or a vector boson is 11\%~\cite{deFlorian:2016spz}.  For the $\PW\PZ$ and $\PZ\PZ$ backgrounds, the overall uncertainty in the cross section is 10\%, with additional uncertainties at large jet multiplicities. Rare SM processes are assigned a 50\% systematic uncertainty. All of the experimental uncertainties described above are evaluated for each  process in all analysis categories.
A 20\% systematic uncertainty is assigned to the charge-misidentified background.
The uncertainty in the nonprompt lepton contribution in the SS dilepton and three-lepton analyses is 30\%, for which the statistical uncertainty in the observed yields in the sideband region is also taken into account.

The theoretical uncertainties for individual backgrounds as well as the systematic uncertainties for the nonprompt background  are uncorrelated, but correlated across the analysis categories.  The different sources of experimental uncertainty are  correlated across the analysis categories and among the background and signal predictions.
The statistical uncertainties from the limited number of events in MC simulation and from the data events in the sideband regions are considered fully uncorrelated.

The impact of different sources of systematic uncertainty is estimated by fixing the nuisance parameter corresponding to each uncertainty one at a time and evaluating the decrease in the total systematic uncertainty. Uncertainties associated with the integrated luminosity, lepton identification, trigger selection efficiencies, nonprompt lepton, and \ttX backgrounds have the greatest effect on both the \ttW and the \ttZ cross section measurements.  The full set of systematic uncertainties is shown in Table~\ref{table:systematics}.

\begin{table}[h!t]
\topcaption{Summary of the sources of uncertainties, their magnitudes, and their effects in the final measurement. The first column indicates the source of the uncertainties, while the second column shows the corresponding input uncertainty on each background source and the signal. The third and fourth columns show the resulting uncertainties in the respective \ttW and \ttZ cross sections. }
\label{table:systematics}
\centering
\resizebox{1.0\linewidth}{!}{
\begin{tabular}{lccc}
 &  Uncertainty from   & Impact on the measured  & Impact on the measured\\
Source & each source (\%)& \ttW cross section (\%)& \ttZ cross section (\%)\\
\hline
Integrated luminosity & 2.5 & 4 & 3 \\
Jet energy scale and resolution & 2--5 & 3 &  3 \\
Trigger & 2--4 & 4--5  & 5 \\
B tagging &  1--5  &  2--5 & 4--5  \\
PU modeling & 1  & 1  & 1 \\
Lepton ID efficiency & 2--7  & 3 & 6--7 \\
Choice in $\mu_\mathrm{R}$ and $\mu_\mathrm{F}$ &  1  &  $<$1 & 1 \\
PDF & 1 & $<$1 & 1 \\
Nonprompt background & 30  & 4  & $<$2 \\
WZ cross section & 10--20   & $<$1  & 2 \\
ZZ cross section & 20   & ---  & 1 \\
Charge misidentification   & 20   & 3     & --- \\
Rare SM background & 50  & 2 & 2 \\
\ttX background & 10--15  & 4 & 3 \\
Stat. unc. in nonprompt  background& 5--50 & 4 & 2 \\
Stat. unc.  in rare SM backgrounds & 20--100 & 1 &  $<$1\\[\cmsTabSkip]
Total systematic uncertainty&  --- & 14 & 12 \\
\end{tabular}
}
\end{table}

\section {Results}
\label{sec:Results}
\begin{figure}[h!t]
  \centering
   \includegraphics[width=0.80\textwidth]{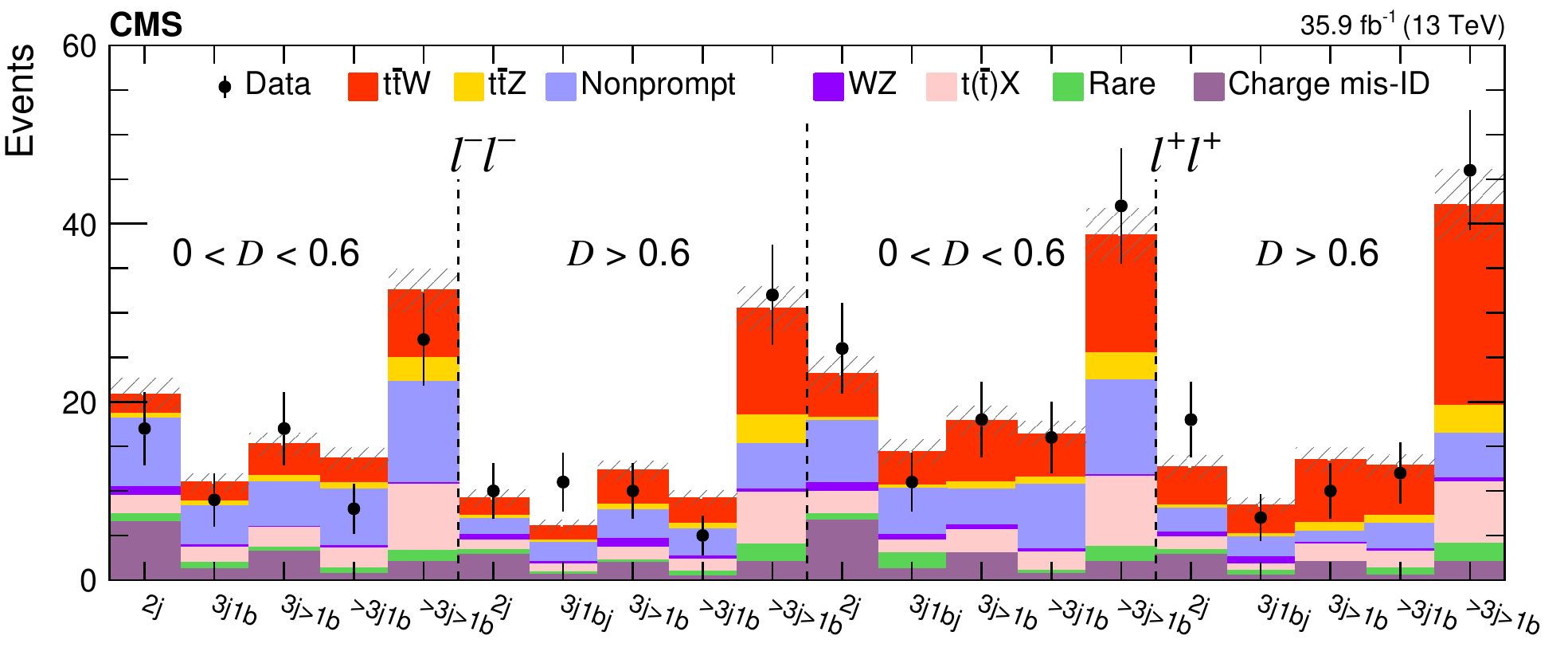}
  \caption{Predicted signal and background yields, as obtained from the fit, compared to observed data in  the SS dilepton analysis. The hatched band shows the total uncertainty associated with the signal and background predictions, as obtained from the fit. }
    \label{fig:SR_ttW}
  \end{figure}

\begin{figure}[h!t]
  \centering
{\includegraphics[width=0.635\textwidth]{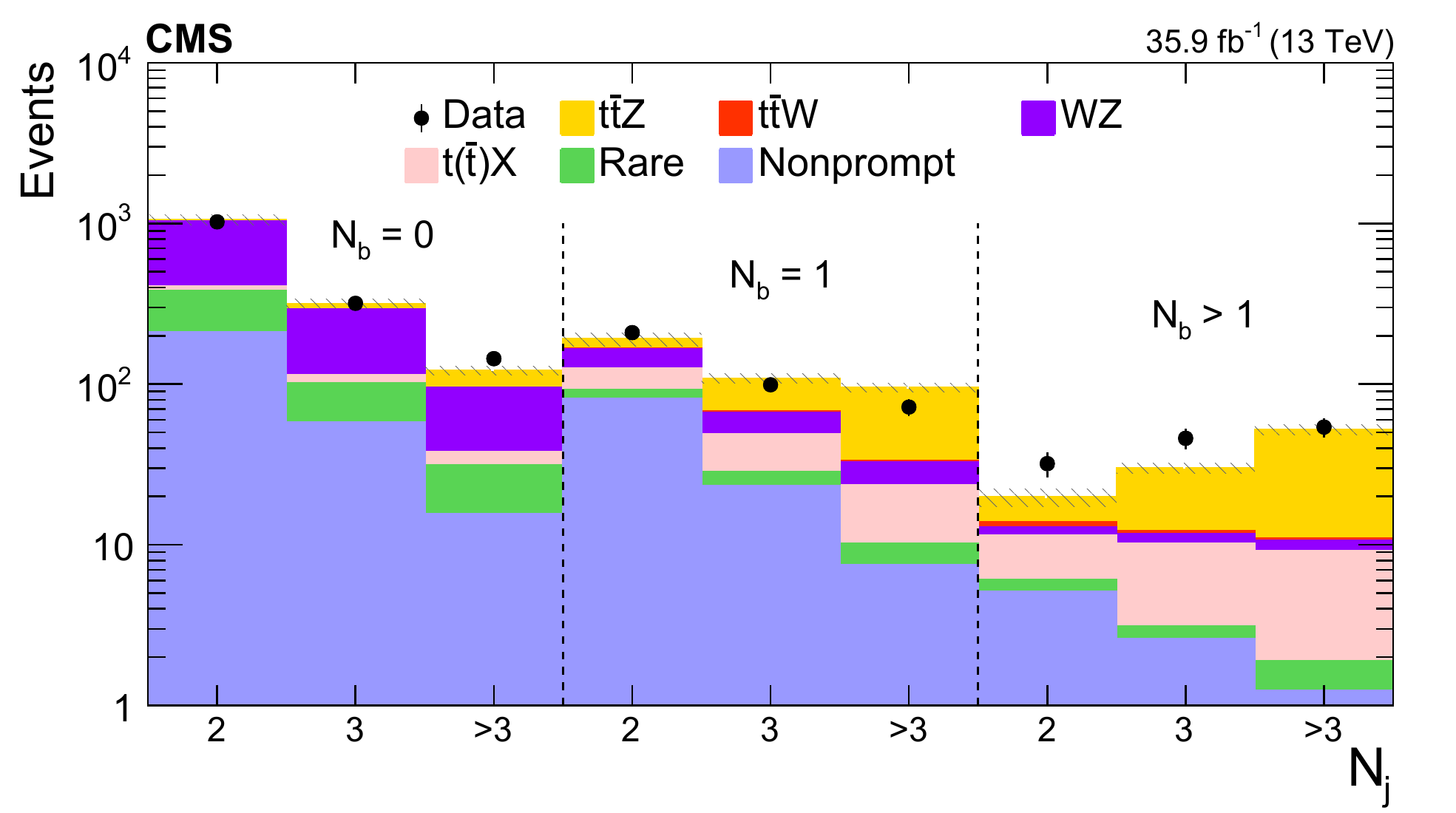}}
{\includegraphics[width=0.31\textwidth]{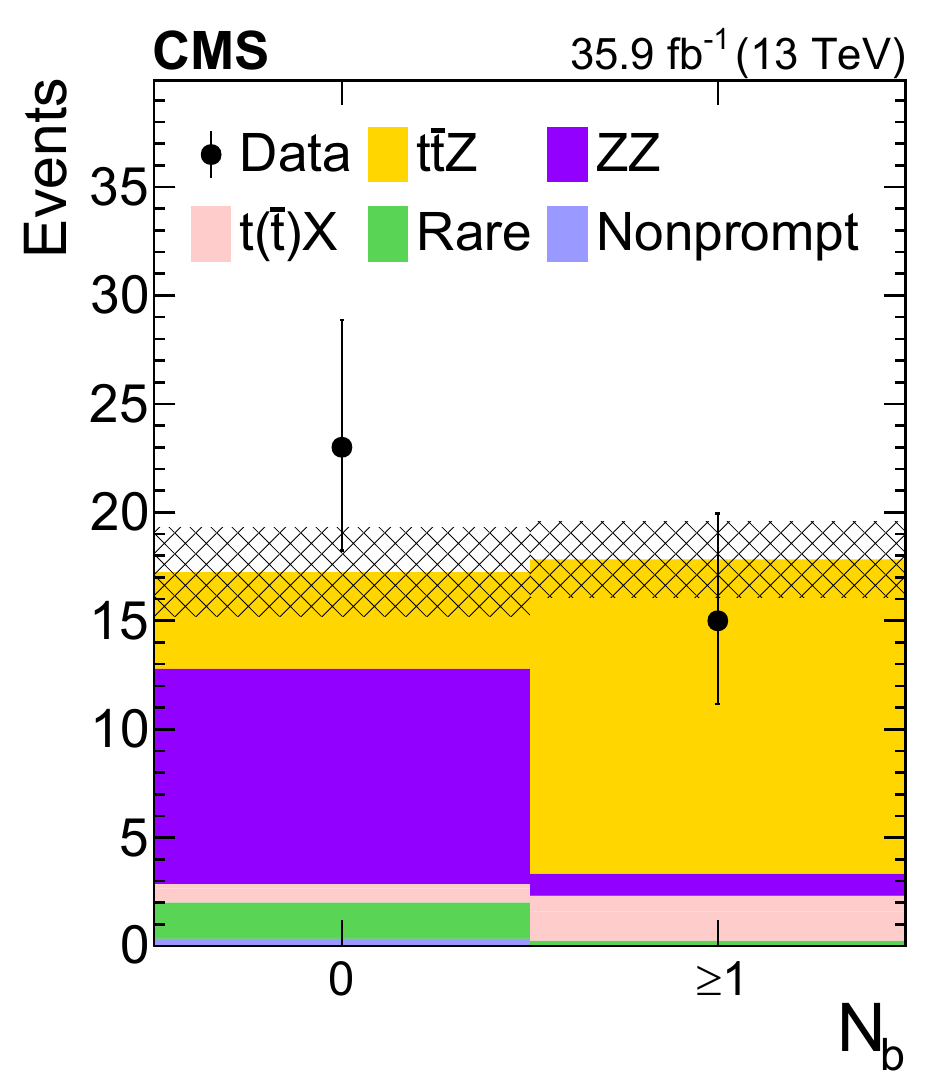}}
  \caption{Predicted signal and background yields, as obtained from the fit, compared to observed data in $\Njets = 2$, 3, and $> 3$ categories in the three-lepton analysis (left), and in $\Nbjets = 0$, 1 categories in the four-lepton analysis (right). The hatched band shows the total uncertainty associated with the signal and background predictions, as obtained from the fit. }
    \label{fig:MoneyPlot1}
  \end{figure}

\begin{figure}[h!t]
\centering
\includegraphics[width=.42\textwidth]{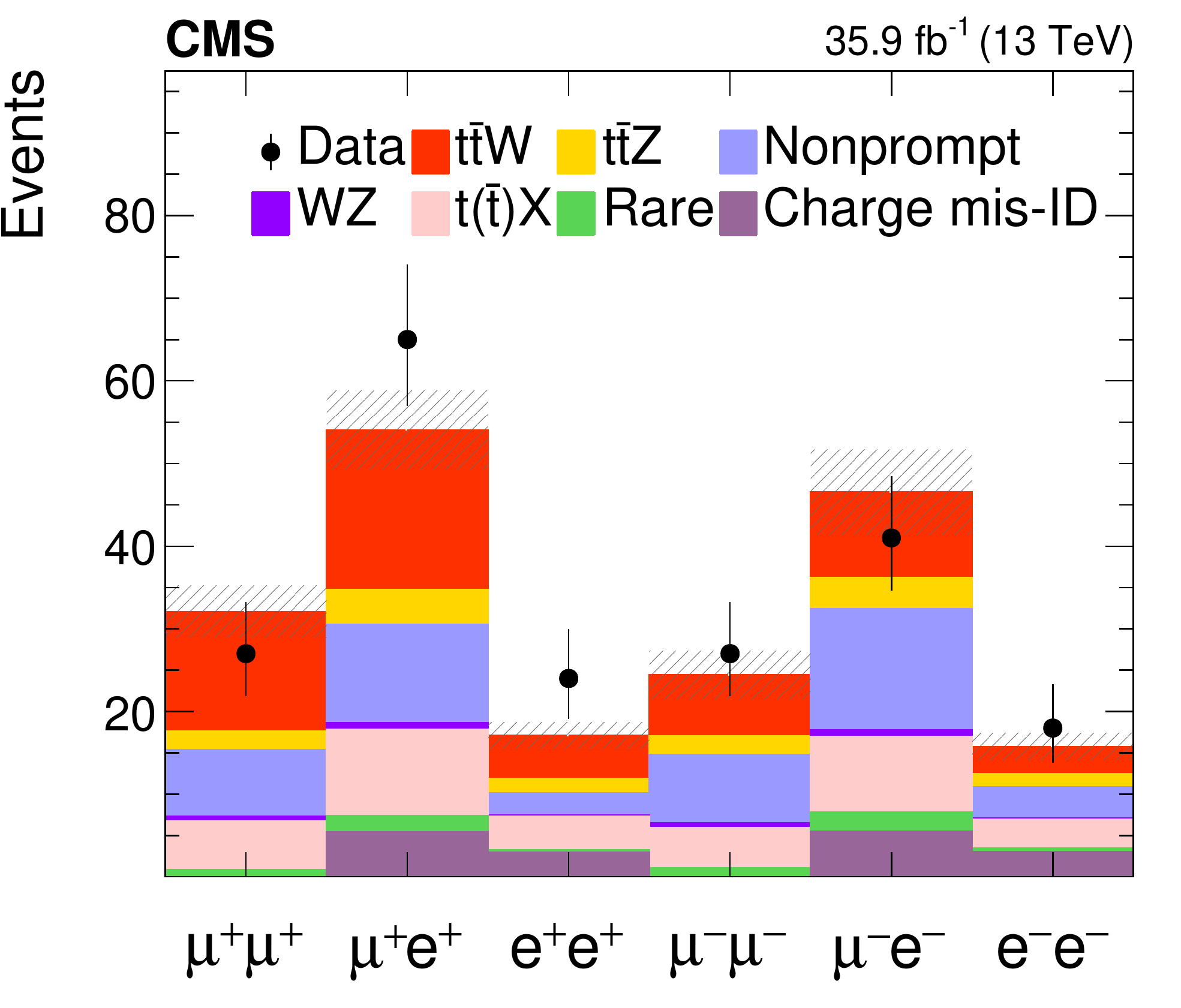}
\includegraphics[width=.42\textwidth]{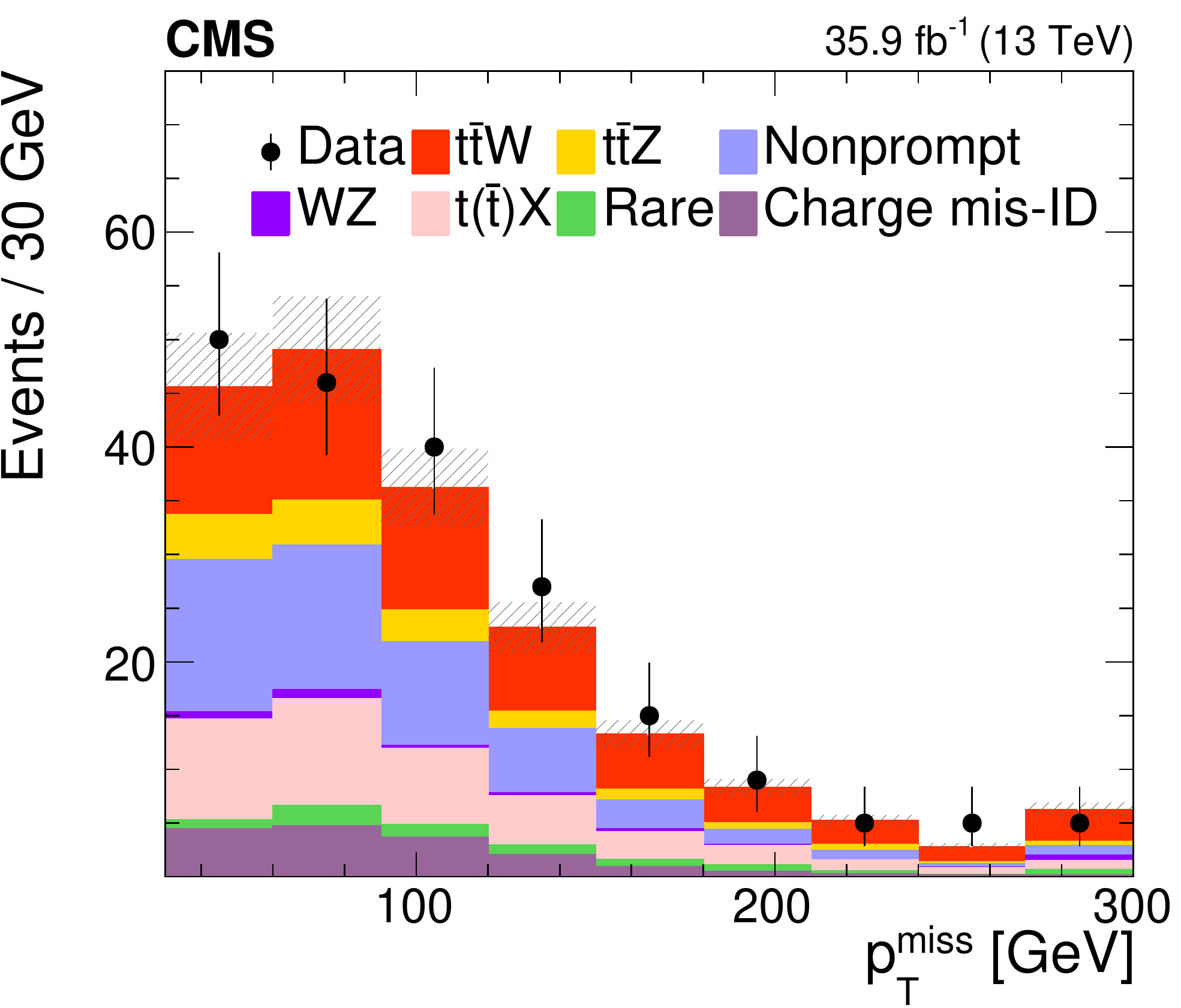}
\includegraphics[width=.42\textwidth]{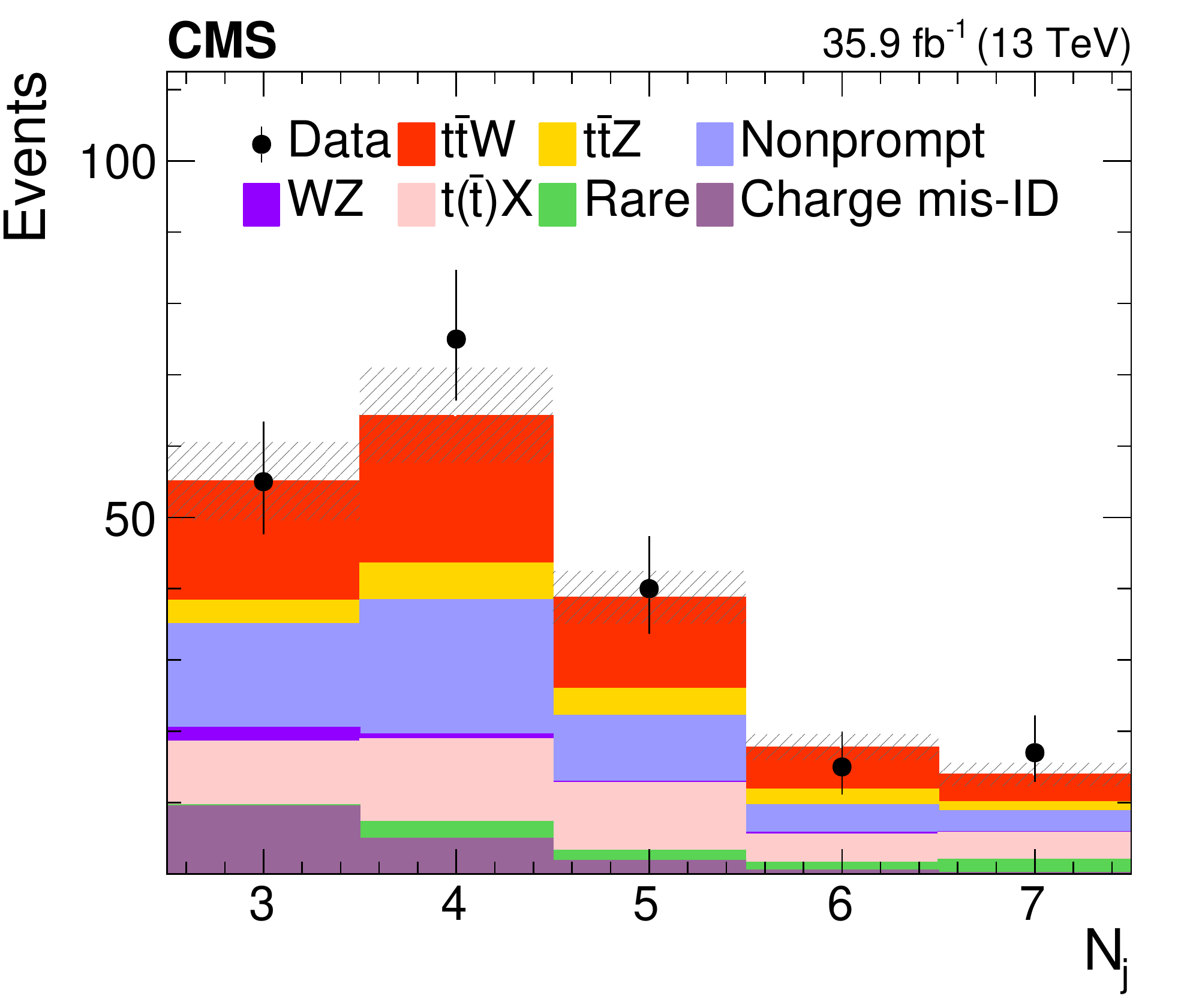}
\includegraphics[width=.42\textwidth]{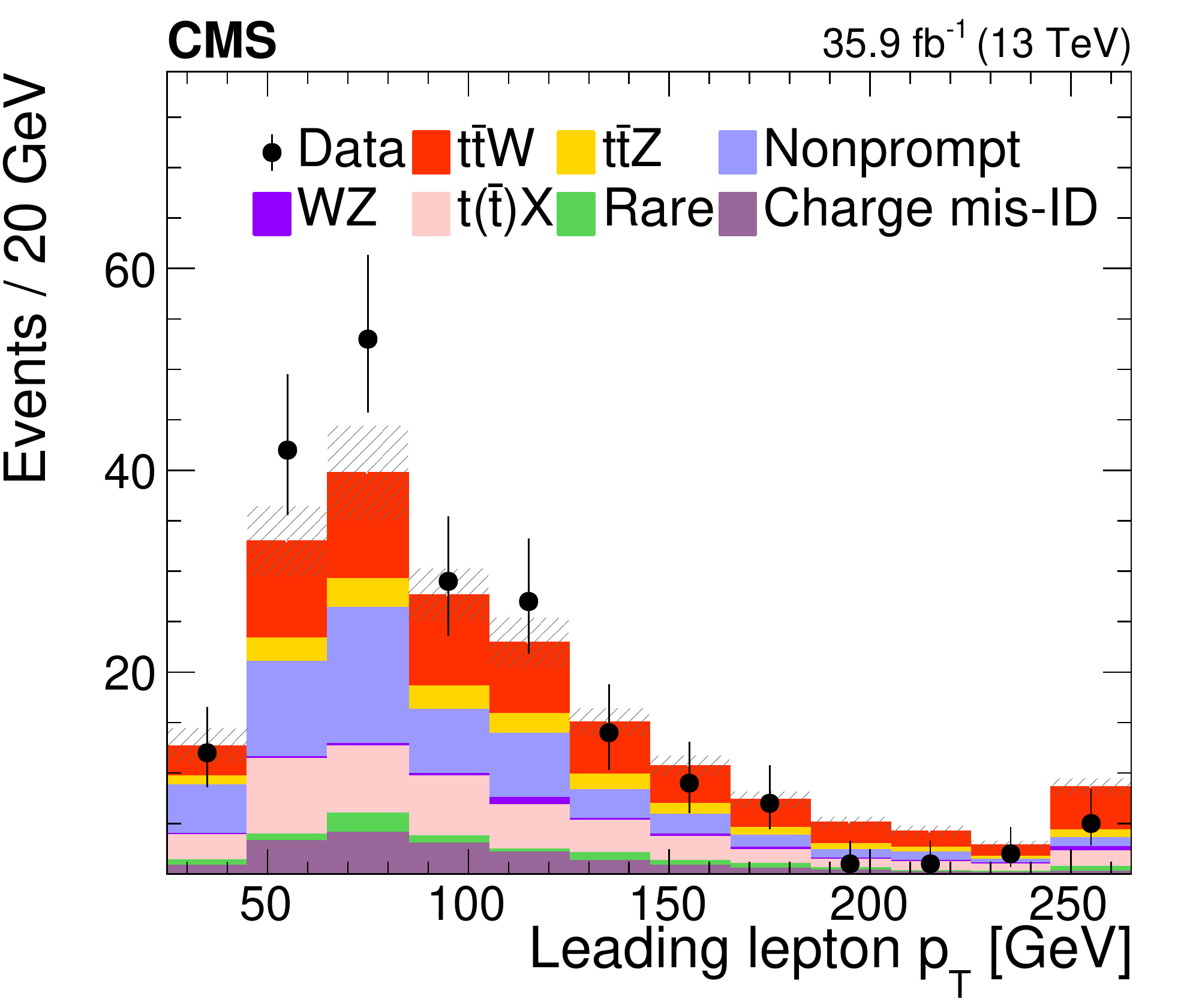}
\caption{Predicted signal and background yields, as obtained from the fit, compared to observed data versus the flavor and the charge combination of leptons (upper left), \ptmiss (upper right),  jet multiplicity (lower left), and the $\pt$ of the leading lepton (lower right) in the SS dilepton channel with at least three jets and at least two b jets. The last bin in each distribution includes the overflow events, and the hatched band shows the total uncertainty associated with the signal and background predictions, as obtained from the fit. }
\label{figures:preselected2L}
\end{figure}

\begin{figure}[h!t]
\centering
\includegraphics[width=.42\textwidth]{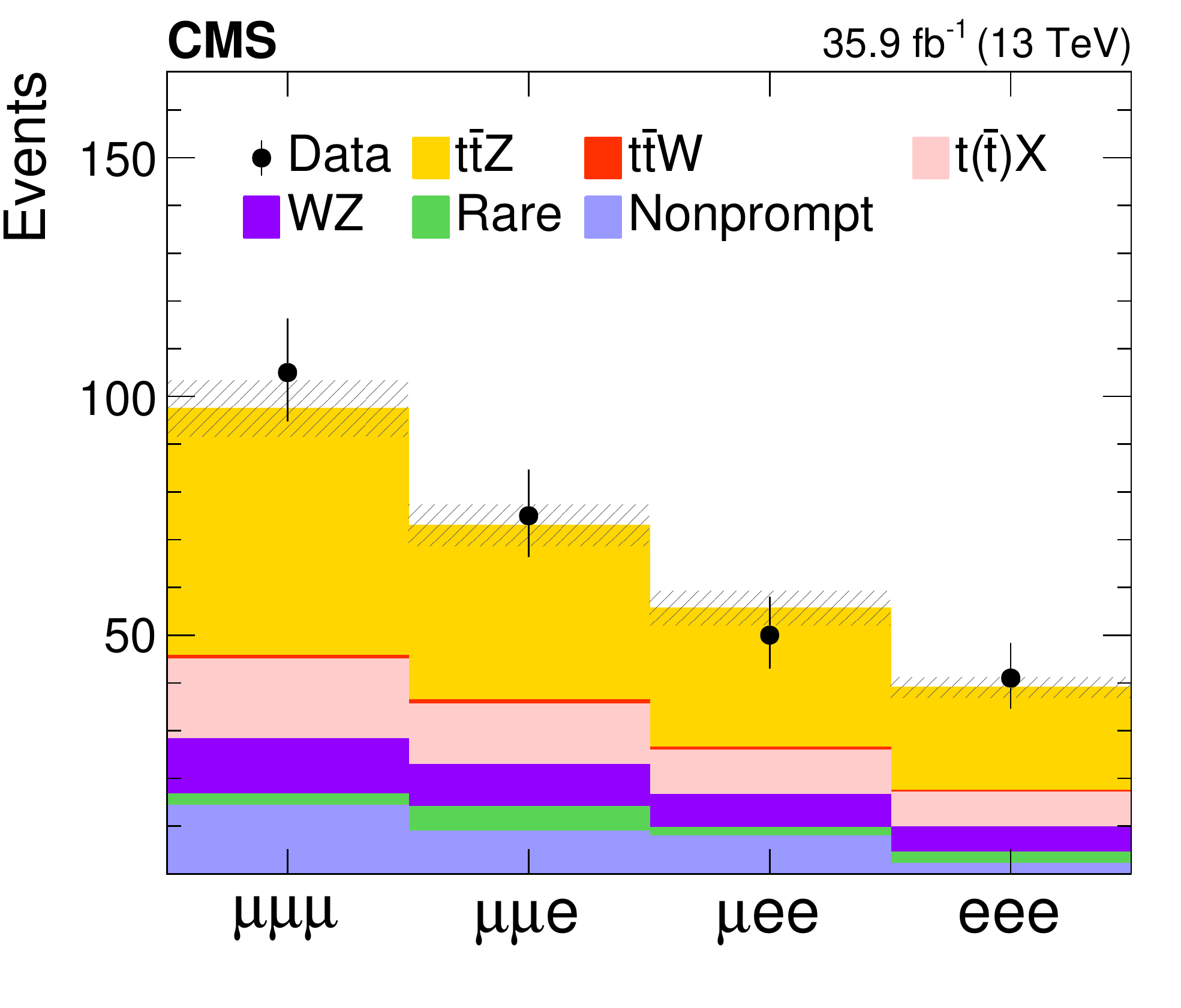}
\includegraphics[width=.42\textwidth]{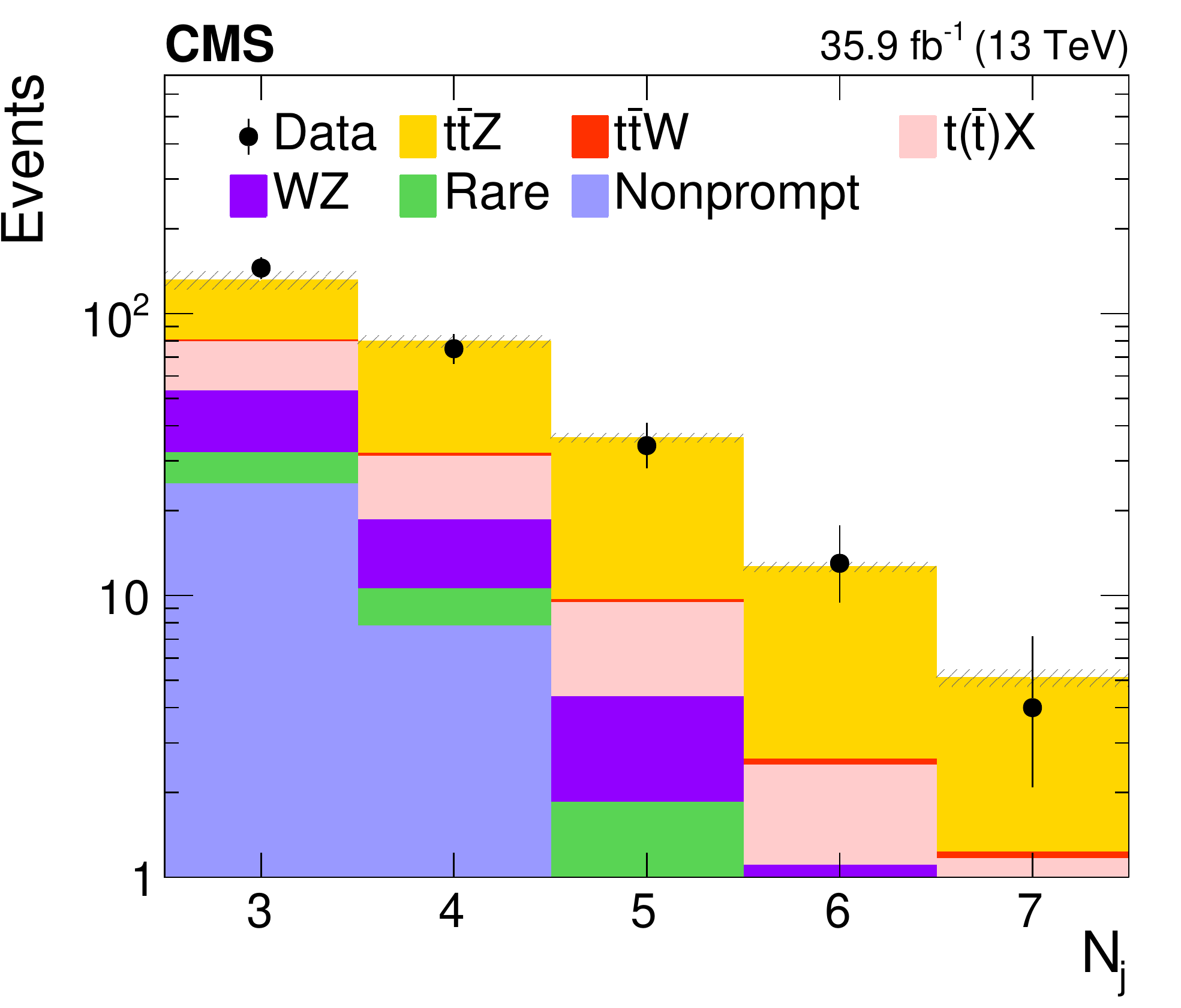}
\includegraphics[width=.42\textwidth]{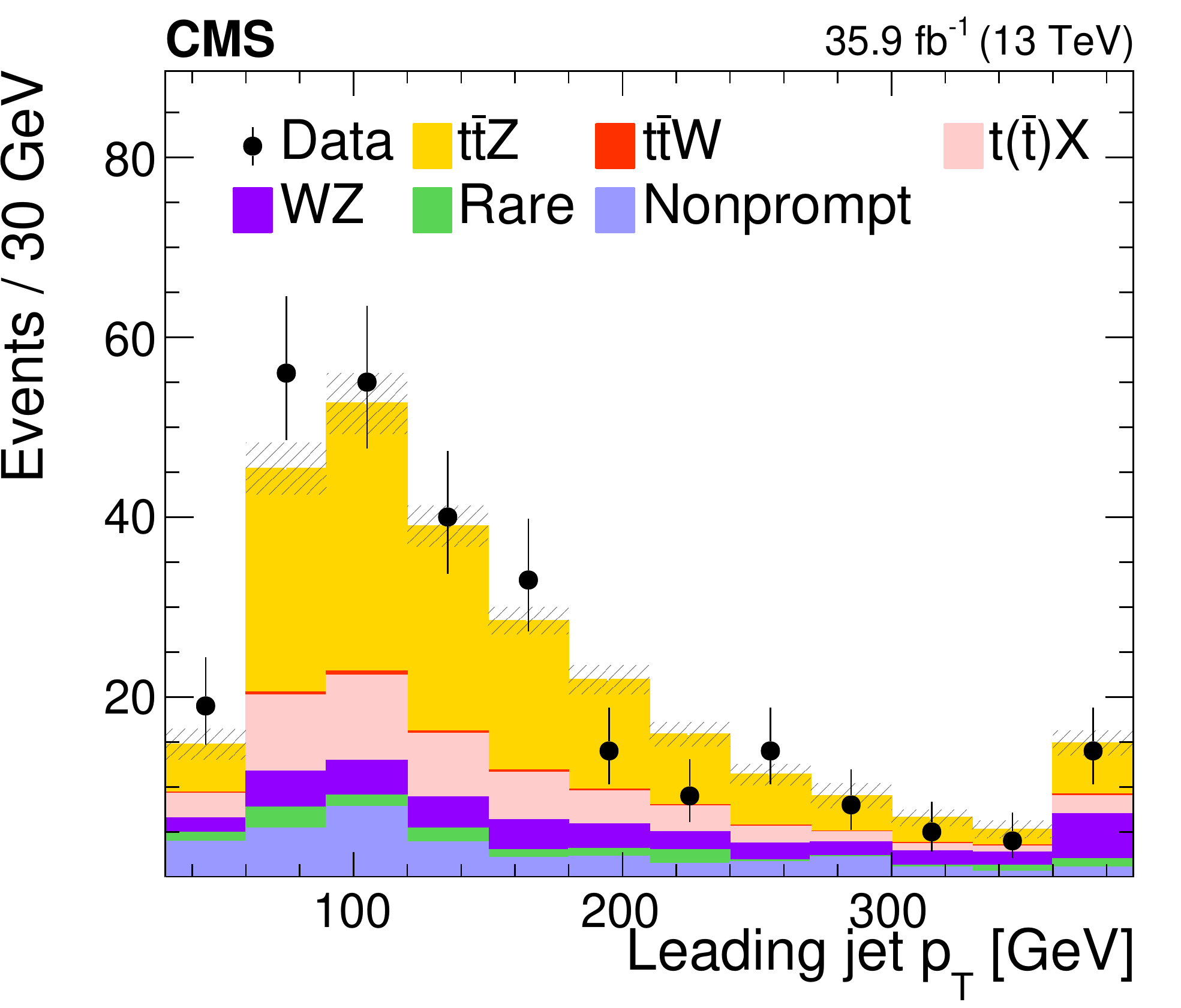}
\includegraphics[width=.42\textwidth]{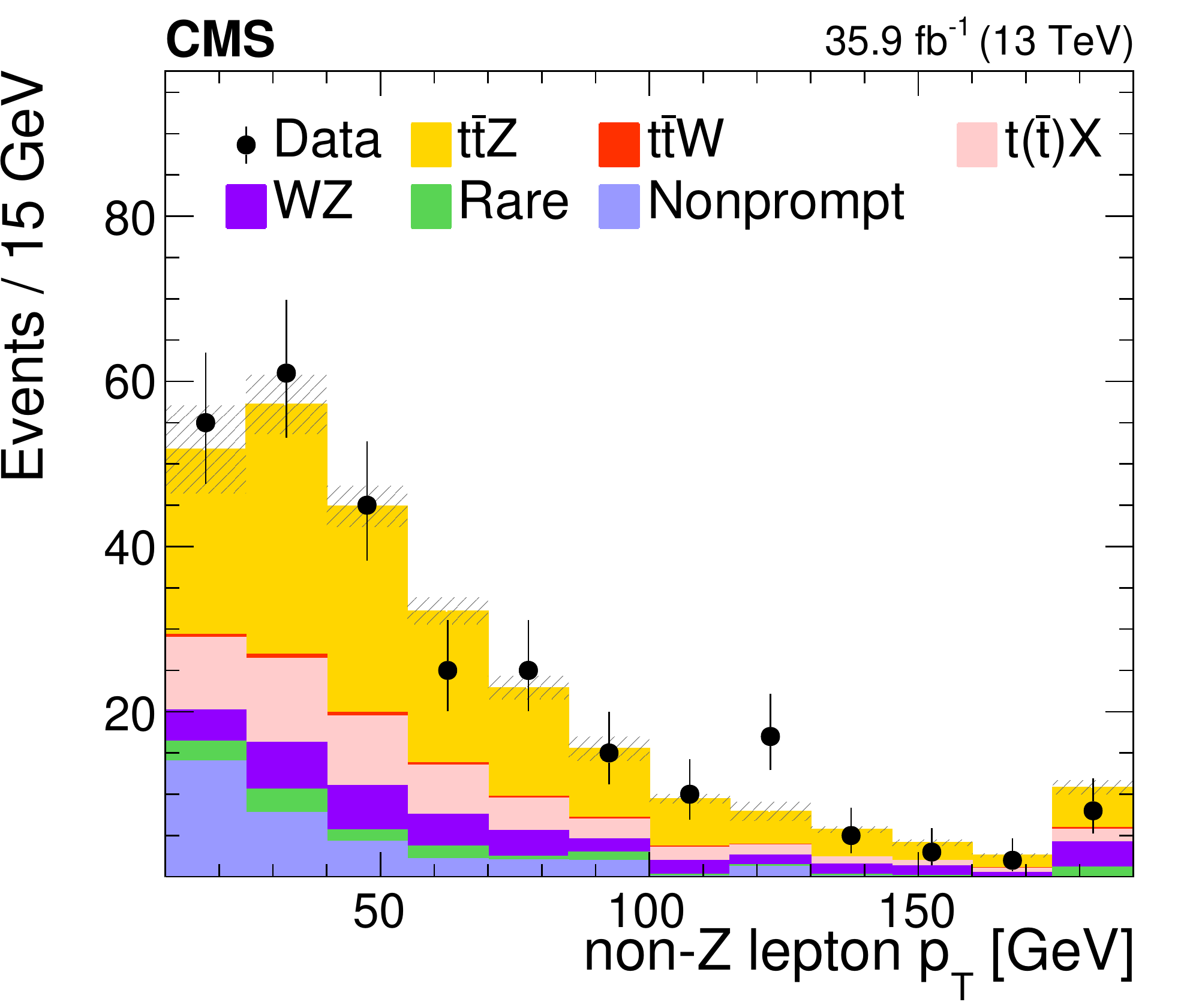}
\includegraphics[width=.42\textwidth]{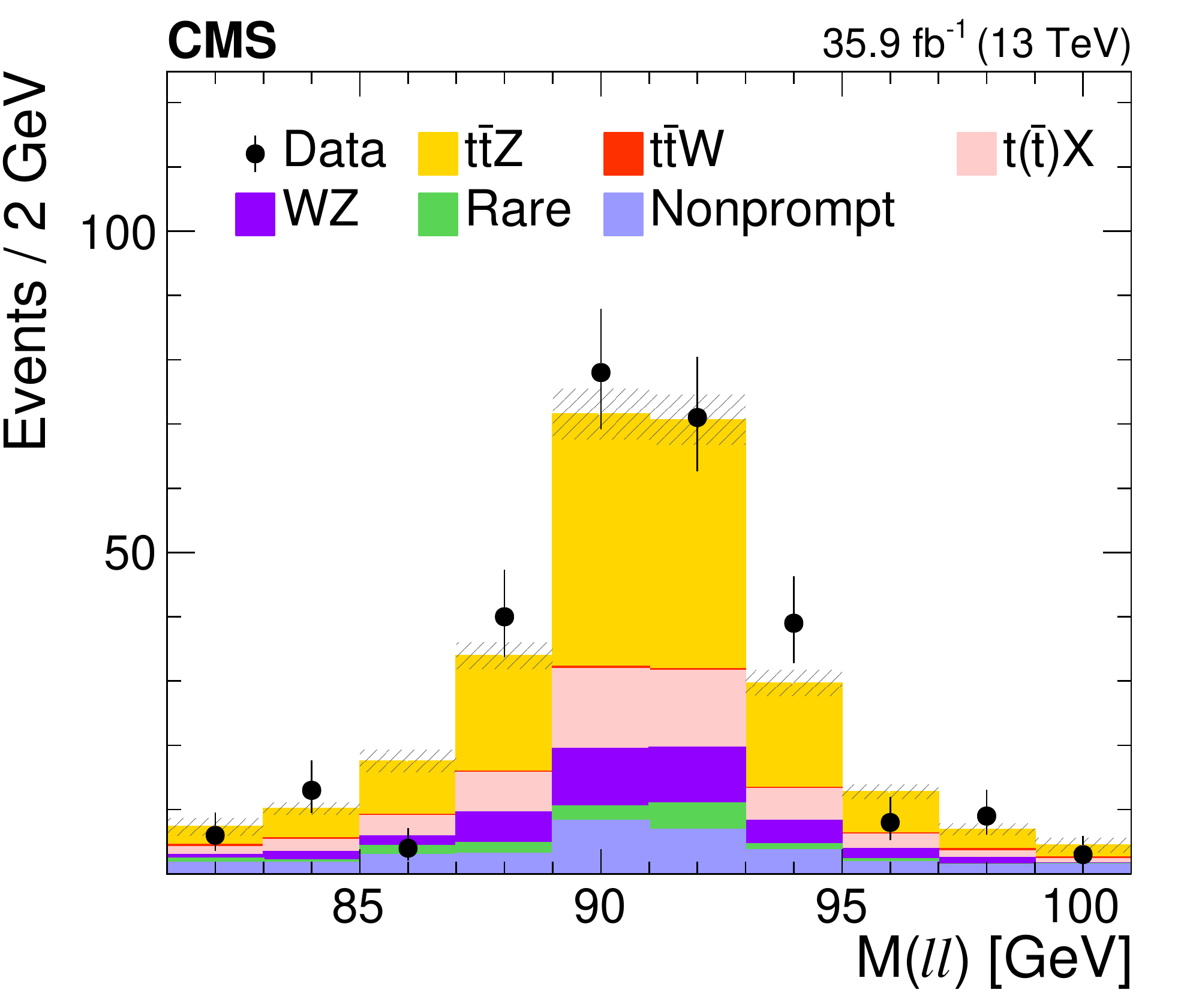}
\includegraphics[width=.42\textwidth]{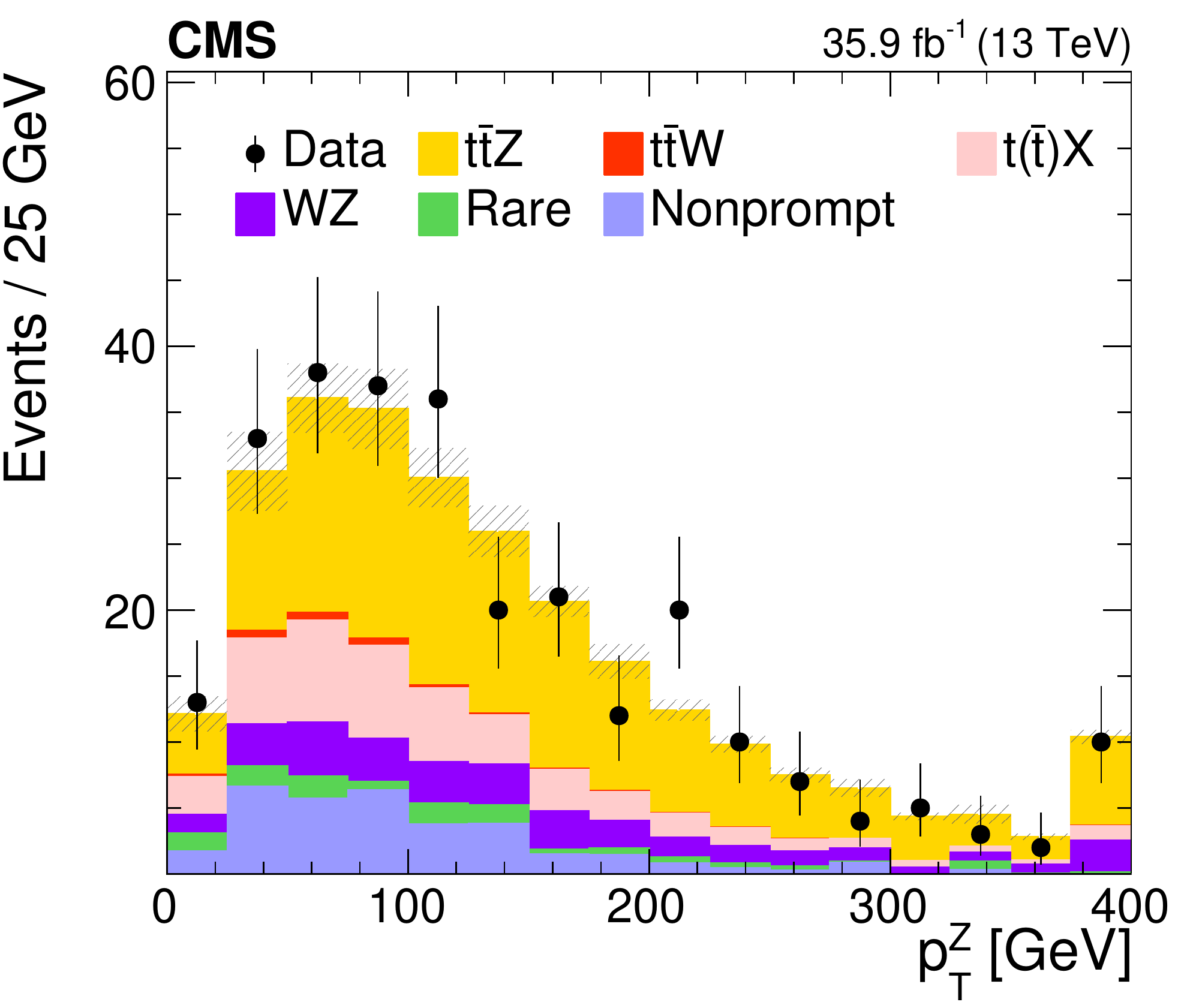}
\caption{Predicted signal and background yields, as obtained from the fit, compared to observed data in the three-lepton channel for events containing at least three jets and at least one b jet. From left to right:  the lepton flavor and jet multiplicity (upper), $\pt$ of the leading jet and the lepton not used to form $\PZ$ (central), and invariant mass of the OSSF lepton pair and $\pt$ of the reconstructed $\PZ$ boson (lower). The last bin in each distribution includes the overflow events, and the hatched band shows the total uncertainty associated with the signal and background predictions, as obtained from the fit.}
\label{figures:preselected3L}
\end{figure}

As described in Section \ref{sec:TwoLepton}, the data are analyzed in three exclusive channels according to the number of leptons in the final state: SS dilepton, three- and four-lepton events. Each channel is further categorized according to the number of jets and b-tagged jets. The predicted SM background and signal yields, and the observed data are shown in Figs.~\ref{fig:SR_ttW} and \ref{fig:MoneyPlot1}, and in Tables \ref{tab:yieldsttW_CR}--\ref{tab:yields4L}, for each of the above categories, respectively. In general, we find good agreement between the predicted yields and the observed data, except for some excess of events accumulated in the $\Njets=2$, 3 and $\Nbjets > 1$ category of the three-lepton channel. Extensive studies were performed to ensure the robustness of the estimated background yields in this region. No hints of a missing or underestimated background were found; therefore, we attribute this excess to a statistical fluctuation in data. In Figs.~\ref{figures:preselected2L} and~\ref{figures:preselected3L}, various kinematic distributions in the predicted and observed yields are presented in \ttW and \ttZ signal-enriched regions: SS dileptons with $\Njets > 2$ and $\Nbjets > 1$, and three-lepton events with $\Njets> 2$ and $\Nbjets> 0$, respectively.

The statistical procedure to extract the cross section is detailed in Refs.~\cite{Junk:1999kv, Read:2002hq, ATL-PHYS-PUB-2011-011, Cowan:2010js}. The observed yields and background estimates in each analysis category, described in Section~\ref{sec:eventselection}, and
the systematic uncertainties described in Section~\ref{sec:Systematic} are used to construct a binned likelihood function $L(r, \theta)$
as a product of Poisson probabilities of all bins. The parameter $r$ is the signal-strength modifier and $\theta$  represents the full suite of nuisance parameters. The signal strength parameter $r = 1$ corresponds to a signal cross section equal to the SM prediction, while $r = 0$ corresponds to the background-only hypothesis.

The test statistic is the profile likelihood ratio, $q(r)=-2L(r,\hat\theta_{\text{r}})/L(\hat{r}, \hat{\theta})$, and asymptotic approximation is used to extract the fitted cross section, the associated uncertainties, and the significance of the observation of the signal process~\cite{Junk:1999kv, Read:2002hq, ATL-PHYS-PUB-2011-011, Cowan:2010js}, where $\hat\theta_{\text{r}}$ reflects the values of the nuisance parameters that maximize the likelihood function for signal strength $r$. The quantities $\hat r$ and $\hat \theta$ are the values that simultaneously maximize $L$.

\begin{table}[h!t]
  \topcaption{ Predicted signal and background yields, as obtained from the fit, compared to observed data in the SS dilepton channel for the $\BDT < 0$ region, \ie the nonprompt lepton control region. The total uncertainty obtained from the fit is also shown.}
  \label{tab:yieldsttW_CR}
  \centering
      \begin{tabular}{lcccccc}
      Process & $\Njets = 2$ & $\Njets = 3 $ & $\Njets > 3$\\
      \hline
      Nonprompt         & $136.5\pm13.9$   & $110.3\pm11.3$   & $57.3\pm6.1$   \\
      Total background  & $192.1\pm15.6$   & $137.7\pm11.7$   & $74.0\pm6.4$   \\[\cmsTabSkip]
      \ttW              & $13.1\pm0.3$   & $17.6\pm0.3$   & $13.8\pm0.3$   \\
      \ttZ              & $1.6\pm0.4$   & $3.1\pm0.7$   & $4.4\pm1.0$   \\[\cmsTabSkip]
      Total             & $206.8\pm15.7$   & $158.4\pm11.8$   & $92.3\pm6.5$   \\
      Observed          & 229 & 144 & 92 \\
      \end{tabular}
\end{table}

\begin{table}[h!t]
\topcaption{ Predicted signal and background yields, as obtained from the fit, compared to observed data in the SS dilepton final state. The total uncertainty obtained from the fit is also shown.}
      \label{tab:yieldsttW}
\centering
\newcolumntype{x}{D{,}{\,\pm\,}{3.2}}
\begin{tabular}{c|crrxxxxr}
\multicolumn{1}{c}{}&\multicolumn{1}{c}{} & \Njets & \Nbjets &  \multicolumn{1}{c}{Background} & \multicolumn{1}{c}{\ttW} & \ttZ & \multicolumn{1}{c}{Total} & Observed \\ \hline
\multirow{10}{*}{$\ell^{-}\ell^{-}$}& \multirow{5}{*}{$0 < \textit{D} < 0.6$}&   2  & $>$0 & 18.1,1.8 & 2.2,0.4&0.5,0.1&20.8,1.9&17\\
 &  &  \multirow{2}{*}{ 3} &  1  & 8.3,0.9 & 2.1,0.4&0.5,0.1&10.9,0.9&9\\
 &  &   & $>$1  & 10.9,1.1 & 3.5,0.6&0.8,0.1&15.2,1.3&17\\
 &  &  \multirow{2}{*}{ $>$3} &  1  & 10.1,1.1 & 2.8,0.5&0.7,0.2&13.7,1.3&8\\
 &  &   & $>$1  & 22.2,2.0 & 7.6,1.2&2.7,0.4&32.5,2.4&27\\ [\cmsTabSkip]
 & \multirow{5}{*}{$\textit{D} > 0.6$}&   2  & $>$0 & 6.8,0.9 & 2.0,0.3&0.4,0.1&9.2,0.9&10\\
 &  &  \multirow{2}{*}{ 3} &  1  & 4.1,0.6 & 1.6,0.3&0.3,0.1&6.1,0.6&11\\
 &  &   & $>$1  & 7.8,0.9 & 3.8,0.6&0.7,0.1&12.3,1.1&10\\
 &  &  \multirow{2}{*}{ $>$3} &  1  & 5.6,0.7 & 2.9,0.5&0.7,0.2&9.2,0.9&5\\
 &  &   & $>$1  & 15.3,1.5 & 12.0,1.9&3.2,0.5&30.5,2.5&32\\ \hline
\multirow{10}{*}{$\ell^{+}\ell^{+}$}& \multirow{5}{*}{$0 < \textit{D} < 0.6$}&   2  & $>$0 & 17.9,1.8 & 4.9,0.8&0.3,0.1&23.1,2.0&26\\
 &  &  \multirow{2}{*}{ 3} &  1  & 10.2,1.3 & 3.7,0.6&0.4,0.1&14.4,1.4&11\\
 &  &   & $>$1  & 10.2,1.2 & 6.9,1.1&0.8,0.2&17.9,1.6&18\\
 &  &  \multirow{2}{*}{ $>$3} &  1  & 10.7,1.2 & 4.9,0.8&0.8,0.2&16.4,1.4&16\\
 &  &   & $>$1  & 22.4,2.0 & 13.3,2.2&3.0,0.5&38.7,3.0&42\\ [\cmsTabSkip]
 & \multirow{5}{*}{$\textit{D} > 0.6$}&   2  & $>$0 & 8.0,1.1 & 4.3,0.7&0.4,0.1&12.7,1.3&18\\
 &  &  \multirow{2}{*}{ 3} & 1  & 4.8,0.7 & 3.2,0.5&0.3,0.1&8.4,0.9&7\\
 &  &   & $>$1  & 5.4,0.7 & 7.1,1.2&1.0,0.2&13.5,1.4&10\\
 &  &  \multirow{2}{*}{ $>$3} &  1  & 6.3,0.8 & 5.6,0.9&0.9,0.2&12.8,1.2&12\\
 &  &   & $>$1  & 16.5,1.5 & 22.5,3.7&3.1,0.5&42.1,4.0&46\\
\end{tabular}
\end{table}

\begin{table}[h!t]
\topcaption{ Predicted signal and background yields, as obtained from the fit, compared to observed data in the three-lepton final state. The total uncertainty obtained from the fit is also shown. }
      \label{tab:yields3LttZ}
      \centering
\newcolumntype{y}{D{,}{\,\pm\,}{5.3}}
\begin{tabular}{rryyyyr}
\Nbjets & \Njets &   \multicolumn{1}{c}{Background}&  \multicolumn{1}{c}{\ttW} &  \multicolumn{1}{c}{\ttZ} &  \multicolumn{1}{c}{Total} &  \multicolumn{1}{c}{Observed} \\ \hline
\multirow{3}{*}{ 0} & 2  & 1032.8,77.1 & 0.9,0.1&18.2,3.2&1051.9,77.2&1022\\
 & 3  & 293.5,21.4 & 0.4,0.1&22.3,3.9&316.3,21.8&318\\
 &$>$3 & 95.4,7.4 & 0.3,0.1&26.1,4.6&121.8,8.7&144\\[\cmsTabSkip]
\multirow{3}{*}{ 1} & 2  & 164.6,17.8 & 1.9,0.3&24.3,4.3&190.7,18.3&209\\
 & 3  & 66.6,6.7 & 0.9,0.2&41.2,7.2&108.7,9.8&99\\
 &$>$3 & 32.8,3.3 & 0.8,0.1&61.3,10.8&94.9,11.3&72\\[\cmsTabSkip]
\multirow{3}{*}{ $>$1} & 2  & 12.9,2.4 & 1.0,0.2&5.9,1.0&19.8,2.6&32\\
 & 3  & 11.6,1.7 & 0.6,0.1&17.9,3.2&30.1,3.6&46\\
 &$>$3 & 10.6,1.6 & 0.4,0.1&41.0,7.2&52.0,7.4&54\\
\end{tabular}
\end{table}

\begin{table}[h!t]
\topcaption{ Predicted signal and background yields, as obtained from the fit, compared to observed data in  the four-lepton final state. The total uncertainty obtained from the fit is also shown. }
\label{tab:yields4L}
\centering
\newcolumntype{z}{D{,}{\,\pm\,}{3.2}}
\begin{tabular}{lzz}
Process                                          &    \multicolumn{1}{c}{ $\Nbjets = 0$}                    &   \multicolumn{1}{c}{  $\Nbjets   >  0$ }        \\ \hline
Total background                         &       12.8,2.0              &        3.3,0.3 \\
\ttZ                                                   &         4.5,0.6                &       14.5,1.8 \\[\cmsTabSkip]
Total                                               &       17.2,2.0                &        17.8,1.8 \\
Observed                                      &       \multicolumn{1}{c}{ 23}                                       &       \multicolumn{1}{c}{ 15}                               \\
\end{tabular}
\end{table}

The measurement of the individual cross sections for \ttW  and \ttZ is performed using the events in the SS dilepton, and the three- and four-lepton categories, respectively, while the $\ttW^{+}$($\ttW^{-}$) signal extraction is performed using the SS dilepton category with $\ell^+\ell^+$($\ell^-\ell^-$). The summary of the expected and observed signal significances for each of these processes is given in Table~\ref{tab:significance}.
We find an expected (observed) signal significance of 4.5 (5.3) standard deviations in the SS dilepton channel, and
4.7 (4.5) standard deviations in the four-lepton channel, while in three-lepton channel both the expected and the observed significances are found to be much larger than 5 standard deviations.
The expected (observed) signal significances for $\ttW^{+}$ and $\ttW^{-}$ processes are calculated as well, being 4.2 (5.5) and 2.4 (2.3), respectively.

\begin{table}[h!t]
\topcaption{ \label{tab:significance} Summary of expected and observed significances (in standard deviations) for \ttW and \ttZ.
  }
\centering
\begin{tabular}{lrr}
Channel & Expected significance & Observed significance \\ \hline
SS dilepton $\ell^-\ell^-$ ($\ttW^{-}$) &  2.4 & 2.3 \\
SS dilepton $\ell^+\ell^+$($\ttW^{+}$) &  4.2 & 5.5 \\
SS dilepton ${\ell^\pm}\ell^{\pm}$  ($\ttW^\pm$) &  4.5 & 5.3 \\
Three-lepton (\ttZ) &    $>$5.0  & $> $5.0\\
Four-lepton (\ttZ)  &  4.7 & 4.5 \\
Three- and four-lepton combined (\ttZ)  & $>$5.0  & $>$5.0\\
\end{tabular}
\end{table}

The measured signal strength parameters are found to be
$1.23^{+0.19}_{-0.18}\stat^{+0.20}_{-0.18}\syst^{+0.13}_{-0.12}\thy$ for \ttW, and  $1.17~^{+0.11}_{-0.10}\stat^{+0.14}_{-0.12}\syst^{+0.11}_{-0.12}\thy$ for \ttZ.
These parameters are used to multiply the corresponding theoretical cross sections for \ttW and \ttZ mentioned in Section~\ref{sec:objects}, to obtain the measured cross sections for \ttW and \ttZ:
\begin{equation*}
\sigma(\Pp\Pp \to \ttW)=0.77^{+0.12}_{-0.11}\stat^{+0.13}_{-0.12}\syst\unit{pb},
\end{equation*}
\begin{equation*}
\sigma(\Pp\Pp \to \ttZ)=0.99^{+0.09}_{-0.08}\stat^{+0.12}_{-0.10}\syst\unit{pb}.\\
\end{equation*}

The measured cross sections for the $\ttW^{+}$ and $\ttW^{-}$ processes are:
\begin{equation*}
    \sigma(\Pp\Pp \to \ttW^{+})=0.58\pm{0.09}\stat^{+0.09}_{-0.08}\syst\unit{pb},\\
\end{equation*}
\begin{equation*}
    \sigma(\Pp\Pp \to \ttW^{-})=0.19\pm{0.07}\stat\pm{0.06}\syst\unit{pb}.
\end{equation*}

\begin{figure}[h]
    \centering
    \includegraphics[width=0.80\textwidth]{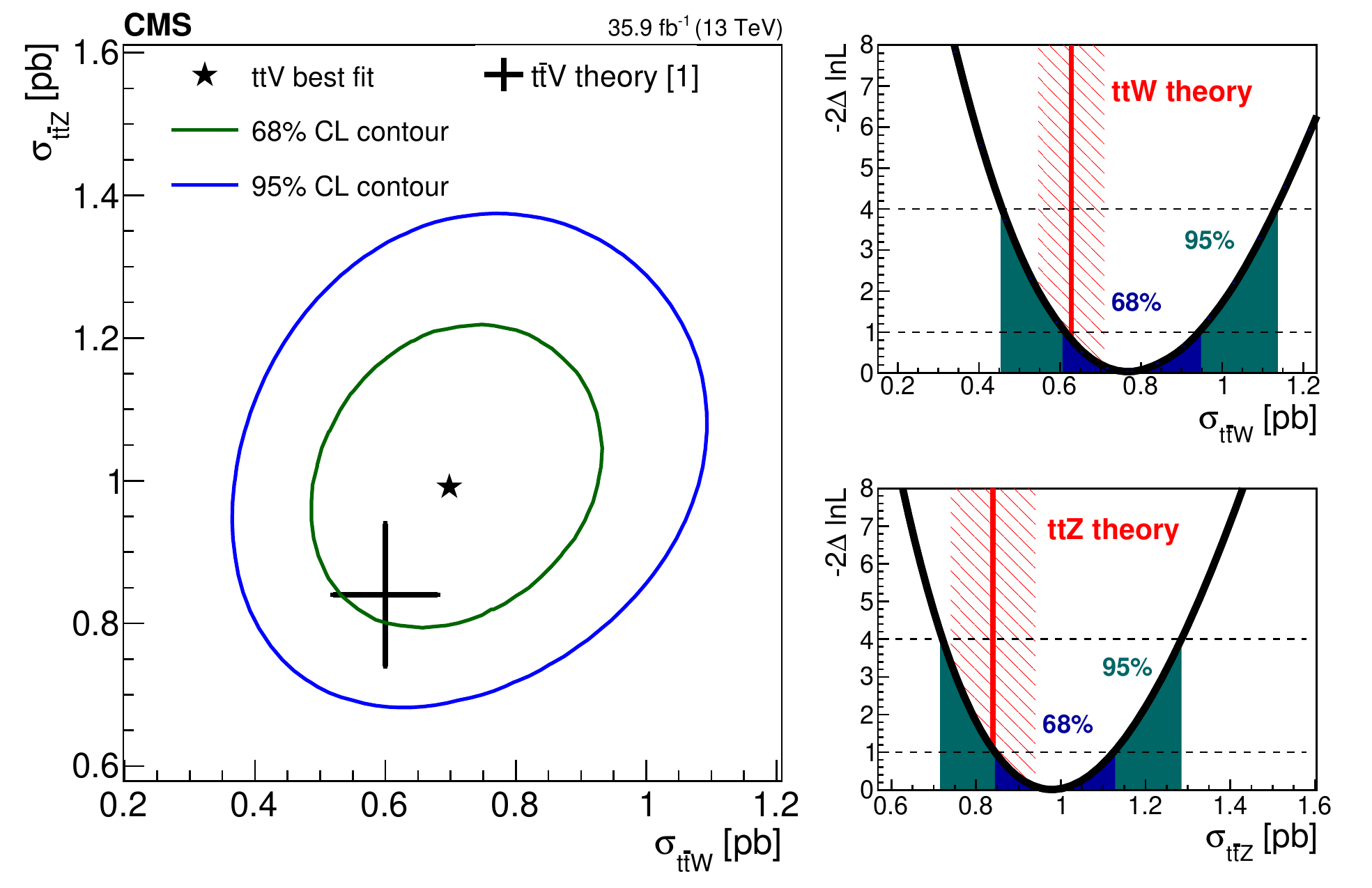}
    \caption{Result of the simultaneous fit for \ttW and \ttZ cross sections (denoted as star), along with its 68 and 95\% CL contours are shown on the left panel. The right panel presents the individual measured cross sections along with the 68 and 95\% CL intervals and the theory prediction~\cite{deFlorian:2016spz} with their respective uncertainties for \ttW and \ttZ. }
    \label{fig:Fit2D}
\end{figure}

The individual measured cross sections for \ttW and \ttZ, as well as the results of a simultaneous fit for these cross sections in all three analysis categories, SS dilepton, three-lepton, and four-lepton, are summarized in Fig.~\ref{fig:Fit2D}. The corresponding 68 and 95\% confidence level (CL)  contours and intervals are also shown. The cross section extracted for \ttZ from the simultaneous fit is identical to the one obtained from the individual measurement, while for \ttW the simultaneous fit result is shifted down by about $ 6\%$, which is smaller than the total systematic uncertainty. This is because the fitted value for the nonprompt background contribution in the three-lepton channel is 9\% higher than the nominal value, so the fitted nonprompt yields in the SS dilepton channel are higher in the combined fit compared to the one in the individual fit.

\section{Effective field theory interpretation}
\label{sec:EFT}
Within the framework of effective field theory, cross section measurements can be used
to search for NP in a model-independent way at energy
scales that are not yet experimentally accessible. Using this
approach, the SM Lagrangian is extended with higher-order operators
that correspond to combinations of SM fields. The extended Lagrangian
is a series expansion in the inverse of the energy scale of the NP,
$1/\Lambda$~\cite{Buchmuller1986}, hence operators are
suppressed as long as $\Lambda$ is large compared with the
experimentally-accessible energy.

The effective Lagrangian is (ignoring the single dimension-five operator, which violates lepton number conservation~\cite{Buchmuller1986})
\begin{linenomath*}
  \begin{equation}
    \mathcal{L}_{\text{eff}} =
    \mathcal{L}_{\text{SM}} +
    \frac{1}{\Lambda^{2}} \sum_ic_i\mathcal{O}_i + \cdots,
  \end{equation}
\end{linenomath*}
where $\mathcal{L}_{\text{SM}}$ is the dimension-four SM Lagrangian,
$\mathcal{O}_i$ are dimension-six
operators, and the ellipsis symbol represents higher-dimension operators. The dimensionless Wilson coefficients $c_i$ parameterize the
strength of the NP interaction.

Assuming baryon and lepton number conservation, there are
fifty-nine independent dimension-six operators~\cite{Grzadkowski2010}.
Thirty-nine of these operators were chosen for study in Ref.~\cite{Alloul2014} because they include at least one Higgs field; the four-fermion operators were omitted.
Constraints on the Wilson coefficients of some dimension-six operators
have been reported in Refs.~\cite{Ellis2014, Whisnant1997, Berger2009, Rontsch2014,
PhysRevD.50.4462, Zhang2012, Tonero2014, JHEP-1601-2016-096, Bylund2016}.

To investigate the effects of NP on any given process, it is necessary
to calculate the expected cross section as a function of the Wilson
coefficients. The matrix element can be written as the sum of SM
and NP components:
\begin{linenomath*}
  \begin{equation}
      \mathcal{M} = \mathcal{M}_0 + \sum_i c_i\mathcal{M}_i.
      \label{matrixelement}
  \end{equation}
\end{linenomath*}

In this work, we consider one operator at a time. The cross section is proportional to the square of the matrix element, and
has the following structure~\cite{Rontsch2014}:
\begin{linenomath*}
\begin{equation}
    \begin{aligned}
      \sigma_{\text{SM+NP}}(c_i) &\propto |\mathcal{M}|^2\\
                                 &\propto s_0 + s_{1i}c_i + s_{2i}c_i^2.
    \label{quadratic_xsec}
    \end{aligned}
\end{equation}
\end{linenomath*}

The coupling structures $s_0$, $s_{1i}$, and $s_{2i}$ are constants which can be determined by evaluating the cross section for at least three values of $c_i$. Note that while $\sigma(c_i)$ is always
quadratic, the minimum is not constrained to appear at the SM value ($c_i=0$), and in cases of destructive interference with the SM, it is possible to have $\sigma_{\text{SM+NP}}(c_i) < \sigma_{\text{SM}}$.

\begin{figure}
  \centering
  \includegraphics[width=0.25\textwidth]{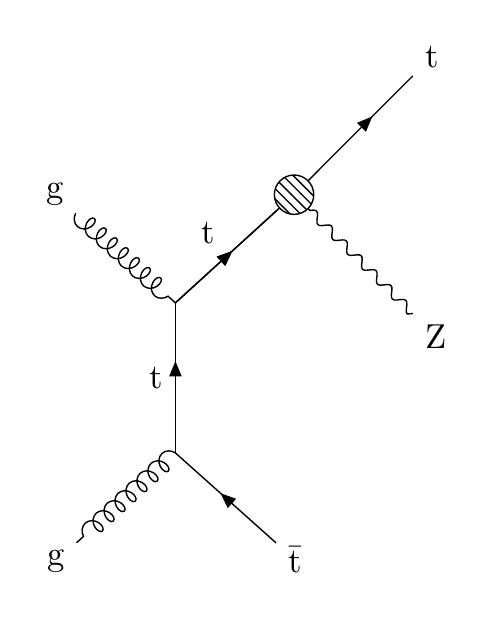}
  \includegraphics[width=0.38\textwidth]{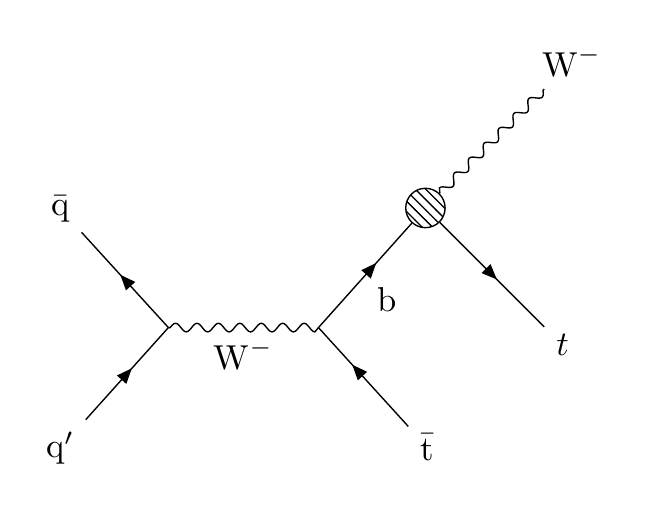}
  \includegraphics[width=0.25\textwidth]{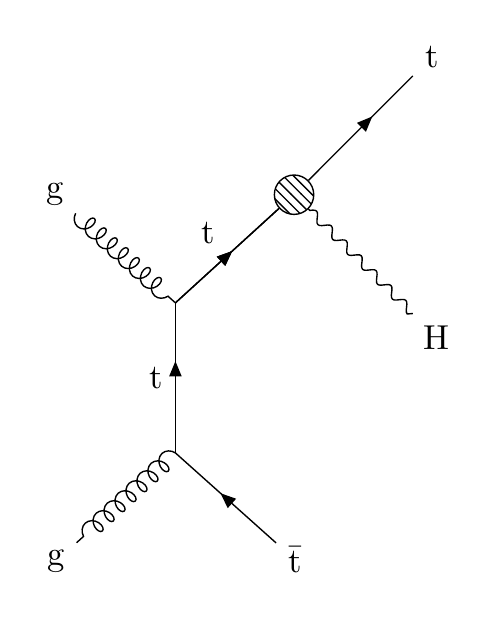}
  \caption{
Feynman diagrams representing some of the most significant NP contributions to the $\ttZ$, $\ttW$, and $\ttH$ processes.}
  \label{fig:diagrams}
\end{figure}

NP effects on $\ttW$ and $\ttZ$ are considered. Because $\ttH$ is sizeable background to $\ttW$, and the NP effects on $\ttH$ are considered as well, as they cannot be disentangled from NP effects on $\ttW$.
The range of Wilson coefficient values to study is chosen
such that $|c_i| < (4\pi)^2$~\cite{DEGRANDE201321}.
The dimension-six operators are encoded using the the
FeynRules~\cite{Alloul2013} implementation from Ref.~\cite{Alloul2014}, and we
follow their notation and operator-naming scheme throughout this
work.  This implementation assumes flavor-independent fermion
couplings. Because the $\PW$ and $\PZ$ boson couplings to light quarks are highly constrained by other measurements, i.e. inclusive W or Z cross section measurements, we removed all NP couplings to the first two generations. This modified implementation
is used in \MGvATNLO~\cite{amcatnlo} to evaluate the cross
section $\sigma_{\text{SM+NP}}$ expected due to both SM and NP
effects at LO, with no constraints on the number of allowed QCD or electroweak vertices, for 30 values of $c_i$, with all other couplings set to
their SM values. We then fit those
points with a quadratic function (see Eq.~(\ref{quadratic_xsec}))
to determine $\sigma_{\text{SM+NP}}(c_i)$.

The signal strength
$r_{\ttZ}(c_i)$ is defined as the ratio of $\sigma_{\text{SM+NP,
\ttZ}}(c_i)$ to $\sigma_{\text{SM+NP, \ttZ}}(0)$, and similarly for
$\ttW$ and $\ttH$.  We use this to construct a profile likelihood test
statistic $q(c_i)$.  The likelihood statistic is maximized to find
the asymptotic best fit $c_i$, similarly to the procedure described in
Section~\ref{sec:Results}.  Each coupling is profiled with the other
couplings set to their SM values.

Of the thirty-nine operators in Ref.~\cite{Alloul2014}, we choose not to consider
operators that do not affect \ttW, \ttZ, or \ttH. The expected 95\% CL interval is calculated for the remaining 24 operators.
We also exclude from consideration operators that produce large effects
in better-measured processes for Wilson coefficient values to which our
measurement is sensitive. To accomplish this, we require that the cross section for each of
$\ttbar$, $\PW\PW$, $\PZ\PZ$, $\PW\PZ$, and inclusive Higgs boson production is not modified
by more than 70\% within our expected 95\% CL interval. Finally, we do not include any operators that produce a significant
effect on background yields (as described in Section~\ref{sec:backgrounds}) other than \ttH, as these can be studied more effectively in
other signal regions.

Eight operators satisfy the above requirements, and constraints on
their Wilson coefficients, \cuW, \cH, \tcthreeG, \cthreeG, \cuG, \cuB, \cHu, and
\ctwoG are reported here.
Feynman diagrams representing some of the most significant NP contributions to the $\ttZ$, $\ttW$, and $\ttH$ processes are shown in Fig.~\ref{fig:diagrams}.

\begin{figure}
\centering
      \includegraphics[width=0.3\textwidth]{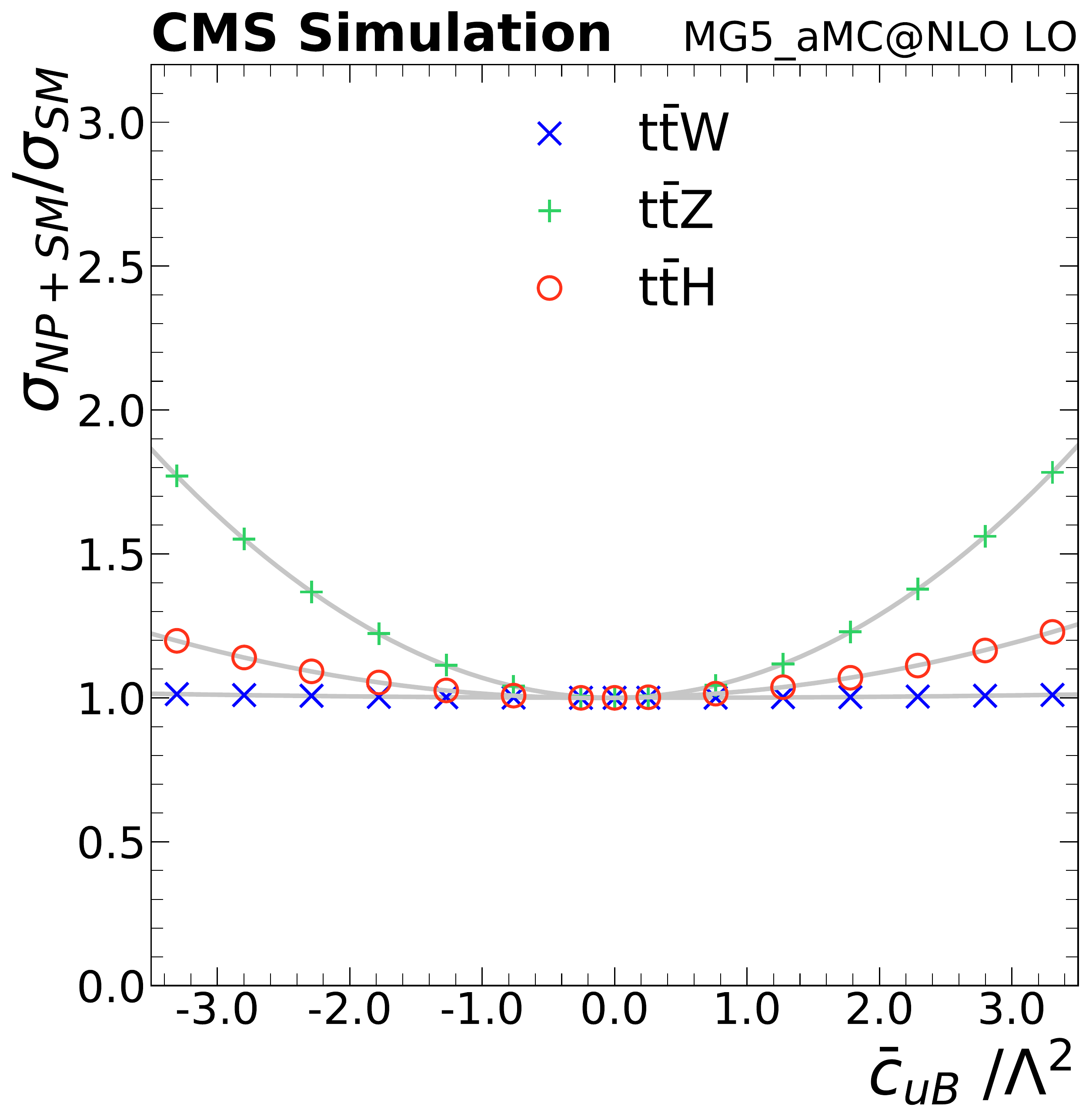}
      \includegraphics[width=0.3\textwidth]{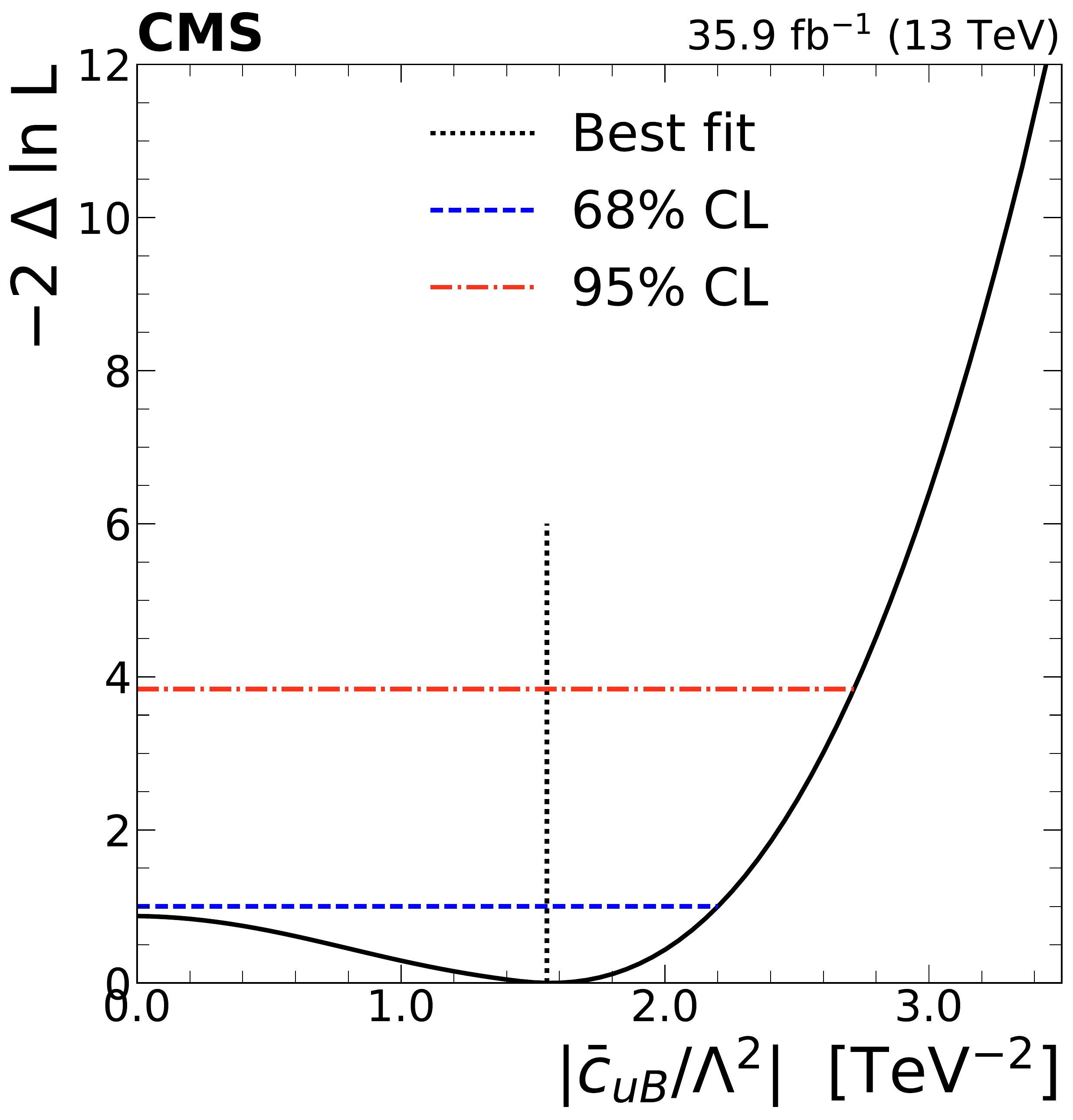}
      \includegraphics[width=0.315\textwidth]{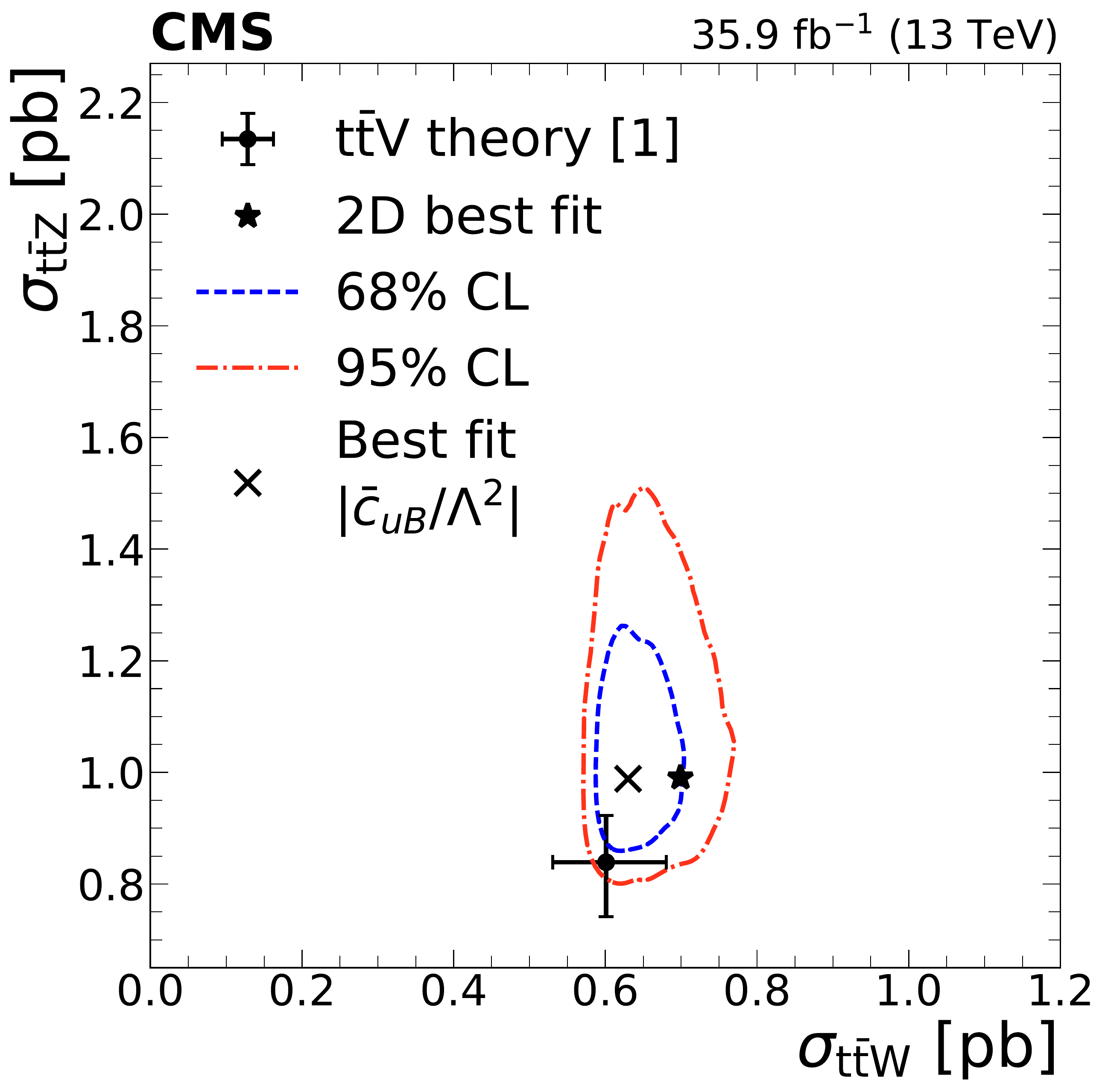}
      \includegraphics[width=0.3\textwidth]{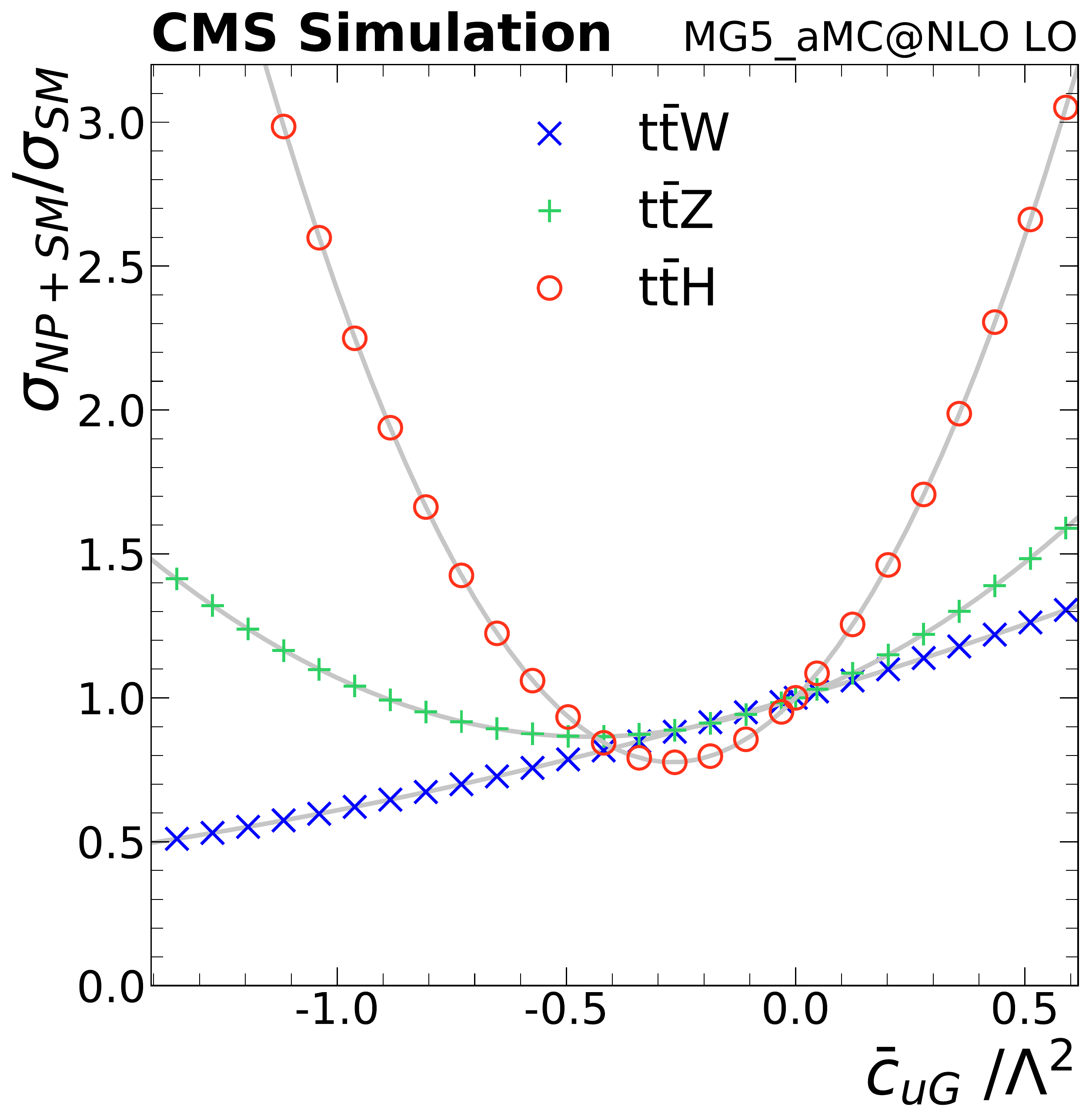}
      \includegraphics[width=0.3\textwidth]{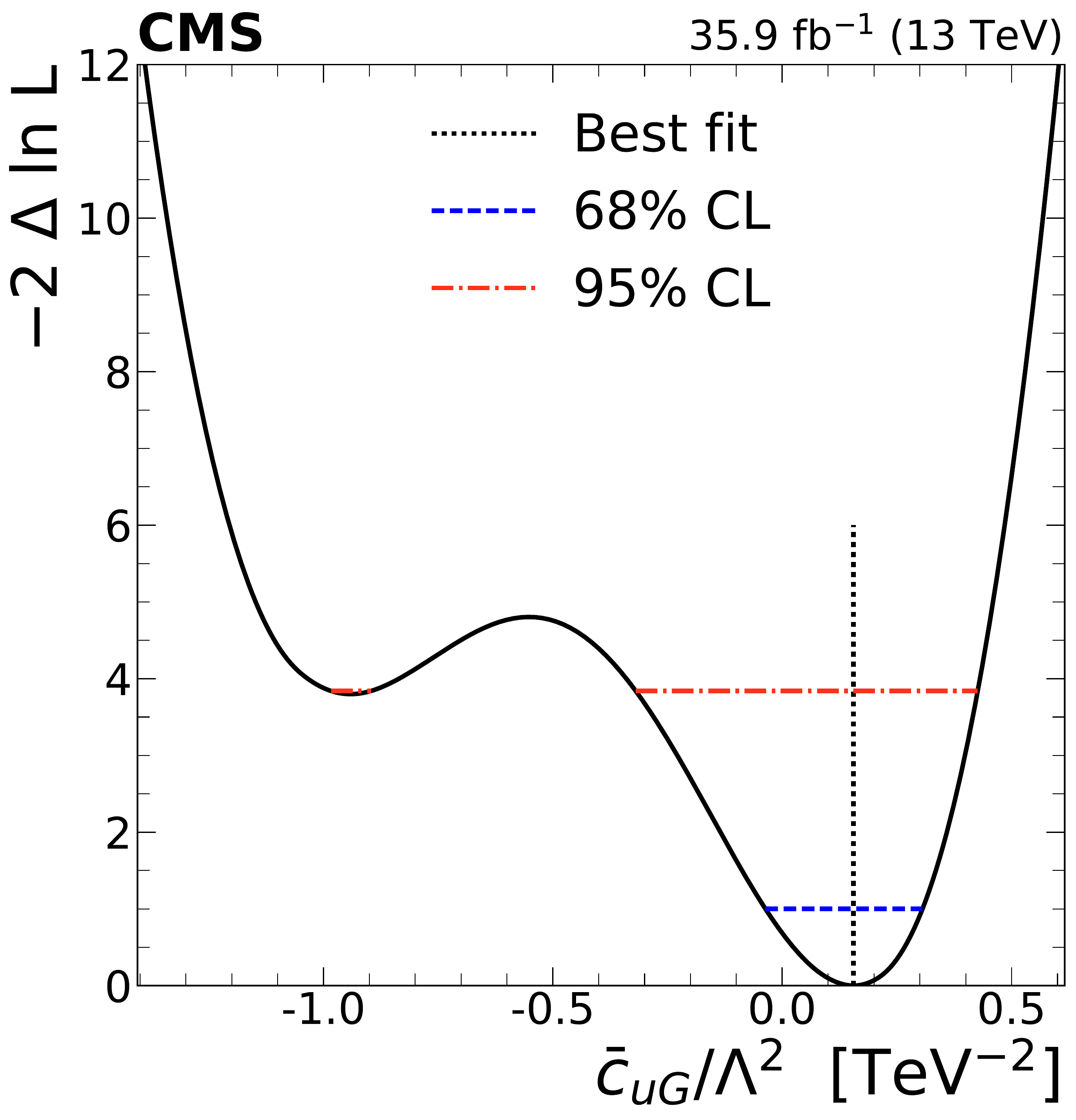}
      \includegraphics[width=0.315\textwidth]{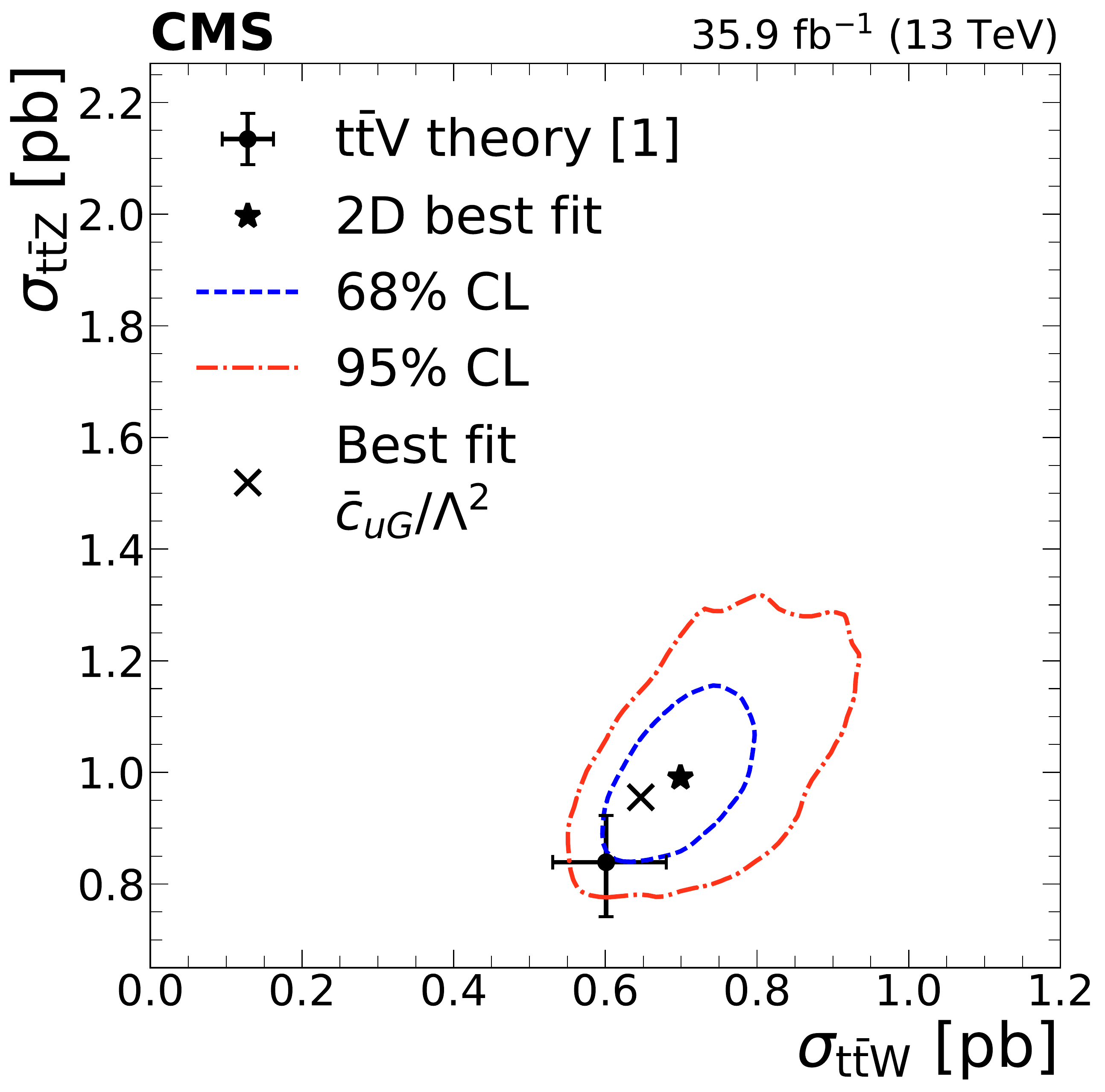}
      \includegraphics[width=0.3\textwidth]{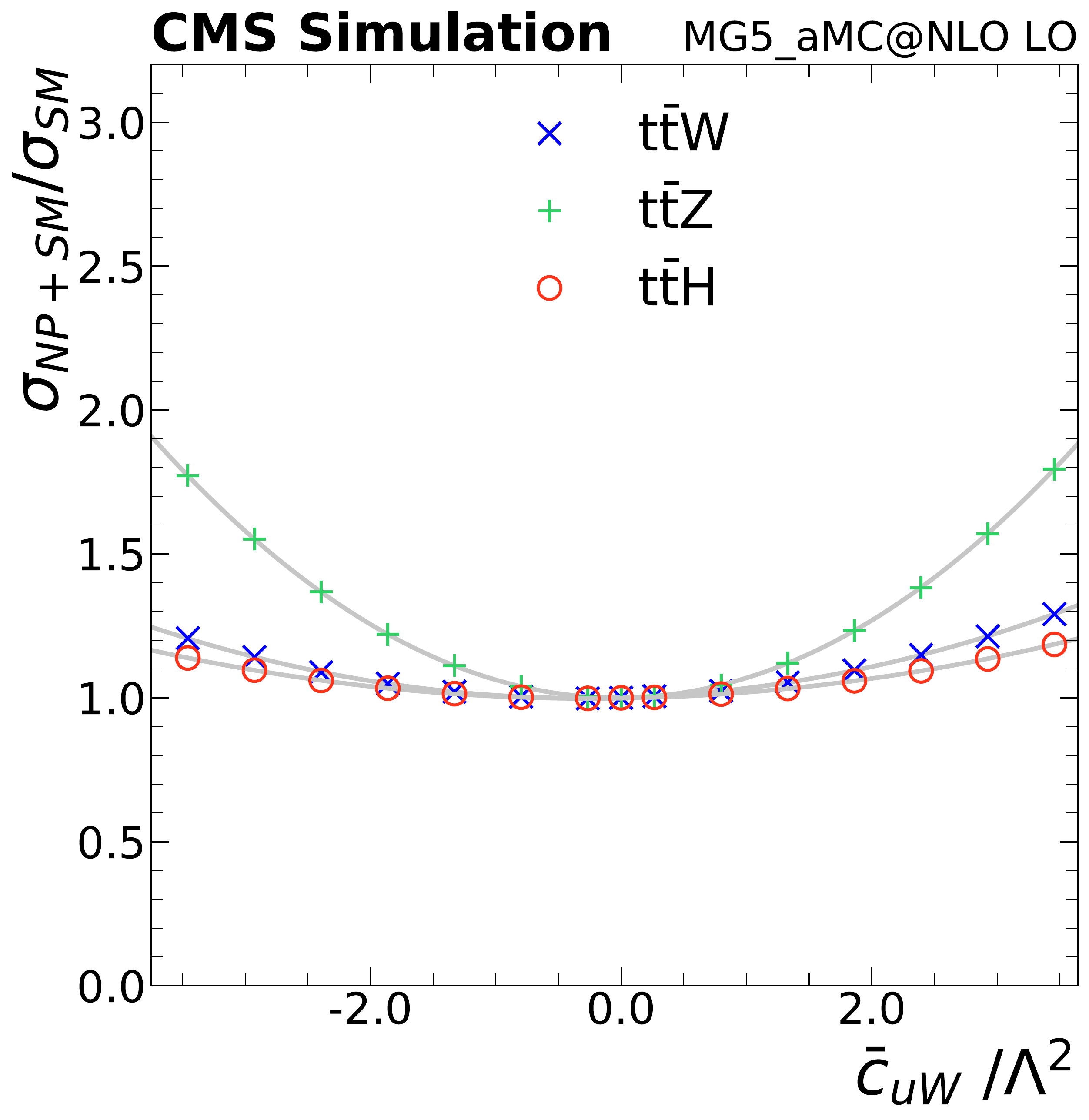}
      \includegraphics[width=0.3\textwidth]{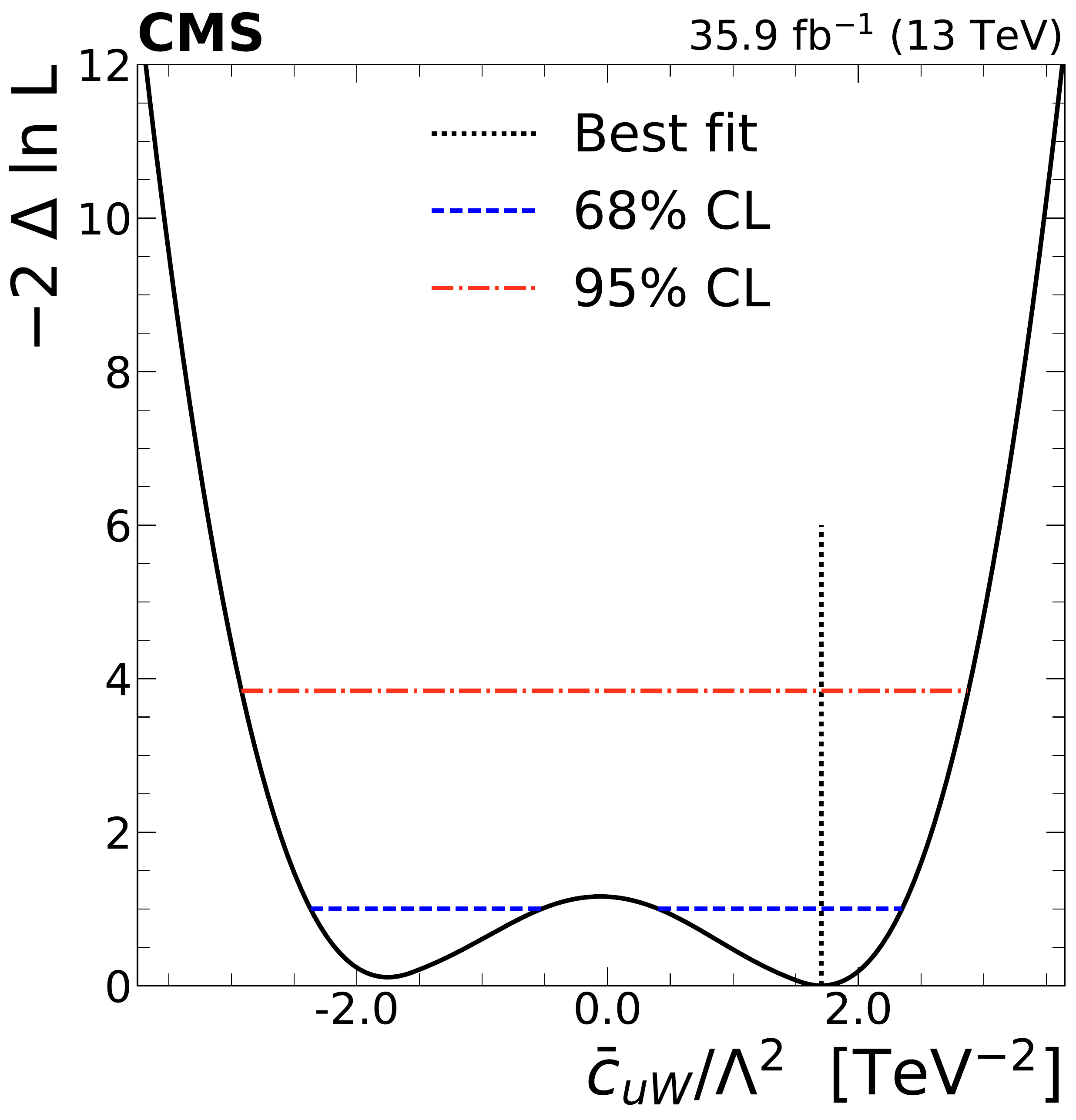}
      \includegraphics[width=0.315\textwidth]{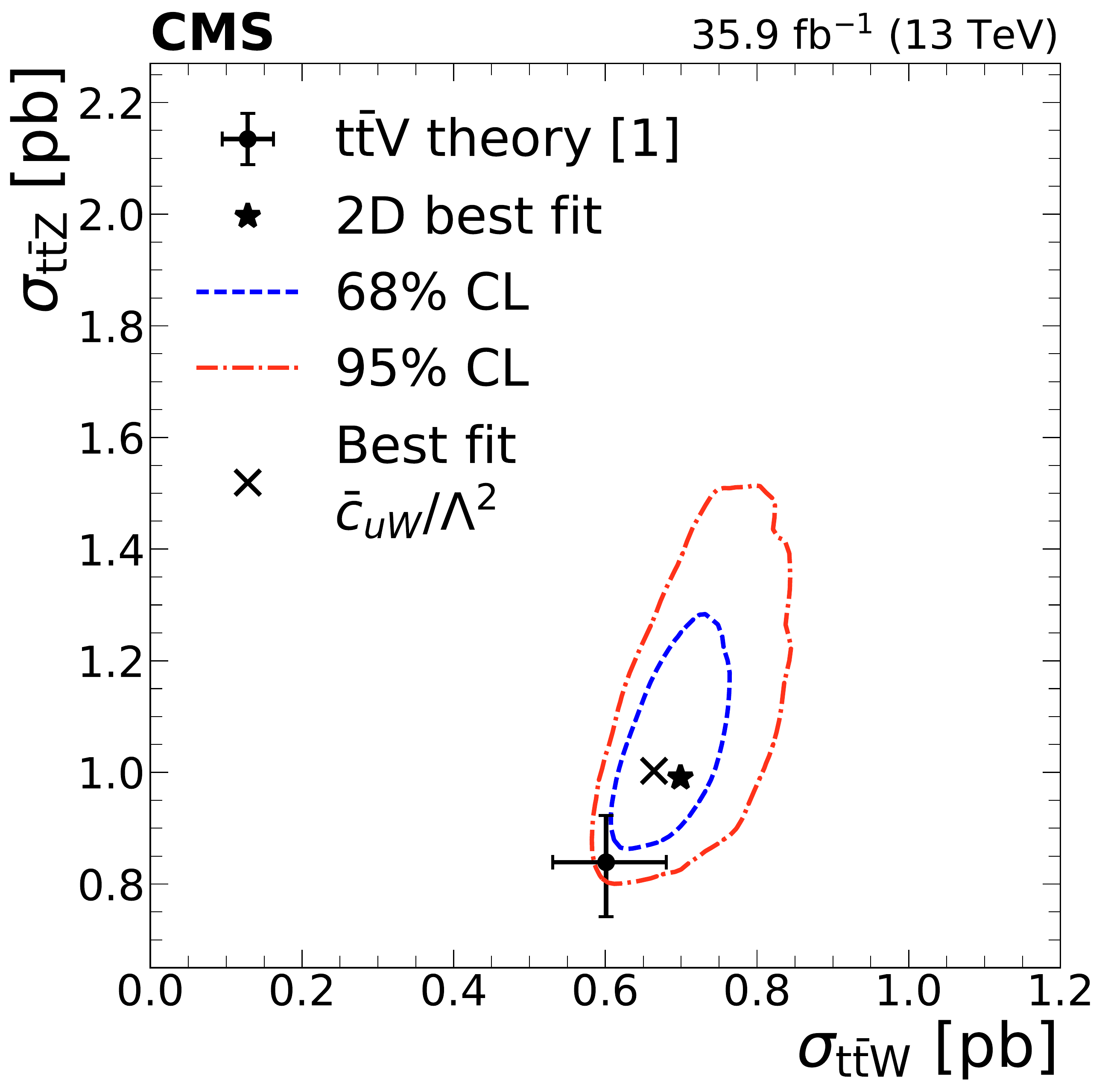}
  \caption{\label{fig:results} Left: signal strength as a function of selected
  Wilson coefficients for $\ttW$ (crosses), $\ttZ$ (pluses), and $\ttH$ (circles). Center: the 1D test statistic $q(c_i)$ scan as a function of $c_i$,  profiling all other nuisance parameters. The global best fit value is indicated by a dotted line. Dashed and dash-dotted lines indicate 68\% and 95\% CL intervals, respectively. Right: The $\ttZ$ and $\ttW$ cross section corresponding to the global best fit $c_i$ value is shown as a cross, along with the corresponding 68\% (dashed) and 95\% (dash-dotted) contours. The two-dimensional best fit to the $\ttW$ and $\ttZ$ cross sections is given by the star. The theory predictions~\cite{deFlorian:2016spz} for \ttW and \ttZ are shown as a dot with bars representing their respective uncertainties.}
\end{figure}

The expected CL intervals for the selected Wilson coefficients are summarized in
Table~\ref{tab:D6_expected_results}. Observed best fit values and CL intervals
are summarized in Table~\ref{tab:D6_results}. For three representative
operators, the calculated signal strengths $r_{\ttZ}(c_i)$, $r_{\ttW}(c_i)$, and
$r_{\ttH}(c_i)$ are shown in the left panels of Fig.~\ref{fig:results}. The profile likelihood scan
is presented in the center panels. In the right panels, results are shown in the
$\sigma_{\ttZ}$ versus $\sigma_{\ttW}$ plane. The 68\% and 95\% contours
are obtained by sampling randomly from the fitted covariance matrix and
extracting the contours which enclose 68.27\% and 95.45\% of the samples.  We remove
any assumptions about the energy scale of the NP made in Ref.~\cite{Alloul2014} and
report the ratio $c_i/\Lambda^2$. In cases where $\sigma_{\text{SM+NP}}(c_i)$ has the same minimum for all three processes, the profile likelihood is symmetric around this point, and we present results for $|c_i - c_{i,\text{min}}|$ to make this symmetry explicit.

\begin{table}
\topcaption{\label{tab:D6_expected_results} Expected 68\% and 95\% CL
  intervals for selected Wilson coefficients.}
\centering
  \begin{tabular}{lll}
    Wilson coefficient                           &   68\% CL $[\TeVns^{-2}]$  & 95\% CL $[\TeVns^{-2}]$  \\
    \hline
 $\bar{c}_{\text{uW}}/\Lambda^2$                             & $[-1.6, 1.5]$                  & $[-2.2, 2.2]$     \\
 $|\bar{c}_{\text{H}}/\Lambda^2 - 16.8\TeV^{-2}|$  & $[3.7, 23.4]$                  & $[0, 28.7]$     \\
 $\widetilde{c}_{\text{3G}}/\Lambda^2$                       & $[-0.5, 0.5]$                  & $[-0.7, 0.7]$     \\
 $\bar{c}_{\text{3G}}/\Lambda^2$                             & $[-0.3, 0.7]$                  & $[-0.5, 0.9]$     \\
 $\bar{c}_{\text{uG}}/\Lambda^2$                             & $[-0.9, -0.8]$ and $[-0.3, 0.2]$ & $[-1.1, 0.3]$     \\
 $|\bar{c}_{\text{uB}}/\Lambda^2|$                           & $[0, 1.5]$     & $[0, 2.1]$      \\
 $\bar{c}_{\text{Hu}}/\Lambda^2$                             & $[-9.2, -6.5]$ and $[-1.6, 1.1]$  & $[-10.1, 2.0]$    \\
 $\bar{c}_{\text{2G}}/\Lambda^2$                             & $[-0.7, 0.4]$                  & $[-0.9, 0.6]$     \\
  \end{tabular}
\end{table}

\begin{table}[ht!b]
\topcaption{\label{tab:D6_results} Observed best fit values for selected Wilson
    coefficients determined from this $\ttW$ and $\ttZ$ measurement, along with corresponding 68\% and 95\% CL intervals.
  In some cases the profile likelihood shows another local minimum that cannot be excluded; the number reported here is the global minimum.}
\centering
\resizebox{1.0\linewidth}{!}{
\begin{tabular}{llll}
Wilson coefficient                                  &   Best fit $[\TeVns^{-2}]$ & 68\% CL $[\TeVns^{-2}]$ & 95\% CL $[\TeVns^{-2}]$                \\
\hline
 $\bar{c}_{\text{uW}}/\Lambda^2$                            &        1.7 & $[-2.4, -0.5]$ and $[0.4, 2.4]$   & $[-2.9, 2.9]$                   \\
 $|\bar{c}_{\text{H}}/\Lambda^2 - 16.8\ \mathrm{TeV}^{-2}|$ &       15.6 & $[0, 23.0]$                   & $[0, 28.5]$                   \\
 $|\widetilde{c}_{\text{3G}}/\Lambda^2|$                    &        0.5 & $[0, 0.7]$                    & $[0, 0.9]$                    \\
 $\bar{c}_{\text{3G}}/\Lambda^2$                            &       $-0.4$ & $[-0.6, 0.1]$ and $[0.4, 0.7]$    & $[-0.7, 1.0]$                   \\
 $\bar{c}_{\text{uG}}/\Lambda^2$                            &        0.2 & $[0, 0.3]$                   & $[-1.0, -0.9]$ and $[-0.3, 0.4]$  \\
 $|\bar{c}_{\text{uB}}/\Lambda^2|$                          &        1.6 & $[0, 2.2]$                    & $[0, 2.7]$                    \\
 $\bar{c}_{\text{Hu}}/\Lambda^2$                            &       $-9.3$ & $[-10.3, -8.0]$ and $[0, 2.1]$ & $[-11.1, -6.5]$ and $[-1.6, 3.0]$ \\
 $\bar{c}_{\text{2G}}/\Lambda^2$                            &        0.4 & $[-0.9, -0.3]$ and $[-0.1, 0.6]$  & $[-1.1, 0.8]$                   \\
\end{tabular}
}
\end{table}

\nocite{Ellis:2016jkw}

\section{Summary}
\label{sec:Conclusions}
A measurement of top quark pair production in association with a $\PW$ or a $\PZ$ boson using proton-proton collisions at 13 TeV is presented. The analysis is performed in the same-sign dilepton final state for \ttW, and the three- and four-lepton final states for \ttZ, and these three final states are used to extract the cross sections for \ttW and \ttZ production. For both processes the observed signal significance exceeds 5 standard deviations. The measured signal strength parameters are
$1.23^{+0.19}_{-0.18}\stat^{+0.20}_{-0.18}\syst^{+0.13}_{-0.12}\thy$ and  $1.17~^{+0.11}_{-0.10}\stat^{+0.14}_{-0.12}\syst^{+0.11}_{-0.12}\thy$ for \ttW and \ttZ, respectively.
The measured cross sections are $\sigma(\ttW)=0.77^{+0.12}_{-0.11}\stat^{+0.13}_{-0.12}\syst\unit{pb}$ and
$\sigma(\ttZ)=0.99^{+0.09}_{-0.08}\stat^{+0.12}_{-0.10}\syst\unit{pb}$, in agreement with the standard model predictions. These results have
been used to set constraints on the Wilson coefficients of dimension-six operators. Eight operators have been identified which are of particular interest because they change the expected cross sections of \ttZ, \ttW, or \ttH without significantly impacting expected background yields. Both \ttZ and \ttH are affected by $\OthreeG$, $\OtthreeG$, $\OtwoG$, and $\OuB$. Only \ttZ is affected by $\OHu$, while $\OH$ affects only \ttH. All three processes \ttZ, \ttW, and \ttH are affected by $\OuG$ and $\OuW$. In cases where new physics beyond the standard model modifies the expected \ttZ cross section, the sensitivity is mainly determined by \ttZ and the fit is able to match the observed excess in data. No operators were identified which provide an independent handle on \ttW. The constraints presented, obtained by considering one operator at a time, are a useful first step toward more global approaches.

\begin{acknowledgments}
We congratulate our colleagues in the CERN accelerator departments for the excellent performance of the LHC and thank the technical and administrative staffs at CERN and at other CMS institutes for their contributions to the success of the CMS effort. In addition, we gratefully acknowledge the computing centers and personnel of the Worldwide LHC Computing Grid for delivering so effectively the computing infrastructure essential to our analyses. Finally, we acknowledge the enduring support for the construction and operation of the LHC and the CMS detector provided by the following funding agencies: BMWFW and FWF (Austria); FNRS and FWO (Belgium); CNPq, CAPES, FAPERJ, and FAPESP (Brazil); MES (Bulgaria); CERN; CAS, MoST, and NSFC (China); COLCIENCIAS (Colombia); MSES and CSF (Croatia); RPF (Cyprus); SENESCYT (Ecuador); MoER, ERC IUT, and ERDF (Estonia); Academy of Finland, MEC, and HIP (Finland); CEA and CNRS/IN2P3 (France); BMBF, DFG, and HGF (Germany); GSRT (Greece); OTKA and NIH (Hungary); DAE and DST (India); IPM (Iran); SFI (Ireland); INFN (Italy); MSIP and NRF (Republic of Korea); LAS (Lithuania); MOE and UM (Malaysia); BUAP, CINVESTAV, CONACYT, LNS, SEP, and UASLP-FAI (Mexico); MBIE (New Zealand); PAEC (Pakistan); MSHE and NSC (Poland); FCT (Portugal); JINR (Dubna); MON, RosAtom, RAS, RFBR and RAEP (Russia); MESTD (Serbia); SEIDI, CPAN, PCTI and FEDER (Spain); Swiss Funding Agencies (Switzerland); MST (Taipei); ThEPCenter, IPST, STAR, and NSTDA (Thailand); TUBITAK and TAEK (Turkey); NASU and SFFR (Ukraine); STFC (United Kingdom); DOE and NSF (USA).

\hyphenation{Rachada-pisek} Individuals have received support from the Marie-Curie program and the European Research Council and Horizon 2020 Grant, contract No. 675440 (European Union); the Leventis Foundation; the A. P. Sloan Foundation; the Alexander von Humboldt Foundation; the Belgian Federal Science Policy Office; the Fonds pour la Formation \`a la Recherche dans l'Industrie et dans l'Agriculture (FRIA-Belgium); the Agentschap voor Innovatie door Wetenschap en Technologie (IWT-Belgium); the Ministry of Education, Youth and Sports (MEYS) of the Czech Republic; the Council of Science and Industrial Research, India; the HOMING PLUS program of the Foundation for Polish Science, cofinanced from European Union, Regional Development Fund, the Mobility Plus program of the Ministry of Science and Higher Education, the National Science Center (Poland), contracts Harmonia 2014/14/M/ST2/00428, Opus 2014/13/B/ST2/02543, 2014/15/B/ST2/03998, and 2015/19/B/ST2/02861, Sonata-bis 2012/07/E/ST2/01406; the National Priorities Research Program by Qatar National Research Fund; the Programa Severo Ochoa del Principado de Asturias; the Thalis and Aristeia programs cofinanced by EU-ESF and the Greek NSRF; the Rachadapisek Sompot Fund for Postdoctoral Fellowship, Chulalongkorn University and the Chulalongkorn Academic into Its 2nd Century Project Advancement Project (Thailand); the Welch Foundation, contract C-1845; and the Weston Havens Foundation (USA).
\end{acknowledgments}

\bibliography{auto_generated}
\cleardoublepage \appendix\section{The CMS Collaboration \label{app:collab}}\begin{sloppypar}\hyphenpenalty=5000\widowpenalty=500\clubpenalty=5000\vskip\cmsinstskip
\textbf{Yerevan~Physics~Institute,~Yerevan,~Armenia}\\*[0pt]
A.M.~Sirunyan, A.~Tumasyan
\vskip\cmsinstskip
\textbf{Institut~f\"{u}r~Hochenergiephysik,~Wien,~Austria}\\*[0pt]
W.~Adam, F.~Ambrogi, E.~Asilar, T.~Bergauer, J.~Brandstetter, E.~Brondolin, M.~Dragicevic, J.~Er\"{o}, A.~Escalante~Del~Valle, M.~Flechl, M.~Friedl, R.~Fr\"{u}hwirth\cmsAuthorMark{1}, V.M.~Ghete, J.~Grossmann, J.~Hrubec, M.~Jeitler\cmsAuthorMark{1}, A.~K\"{o}nig, N.~Krammer, I.~Kr\"{a}tschmer, D.~Liko, T.~Madlener, I.~Mikulec, E.~Pree, N.~Rad, H.~Rohringer, J.~Schieck\cmsAuthorMark{1}, R.~Sch\"{o}fbeck, M.~Spanring, D.~Spitzbart, W.~Waltenberger, J.~Wittmann, C.-E.~Wulz\cmsAuthorMark{1}, M.~Zarucki
% ...
\vskip\cmsinstskip
% ...
1:~Also at~Vienna~University~of~Technology,~Vienna,~Austria\\
% ...
\end{sloppypar}
\end{document}